%% file: ms.tex
\documentclass[a4paper,fleqn,usenatbib]{mnras}

\usepackage{graphicx}	
\usepackage{amsfonts}
\usepackage{mathtools}
\usepackage{dsfont}
\usepackage{amsmath}	
\usepackage{amssymb}	
\usepackage{multicol}        
\usepackage{bm}		
\usepackage{pdflscape}	
\usepackage{dcolumn}   
\usepackage{float} 
\usepackage{grffile}
\usepackage{color}
\usepackage[dvipsnames]{xcolor}
\usepackage{mathtools}

\usepackage[T1]{fontenc}
\usepackage{ae,aecompl}

\usepackage{newtxtext}

\usepackage{tikz}

\input{defs.tex}

\title[Fast estimation of aperture-mass statistics I] {Fast estimation
  of aperture mass statistics I: aperture mass variance and an application to the \cfhtls data}

      \author[Porth et al.]{
        Lucas Porth$^{1}$\thanks{l.porth@sussex.ac.uk}, 
        Robert E. Smith$^{1}$\thanks{r.e.smith@sussex.ac.uk},
        Patrick Simon$^{2}$,
        Laura Marian$^{1}$ and
        Stefan Hilbert$^{3}$
\\
$^{1}$Astronomy Centre, Department of Physics \& Astronomy, University of Sussex, Brighton, BN1 9RH, UK\\
$^{2}$Argelander-Institut f\"ur Astronomie, Universit\"at Bonn, Auf dem H\"ugel 71, 53121 Bonn, Germany\\
$^{3}$Max-Planck-Institut f\"ur Astrophysik, Karl-Schwarzschild-Str. 1, 85748 Garching, Germany\\
}
      
\date{\today}

\pubyear{\today}

\begin{document}

\label{firstpage}

\pagerange{\pageref{firstpage}--\pageref{lastpage}}

\maketitle


\begin{abstract}
  We explore an alternative method to the usual shear correlation
  function approach for the estimation of aperture mass statistics in
  weak lensing survey data. Our approach builds on the direct
  estimator method of \citet{Schneideretal1998}. In this paper, to test and
  validate the methodology, we focus on the aperture mass
  dispersion. After computing the signal and noise for a weighted set
  of measured ellipticites we show how the direct estimator can be
  made into a linear order algorithm that enables a fast and efficient
  computation. We then investigate the applicability of the direct estimator approach in the presence of a real survey mask with holes and chip gaps. For this we use a large ensemble of full ray-tracing mock simulations. By using various weighting schemes for combining information from different apertures we find that inverse variance weighting the individual aperture estimates with an aperture completeness greater than 70\% yields an answer
  that is in close agreement with the standard correlation function
  approach. We then apply this approach to the \cfhtls 
  as pilot scheme and find that our method recovers to high accuracy
  the \citet{Kilbingeretal2013} result for the variance of both the E and
  B mode signal, after we correct the catalogue for the shear bias in
  the \lensfit algorithm for pairs closer than 9''. We then explore the
  cosmological information content of the direct estimator using the
  Fisher information approach. We show that there is a only modest
  loss in cosmological information from the rejection of apertures
  that are of low completeness. This method unlocks the door to fast
  and efficient methods for recovering higher order aperture mass
  statistics in linear order operations.

\end{abstract}




\section{Introduction}\label{sec:intro}

Weak gravitational lensing of the light from galaxies is a key tool
for constraining the cosmological parameters and distinguishing
between competing models of the Universe
\citep{Blandfordetal1991,Kaiser1998,ZhangPetal2007}.  The first
measurements of the correlations in the shapes of distant background
galaxy images are now over two decades old
\citep{Baconetal2000,Kaiseretal2000,VanWaerbekeetal2000,Wittmanetal2000}
and the field of cosmic shear has rapidly matured from these early
pioneering studies that mapped of the order a square degree, to the
modern surveys \kids\footnote{{\tt
    kids.strw.leidenuniv.nl}}, DES\footnote{{\tt
    www.darkenergysurvey.org}}  and HSC\footnote{{\tt
    hsc.mtk.nao.ac.jp/ssp/}}, which are mapping thousands of square
degrees
\citep{Hildebrandtetal2017,Troxeletal2018,HSC2018,Hikageetal2019}.
The next decade will herald in new surveys like \euclid\footnote{{\tt
    www.cosmos.esa.int/web/euclid}} and LSST\footnote{{\tt
    www.lsst.org}} that will map volumes close to the entire physical
volume of our observable Universe \citep{Euclid2011,LSST2009}.  This
will mean that our ability to extract information from such rich data
sets will depend almost entirely on our ability to understand and
model the complex nonlinear physics involved and our ability to
optimally correct or mitigate the systematic errors.

In the last decade, much effort has been invested in extracting
cosmological information from the two-point shear correlation
functions, and attempts have been made to carefully account for all
systematic effects, such as PSF corrections, bias in the ellipticiy
estimator, intrinsic alignments 
\citep{Schneider2006p3,Masseyetal2013,TroxelIshak2015}. The two-point
shear correlation functions are the lowest order statistics that are
of interest and if the convergence field were a Gaussian random field,
then they would contain a complete description of the statistical
properties of the cosmic shear signal. However, the distribution of
observed galaxy ellipticities are non-Gaussian due to various effects:
firstly, the nonlinear growth of large-scale structure induces the
coupling of density modes on different scales
\citep{Schneideretal1998}; secondly, the estimator for shear from
ellipticity is a nonlinear mapping; thirdly, the violation of the Born
approximation and the lens-lens coupling also lead to non-Gaussianity
in the shear maps. This all leads to a `flow' of information into the
higher order statistics \citep{TaylorWatts2001}. A consequence of this
is that the errors on measurements of the convergence power spectrum
become highly correlated on small scales, limiting the amount of
additional information that can be recovered by pushing down to
smaller scales
\citep{Satoetal2011,Hilbertetal2012,Kayoetal2013,Marianetal2013}.

The need to go beyond the simple two-point analysis of the data has
been highlighted by a number of authors \citep[see for
  example][]{Sefusattietal2006,Byunetal2017}. For example, it is well
known that $\xi_+$ and $\xi_-$ exhibit a degeneracy between the
amplitude of matter fluctuations $\sigma_8$ and the matter density
parameter $\Omega_{\rm m}$, which scales as $\sigma_8\Omega_{\rm
  m}^{0.5}$. One way to break this degeneracy is by combining the
information from the 2-point and 3-point shear correlation functions
\citep{KilbingerSchneider2005,Sembolonietal2011,Fuetal2014}; another
way is through adding in the information found in the statistical
properties of the peaks in the shear field
\citep{Marianetal2013,Kacprzaketal2016}. Given the potential of the
non-Gaussian probes to tighten constraints on cosmology and break
model and nuisance parameter degeneracies, it is important to study
how to optimally measure them, determine how systematics affect them,
and to improve the modelling of them. This work will be essential to
undertake, if we are to take full advantage of surveys like \kids, DES,
HSC, \euclid \ and LSST.

One of the bottlenecks for accessing the information in the
higher-order statistics is that they are challenging quantities to work
with. For example, owing to the fact that the shear is a spin-2 field,
there are in principle $2^{n}$ correlation functions to measure for
each $n$th order cumulant
\citep{SchneiderLombardi2003,TakadaJain2003,Jarvisetal2004,KilbingerSchneider2005}.
Building the necessary computational tools to measure the 3- and
4-point shear correlation functions is technically challenging and
will require large amounts of CPU time to compute all possible
configurations \citep{Jarvisetal2004,Kilbingeretal2014}. This is
especially true if measurement-noise covariance matrices are to be derived from mock
catalogues.  In addition, the shear correlation functions are not
necessarily the best quantity to measure since they are not E/B mode
decomposed
\citep{Schneideretal2002a,SchneiderKilbinger2007}.

A powerful method to disentangle systematic effects from cosmic shear
signals is the E/B decomposition
\citep{Crittendenetal2001,Schneideretal2002a}. At leading order, pure
weak lensing signals are sourced by a scalar lensing potential, which
means that their deflection fields are curl free.  Equivalently, the
ring-averaged cross component of the shear is expected to be zero (the
B mode), while the tangential one contains all the lensing signal (the
E mode). Thus B modes enable a robust test for the presence of
systematic errors. One method to take advantage of this E/B
decomposition is the so-called `aperture mass statistics' first
introduced by \citet{Schneideretal1998}. `Aperture mass' ($\Map$) and
`Map-Cross' $(\Mx)$ are obtained by convolving the tangential
and cross shear with an isotropic filter function. Therefore
by construction they are E/B-decomposed. Taking the second moment
leads to the variance of aperture mass, the third to its skewness etc.

The standard approach for measuring the aperture mass statistics in
data utilises the fact that, for the flat sky, any $n$-point moment
can be expressed in terms of integrals over the $n$-point shear
correlation functions, modulo a kernel function
\citep{Schneideretal2002a,Jarvisetal2004}. The reason for adopting
this strategy stems from the fact that for a real weak lensing survey,
the survey mask is a very complicated function: firstly there are
survey edges; next, due to the fact that bright stars and their
diffraction halo need to be drilled out, chip gaps, if not accounted
for in the survey dither pattern, can lead to additional holes. This
small-scale structure in the survey mask means that in order to make
the most of the survey data one should measure the correlation
functions.  However, this approach is not without issue: for example,
for the correlation function estimator of the aperture mass dispersion to be accurate and
E/B decomposed, one needs to measure $\xi_+$ and $\xi_-$ in angular
bins sufficiently fine for the discretisation of the integrals to be
reliable \citep{Fuetal2014}.  Further, one also needs to measure the
correlation function on scales $\vartheta\in[0,2\vartheta]$ for the
polynomial filter function of \citet{Schneideretal1998}. Owing to
galaxy image blending, signal-to-noise issues and the finite size of
the survey, the lower bound is never possible and the upper bound
means that biases can occur due to edge effects. This leads to so
called E/B leakage \citep{KilbingerSchneider2005}. In addition, while
the mean estimate is unbiased, the covariance matrix does require one
to carefully account for the mask
\citep{Schneideretal2002b,Friedrichetal2016}. More recent developments that
also make use of the shear correlation functions, while circumventing the 
issues of E/B leakage on small scales are the ring statistics and COSEBIs
\citep{SchneiderKilbinger2007,Schneideretal2010}.

In this paper, we take a different approach and explore the direct
estimators of the aperture mass statistics, which were first proposed
in \citet{Schneideretal1998}. Rather than measuring the correlation
functions of the shear polar, only to reduce them by integration to a
scalar, we instead directly measure $\Map$ for a set of apertures and
then use an optimised weighting scheme to average the estimates. As we
will show in what follows, this approach has some significant
advantages over the correlation function approach.  In addition to the
variance, one can also measure higher order statistics, such as the skewness and
kurtosis, with very little additional computational complexity, code
modification or CPU expense (see Porth et al. in preparation).  These
efficiencies will also potentially enable fast computation of
covariance matrices and thus rapid exploration of the likelihood
surface for such statistics.  The possible down sides to this
approach, which we explore, are the potential loss of cosmological
information arising due to the fact that some incomplete apertures
will be rejected. On the other hand, we will also explore the
possibility of not rejecting all incomplete apertures, but
accepting/weighting apertures based on criteria such as coverage
factor and the signal-to-noise. This will lead to E/B leakage,
however, as we will show the levels of leakage can be made
sufficiently small so that the statistic is accurate within the
required errors. As a practical
demonstration of this approach we apply it to the \cfhtls data and
present a careful comparison of it with the two-point correlation
function method. Lastly, we make use of a large suite of mock
catalogues to study the cosmological information content of the two
methods for a nominal \cfhtls like survey and show that there is no
substantial loss of information. 

The paper breaks down as follows: In \S\ref{sec:theory} we define the
key theoretical concepts for weak shear and introduce our notation. In
\S\ref{sec:aperturemass} we define the aperture mass and give
expressions for the aperture mass variance in terms of the matter
power spectrum, we also give the alternative relation between it and
the shear correlation functions. In \S\ref{sec:est} we develop the
direct estimator methodology, giving an explicit computation for the mean
and variance in the presence of ellipticity weights and also show how
the direct estimator can be accelerated and made effectively linear
order in the number of galaxies and number of apertures. We discuss
various strategies for combining estimates from an ensemble of
apertures that give both, high signal-to-noise and a small bias induced by including incomplete apertures. In \S\ref{sec:cfhtlens} we turn to the
analysis of the \cfhtls data. We give an overview of the data we use
and also the mock catalogues that we generate to test for systematic
errors.  As a preliminary analysis we present the aperture mass maps
for the survey. In \S\ref{sec:results} we investigate the bias of the direct estimator induced by the \cfhtls mask through measuring the aperture mass variance on the mock catalogues and comparing it to the results obtained when using the correlation function method. After determining the weighting scheme that induces the smallest bias we use it to measure the aperture mass variance on the true \cfhtls data and compare it to the analysis presented in \citet{Kilbingeretal2013}. We also check how the results change when removing blended sources from the data. In \S\ref{sec:fish} we use the mock catalogues to investigate the cosmology dependence and the information content of both estimators via the Fisher information. Finally, in \S\ref{sec:conclusions} we summarise our
findings, conclude and discuss future work.


\section{Theoretical background}\label{sec:theory}


\subsection{Basic cosmic shear concepts}\label{ssec:basics}
In this paper we are principally concerned with the weak lensing of
distant background galaxy shapes by the intervening large-scale
structure \citep{Blandfordetal1991,Kaiser1998,Seitzetal1994,JainSeljak1997,Schneideretal1998}; see \citet{BartelmannSchneider2001,Dodelson2003,Dodelson2017} for reviews. The two fundamental quantities describing this mapping from true to observed galaxy images are the convergence $\kappa$ and the shear $\gamma$ which, assuming a metric theory of gravity, are both derived from an underlying scalar lensing potential. In a cosmological setting the convergence at angular position $\bm \theta$ and radial distance $\chi$ can be connected to the density contrast $\delta(\chi\bm\theta,\chi)$ as:
\begin{align}
  \!\!\kappa(\bm\theta,\chi)  & = 
\frac{3}{2}\Omega_{\rm m,0}\left(\frac{H_0}{c}\right)^2 \int_0^{\chi} \diff\chi'
 \frac{(\chi-\chi')\chi'}{\chi \ a(\chi')}
 \delta(\chi'\bthet,\chi') \ , \label{eq:kappa}
 \end{align}
where $\Omega_{\rm m,0}$ is the total matter density, $H_0$ denotes the Hubble constant, $a$ is the scale factor and $c$ the speed of light.

In a real survey we will not necessarily have access to the precise
redshifts of each source galaxy. Instead, we will typically have the
redshift distribution of sources determined through photometric
redshift estimates. Hence, the effective convergence will be obtained
by averaging over the source population $p_\chi$:
\begin{align}
  \kappa(\bthet) & = \int_0^{\chiH} \diff\chi \ p_\chi(\chi) \kappa(\bthet,\chi) \nn \\
  & = \frac{3}{2}\Omega_{\rm m,0}\left(\frac{H_0}{c}\right)^2
  \int_0^{\chiH} \diff\chi' \frac{\chi'}{a(\chi')} g(\chi') \delta(\chi'\bthet,\chi') \ ,\label{eq:kappaeff}
\end{align}
where $\chiH$ is the comoving distance to the horizon and the weight function $g(\chi)$ is defined as
\begin{align}
  g(\chi') &\equiv \int_{\chi'}^{\chiH} \diff \chi \ p_\chi(\chi)
  \frac{\chi-\chi'}{\chi}
  \nn \\
 &= \frac{1}{N_{\rm TOT}}\int_{z(\chi')}^{\zH} \diff z \ 
   \frac{\diff N(z)}{\diff z} 
  \frac{(\chi(z)-\chi')}{\chi(z)} \ , \label{eq:gchi}
\end{align} 
where we used that the weight function can be equivalently
written in terms of the differential number counts by noting
that $p_\chi(\chi)\diff \chi = p_z(z)\diff z = (\diff N/\diff z)/N_{\rm TOT}$.

\begin{figure*}
  \centering {
    \includegraphics[angle=0,width=8cm]{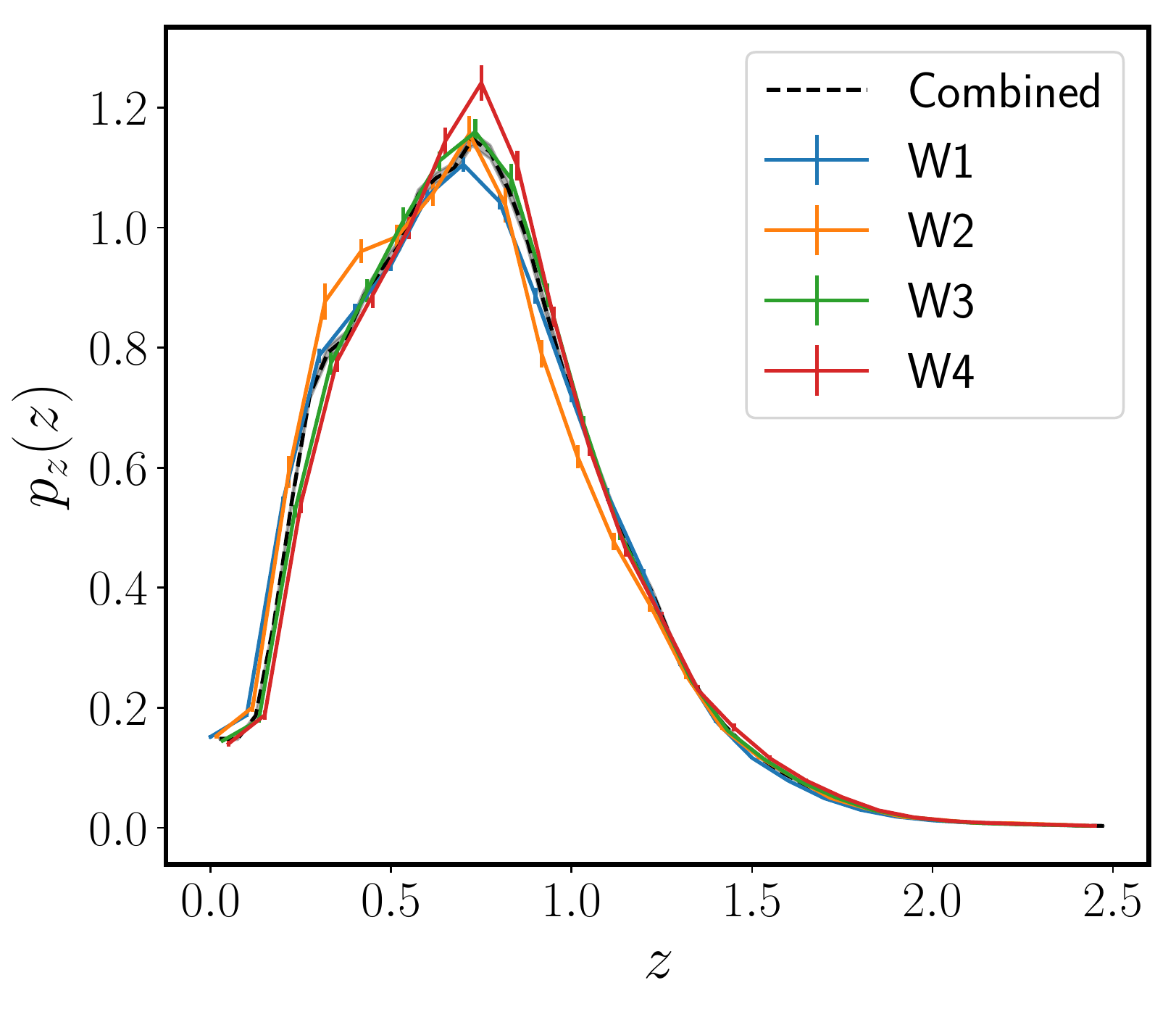}
    \includegraphics[angle=0,width=8cm]{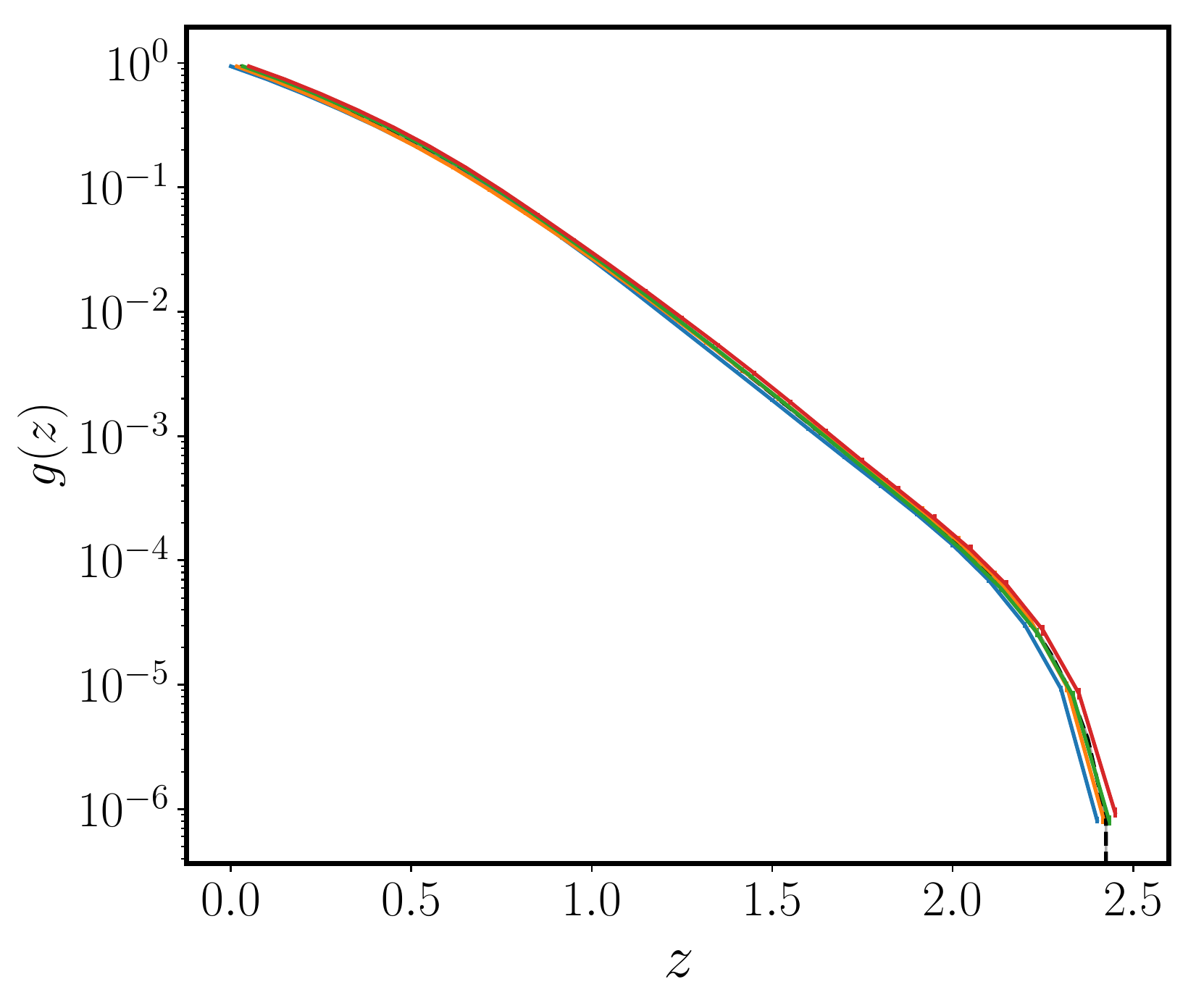} }
  \caption{\small{{\bf Left panel}: Redshift probability density
      distribution of galaxies in the W1, W2, W3 and W4 fields of
      the \cfhtls as a function of redshift
      (i.e. $dN/dz/N_{\rm TOT}$ versus $z$). The coloured solid lines
      show the results for the individual fields and the dashed black
      line shows the total for the combined fields. error bars and the grey shaded
      band show the 1$\sigma$ confidence regions, obtained from a
      jackknife resampling of the survey area.  {\bf Right panel}: The
      weight function $g(\chi(z))$ given in \Eqn{eq:gchi} as a function
      of redshift for the four \cfhtl fields and the combined
      set. Line styles and colours are as in the left panel. }
    \label{fig:reddist}}
\end{figure*}

In the left panel of Figure~\ref{fig:reddist} we show the redshift distribution of galaxies in the \cfhtls for the four
fields W1, W2, W3 and W4 and the total obtained for the combination of
all fields. They were obtained by averaging over the BPZ posterior including the lens weights. One
can clearly see that there are significant field to field variations
in the redshift distributions, with the W2 and W4 fields showing the
largest deviations from the mean in the range $z\in[0.2,0.4]$ for W2
and $z\in[0.5,0.7]$ for W4, respectively. The shaded region in the
plot shows the standard error region on the mean. This was estimated
using a jackknife resampling of the data.

The right panel of Figure~\ref{fig:reddist} shows the lensing weight
function $g(z)$ computed using the estimated distribution function
$p(z)$ shown in the left panel for the \cfhtl. We see that while
there are features in the $p(z)$ distribution, these are effectively
washed out when computing the lensing weights for the population.  In
fact, the most significant outlier is the W2 field, which appears to
have a slightly high amplitude at for redshifts $z>1.5$.


\subsection{Tangential and cross shear}\label{ssec:tanshear}

Owing to the fact that gravity only `excites' certain shear patterns
we wish to rotate the shear into the frame where we can more easily
separate out the these modes. This is done by decomposing the
shear into `tangential' and `cross' components. Consider the shear
field at a position vector $\bthet+\bthet_0$, where $\bthet_0$ is an
arbitrary location and $\bthet$ is a radial vector centred on
$\bthet_0$. We may rotate the shear field by the polar angle of the
separation vector $\bthet$ to obtain the tangential and cross
components \citep{BartelmannSchneider2001}:
\begin{align}
\gamma_{\rm t}(\bthet;\bthet_0) & \equiv  -\Re\left[\gamma(\bm\theta+\bthet_0)\ex^{-2i\phi} \right] \label{eq:gt1} \ ;\\
\gamma_{\times}(\bthet;\bthet_0) & \equiv -\Im\left[\gamma(\bm\theta+\bthet_0)\ex^{-2i\phi} \right] \ ,
\label{eq:gx1}
\end{align}
where $\phi$ is the polar angle associated with the vector
$\bthet$. 
The main advantage of this transformation is that for an
axially-symmetric mass distribution, the shear is always tangentially
aligned relative to the direction towards the origin of the mass
distribution and the cross component will vanish. This result is not
true for any randomly selected point for the origin. However, if we
average the tangential shear over a ring it can be related to the
enclosed surface mass density $\overline{\kappa}$:
$\left<\gamma_{\rm t}(\bthet;\bthet_0)\right>_{\rm
  circ}=\overline{\kappa}(\theta;\bthet_0)-\left<\kappa(\theta;\bthet_0)\right>_{\rm
  circ}$. On the other hand, if we ring average the cross-shear it
will vanish: $\left<\gamma_\times(\bthet;\bthet_0)\right>_{\rm
  circ}=0$ \citep{Kaiser1995,Schneider1996}.


\section{Aperture mass measures for cosmic shear}\label{sec:aperturemass}

In this paper we are primarily concerned with the statistical
properties of the ring averaged tangential shear integrated over a
filter function with compact support -- the aperture mass.


\subsection{Aperture Mass}\label{ssec:apmass}

Aperture mass was developed by \citet{Schneider1996} as a technique
for using a weighted set of measured shears within a circular region
to estimate the enclosed projected mass overdensity. It can be defined as follows:
consider an angular position vector in the survey $\bthet_0$, and let
us compute the tangential shear field around this point. Aperture mass
is now defined as the convolution of the tangential shear with a
 circularly-symmetric filter function $\wg$, with a
characteristic scale $\thetc$, above which the filter functions are
typically set to zero. It can be expressed as:
\be 
\Map(\bthet_0;\thetc) \equiv \int_{\RR} \dthet_1 \gamma_{{\rm t}}(\bthet_1;\bthet_0)
\wg(|\bthet_{1}-\bthet_0|;\thetc)\ . \label{eq:MapShear1}
\ee
In a similar vain one can also define the cross component of aperture
mass, which we refer to as `map-cross':
\be 
\Mx (\bthet_0;\thetc) \equiv \int_{\RR} \dthet_1 \gamma_{\times}(\bthet_1;\bthet_0)
\wg(|\bthet_{1}-\bthet_0|;\thetc) \ . \label{eq:MxShear1}
\ee
In the absence of systematic errors (B-modes) in the lensing data,
map-cross should vanish. $\Map$ and $\Mx$ are therefore said to be E/B
decomposed \citep{Schneideretal2002a}.
 
As was proven by \citet{Schneider1996}, owing to the fact that the
shear and convergence are sourced by the same scalar potential, one
can derive an equivalent relation to that above, but computed by
convolving the convergence $\kappa$ with a different filter function
$\wk$:
\be 
\Map(\bthet_0;\thetc) = \int_{\RR} \dthet_1 \kappa(\bthet_1) 
\wk(\left|\bthet_1-\bthet_0\right|;\thetc) \ . \label{eq:MapKap1} 
\ee
It is important to note that the filter functions $\wg$ and $\wk$ are
not independent of one another, but are related
\citep{Schneider1996}:
\begin{align}
\wg(\theta;\thetc) & = 
\frac{2}{\theta^2} \int_0^{\theta} \diff\theta'\theta' \wk(\theta';\thetc)
- \wk(\theta;\thetc) \ ; \label{eq:wQ}\\ 
\wk(\theta;\thetc) & = 2\int_{\theta}^{\infty} \frac{\diff\theta'}{\theta'}
\wg(\theta';\thetc)-\wg(\theta;\thetc) \label{eq:wU}  \ .
\end{align}
Also, it is worth noting that the $U$ filter is a compensated function \citep{BartelmannSchneider2001}.

For this work we will be using a polynominal filter function introduced in \citet{Schneideretal1998}:
\be 
Q(\theta;\thetc) = \frac{6}{\pi \thetc^2}
\left(\frac{\theta}{\thetc}\right)^2
\left[1-\left(\frac{\theta}{\thetc}\right)^2\right]
\mathcal{H}(\thetc - \theta) \label{eq:schneiderQ} \ ,
\ee
where ${\mathcal H}(x)$ is the Heaviside function.


\subsection{Aperture mass variance}\label{sec:apmass}
For cosmic shear, the expectation of the aperture mass around a
randomly selected point vanishes, since
$\left<\kappa\right>=\left<\gamma_{\rm t}\right>=0$. Thus, the lowest order
non-zero quantity of interest is the variance. Using
\Eqn{eq:MapShear1} the variance of the aperture mass can be written
as:
\begin{align}
\MapSq(\bthet_0;\thetc) &= 
\int_{\RR} \dthet_1 \dthet_2 
\left<\gamma_{{\rm t}}(\bthet_1;\bthet_0)\gamma_{{\rm t}}(\bthet_2;\bthet_0)\right> \nn \\
& \times \wg(|\bthet_1-\bthet_0|;\thetc) \wg(|\bthet_2-\bthet_0|;\thetc) \ .
\label{eq:MapVar1}
\end{align}
Using \Eqn{eq:MapKap1} we see that this can be equivalently written as:
\begin{align} 
\MapSq(\bthet_0;\thetc) & =
\int_{\RR}  \dthet_1 \dthet_2 \left<\kappa(\bthet_1)\kappa(\bthet_2)\right>  \nn \\
& \times \wk(\left|\bthet_1-\bthet_0\right|;\thetc)\wk(\left|\bthet_2-\bthet_0\right|;\thetc) \ .
\label{eq:MapVar2}
\end{align}
The Fourier transform of the convergence, $\tilde{\kappa}$, is defined as follows:
\begin{align}
\kappa(\bthet)  &\equiv    \int_{\RR} \frac{\dell}{(2\pi)^2} 
\tilde{\kappa}(\bell) \ex^{-i\bell\cdot\bthet}
\label{eq:kapFT} .
\end{align}
On using the above transform in \Eqn{eq:MapVar2} we find:
\begin{align}
\MapSq(\bthet_0;\thetc) & =
\int_{\RR}  \dthet_1 \dthet_2
\int_{\RR} \frac{\dell_1}{(2\pi)^2} 
 \frac{\dell_2}{(2\pi)^2}
\left<\tilde{\kappa}(\bell_1)\tilde{\kappa}(\bell_2)\right> \nn \\
& \hspace{-1.5cm}\times \ex^{-i(\bell_1\cdot\bthet_1+\bell_2\cdot\bthet_2)}
\wk(\left|\bthet_1-\bthet_0\right|;\thetc)\wk(\left|\bthet_2-\bthet_0\right|;\thetc) \ .
\label{eq:MapVar3}
\end{align}
We next use the statistical homogeneity and isotropy of the
correlations of $\kappa(\bthet)$ to define the convergence power
spectrum:
\be \left<\tilde{\kappa}(\bell_1)\tilde{\kappa}(\bell_2)\right> = (2\pi)^2
\dirac(\bell_1+\bell_2) C_{\kappa\kappa}(\ell_1) \ .\ee
On inserting this into \Eqn{eq:MapVar3} and integrating over the Dirac
delta function we see that the aperture mass variance can be written:
\begin{align}
\MapSq(\thetc) & =
\int_{\RR} \frac{\dell}{(2\pi)^2} C_{\kappa\kappa}(\ell) 
\left|\tilde{U}(\bell;\thetc)\right|^2
\ \label{eq:MapVar4} .
\end{align}
where
\be \tilde{U}(\bell;\thetc)=
\int_{\RR}  \dSq{y} \ \ex^{i\bell\cdot\by}  \wk(\left|\by\right|;\thetc)\ .
\ee
To progress we need to relate the convergence power to the matter
power  spectrum  $P$ that, in the
small-scale limit and under the Limber approximation, can be related to the convergence power spectrum as:
\begin{align}
  C_{\kappa}(\ell)
  & =\frac{9}{4}\Omega_{\rm m,0}^2\left(\frac{H_0}{c}\right)^4
  \int_0^{\chiH} \diff\chi \frac{g^2(\chi)}{a^2(\chi)}
  P\left(\frac{\ell}{\chi},\chi\right) \label{eq:Limb2Corr} \ .
  \end{align}
On inserting this relation into \Eqn{eq:MapVar4} and using the Schneider polynomial filter function \eqn{eq:schneiderQ} such that $\tilde{U}(\ell; \thetc)= 24
J_4(\ell\thetc)/(\ell\thetc)^2$ we have \citep{Schneideretal1998}:
\begin{align}
\MapSq(\bthet_0;\thetc) 
& =
\frac{1}{\thetc^2}\frac{24^2}{2\pi}\frac{9}{4}\Omega_{\rm m,0}^2
\left(\frac{H_0}{c}\right)^4
\int_0^{\chiH} \diff\chi \frac{g^2(\chi)}{a^2(\chi)}\nn \\
& \times \int_0^{\infty} \diff y \frac{J_4^2(y)}{y^3} P\left(\frac{y}{\thetc\chi},\chi\right)
\ \label{eq:MapVar} .
\end{align}
%


\subsection{Aperture mass variance from shear correlation functions}\label{sec:shearcorr}

As discussed earlier, the standard method for estimating the aperture
mass variance is through the two-point shear correlation
functions. Let us make that connection explicit. The complex shear
field has two non-vanishing two-point correlation functions that can
be written in terms of its tangential and cross-components as
\citep{Schneideretal2002a}:
\begin{align}
\xi_+(\theta) & \equiv \left<\gamma_{\rm t}(\bthet_1;\bthet_1)\gamma_{\rm t}(\bthet_2;\bthet_1)\right> +
\left<\gamma_\times(\bthet_1;\bthet_1)\gamma_\times(\bthet_2;\bthet_1)\right> \ ; \label{eq:xi+_2} \\
\xi_-(\theta) & \equiv \left<\gamma_{\rm t}(\bthet_1;\bthet_1)\gamma_{\rm t}(\bthet_2;\bthet_1)\right> -
\left<\gamma_\times(\bthet_1;\bthet_1)\gamma_\times(\bthet_2;\bthet_1)\right>   \label{eq:xi-_2} \ ,
\end{align}
where in this subsection $\theta\equiv|\bthet_1-\bthet_2|$. It can be shown that
$\xi_+$ and $\xi_-$ can be written in terms of the convergence power
spectrum as:
\begin{align}
\xi_+(\theta) & = \int_{\RR} \frac{\dell}{(2\pi)^2} C_{\kappa\kappa}(\ell) J_0(\ell\theta)  \ ;\\
\xi_-(\theta) & = \int_{\RR} \frac{\dell}{(2\pi)^2} C_{\kappa\kappa}(\ell) J_4(\ell\theta)\ .
\end{align}
Using the orthogonality of the Bessel functions we can invert the
above expressions to obtain the convergence power spectrum:
\begin{align}
  C_{\kappa\kappa}(\ell)
& = 2\pi \int_0^{\infty} \diff \theta \theta \xi_+(\theta) J_0(\ell\theta) \ ; \label{eq:CkXi+}\\ 
& = 2\pi \int_0^{\infty} \diff \theta \theta \xi_-(\theta) J_4(\ell\theta) \ . \label{eq:CkXi-}
\end{align}
The important consequence of the above relations is that we can now
rewrite the aperture mass variance using the shear correlation
functions. On substitution of \Eqns{eq:CkXi+}{eq:CkXi-} into
\Eqn{eq:MapVar4} one finds \citep{Schneideretal2002a}:
\begin{align}
\MapSq(\bthet_0;\thetc) & = 
\int_{\RR} \frac{\dell}{(2\pi)^2}\left|\tilde{U}(\bell;\thetc)\right|^2\pi
\left[\int_0^{\infty} \diff \theta \theta \xi_+(\theta) J_0(\ell\theta)\right. \nn \\
& \left.   +\int_0^{\infty} \diff \theta \theta \xi_-(\theta) J_4(\ell\theta)\right]
\ \label{eq:MapVar7} .
\end{align}
On reordering the integrals over $\ell$ and $\theta$, we see that the
above can be written more compactly as:
\begin{align}
\MapSq(\bthet_0;\thetc) & = 
\frac{1}{2\thetc^2}\int_0^{\infty} \diff\theta \theta \bigg[ \xi_+(\theta) T_+(\theta|\thetc) \nn \\
& +\xi_-(\theta) T_-(\theta|\thetc)\bigg]\ ,\label{eq:MapVar8} 
\end{align}
where
\begin{align}
  T_+(\theta|\thetc) &
  \equiv  \int_{0}^{\infty} \diff\ell \ell \left|\tilde{U}(\ell;\thetc)\right|^2 J_0(\ell\theta) 
\ \label{eq:T+} \ ,\\
  T_-(\theta|\thetc) & 
  \equiv \int_{0}^{\infty} \diff\ell \ell \left|\tilde{U}(\ell;\thetc)\right|^2 J_4(\ell\theta) 
\ \label{eq:T-} .
\end{align}
Once again, on adopting the Schneider polynomial filter \eqn{eq:schneiderQ} we see that the above kernels have an analytic form
\citep{Schneideretal2002a}:
\begin{align}
T_+(y) & =\frac{{\mathcal H}(2-y)}{100\pi}
\bigg[240 \left(2-15 y^2\right) \cos ^{-1}\left(\frac{y}{2}\right) + y \sqrt{4-y^2}\nn \\
  & \times
  \left(120+2320 y^2 -754 y^4 +132 y^6 -9y^8\right)
  \bigg]  \ ; \label{eq:T+S} \\
T_-(y) & = \frac{192}{35\pi} y^3 \left(1-\frac{y^2}{4}\right)^{7/2} {\mathcal H}(2-y)  \label{eq:T-} \ ,
\end{align}
where in the above $y\equiv \theta/\thetc$.

There are several important things to note about this: first, for the
case of the Schneider polynomial filter function, one needs to measure
$\xi_+$ and $\xi_-$ over the range $\theta\in[0,2\thetc]$, meaning
that we need information from scales close to zero separation. The correlations on small scales
can not be accurately measured and will be dominated by image blending
issues and shape noise \citep{Kilbingeretal2006}. Second, the
integration to obtain the variance from \Eqn{eq:MapVar8} can only be
approximately done using a set of discrete bins which need to be sufficiently dense and non empty.  The result of all of this is that there
will be some amount of E/B leakage, which will lead to a suppression of
the signal on small scales \citep{Kilbingeretal2006}. The first issue is also a problem for the direct estimator, but
the second is not.


\section{Estimating the aperture mass statistics}\label{sec:est}

As discussed in the previous section, there are two approaches to
estimating the aperture mass statistics. The correlation function
approach outlined in the previous section has been studied in great
detail. The direct estimator approach that we explore in this work has
not been as well explored, we therefore now describe our extension of
this approach in some detail.


\subsection{The direct estimator for the aperture mass dispersion
  for a single field -- including the source weights}\label{ssec:direct}
Here we follow \citet{Schneideretal1998}, but extend the work to
include a set of arbitrary weights for each source galaxy. Let us
first introduce the direct estimator of the aperture mass dispersion
for a single field.  Consider an aperture of angular radius $\thetc$,
centred on the position $\bthet_{0,k}$. The aperture contains $N_k$
galaxies with positions $\bthet_i$ with complex ellipticities
$\epsilon_i$. For the case of weak lensing the observed ellipticities
and intrinsic ellipticities $\epsilon_i^{(\rm s)}$ are approximately
related though $\epsilon_i=\gamma_i+\epsilon_i^{(\rm s)}$. In complete
analogy to the definition of tangential and cross shear defined in
\Eqns{eq:gt1}{eq:gx1} we define the same quantities for the tangential
and cross components of ellipticity: $\epsilon_{\rm t}
=-\Re\left[\epsilon \ \ex^{-2i\phi} \right]$ and $\epsilon_\times
=-\Im\left[\epsilon \ \ex^{-2i\phi} \right]$, where the polar angle
$\phi$ is relative to the origin $\bthet_{0,k}$. Our estimator for the
aperture mass variance is defined as:
\begin{align}
  \widehat{{\mathcal M}_{\rm ap}^2(\thetc|\btheta_{0,k})} & := (\pi\thetc^2)^2 \
  \frac{\sum_{i}^{N_{k}}\sum_{j\ne i}^{N_{k}} w_i \ w_j \ Q_i \ Q_j \epsilon_{{\rm t},i} \ \epsilon_{{\rm t},j}}{\sum_{i}^{N_{k}}\sum_{j\ne i}^{N_{k}}
     w_i w_j}
  \label{eq:est1}\ ,
\end{align}
where $w_i$ are weights assigned to the $i$th galaxy, the $Q_i\equiv
Q(|\bthet_i|;\thetc)$ and where $\epsilon_{{\rm t},i}$ is the observed
tangential ellipticity of the $i$th galaxy measured with respect to
the origin $\btheta_{0,k}$. Note that since the double sum will occur
repeatedly, we will use the short-hand notation
$\sum_{i}^{N_{k}}\sum_{j\ne i}^{N_{k}}\rightarrow \sum_{i\ne j}$ for
brevity. We will also suppress the origin $\btheta_{0,k}$ and also
take $N_k=N$.

We show that this provides an unbiased estimator for the true aperture
mass dispersion. This can be done through applying three averaging
processes: averaging over the intrinsic ellipticity distributions $A$;
then the source galaxy positions $P$; and then the ensemble average
over the cosmic fields $E$ \citep[following the notation
  of][]{Schneideretal1998}. Ignoring the prefactor and the denominator
for a moment, if we perform the $A$ average then we get:
\begin{align}
  A\left( \sum_{i\neq j} w_iw_jQ_iQ_j\tang{\epsilon}{i}\tang{\epsilon}{j} \right)
  & = A\left( \sum_{i,j} w_iw_jQ_iQ_j\tang{\epsilon}{i}\tang{\epsilon}{j}\right) \nn \\
  & \hspace{-2.5cm}- A\left(\sum_i w_i^2Q_i^2\tang{\epsilon}{i}^2  \right) \nn \\
  & \hspace{-3.5cm} = \sum_{i,j} w_iw_jQ_iQ_j \bigg\{ \tang{\gamma}{i}\tang{\gamma}{j}
  + 2 \tang{\gamma}{i} \ A\left( \tang{\epsilon}{j}^{(s)}\right) + A\left( \tang{\epsilon}{i}^{(s)}\tang{\epsilon}{j}^{(s)} \right)   \bigg\} \nn\\
  & \hspace{-2.5cm}- \sum_i w_i^2 Q_i^2 \left\{ \tang{\gamma}{i}^2 + 2 \ \tang{\gamma}{i} \ A\left( \tang{\epsilon}{i}^{(s)}\right)
  + A\left( [\tang{\epsilon}{i}^{(s)}]^2\right)   \right\} \nn \\
  & \hspace{-3.5cm} = \sum_{i,j} w_iw_jQ_iQ_j \left\{ \tang{\gamma}{i} \tang{\gamma}{j} + \frac{\sigma_\epsilon^2}{2} \delta^K_{ij} \right\}\nn \\
  & \hspace{-2.5cm}-  \sum_i w_i^2 Q_i^2 \left\{ \tang{\gamma}{i}^2  +\frac{\sigma_\epsilon^2}{2} \right\} \nn \\
  &\hspace{-3.5cm} = \sum_{i\neq j} w_iw_jQ_iQ_j\tang{\gamma}{i}\tang{\gamma}{j}\ .\label{eq:est2}
\end{align}
Note that in the above we assumed that each galaxies' intrinsic ellipticity is indiviually drawn from the same Gaussian distribution $\mathcal{G}\left(0, \sigma^2_\epsilon \right)$ with zero mean and the shape noise $\sigma^2_\epsilon$ as variance, i.e. no intrinsic alignments. Next, we perform the average over the spatial positions of the source galaxies:
\begin{align}
  P\left( \sum_{i\neq j} w_iw_jQ_iQ_j\gamma_{{\rm t},i} \gamma_{{\rm t},j} \right)
  &= \prod_{\alpha=1}^{N} \left\{\frac{\diff^2 {\bm \theta}_\alpha}{\pi\thetc^2}\right\} \nn \\
  & \hspace{-2.5cm}\times \sum_{i\neq j} w_iw_j Q_i Q_j \gamma_{{\rm t},i} \gamma_{{\rm t},j} \nn \\
  &\hspace{-2.5cm}= \frac{\sum_{i\neq j} w_iw_j}{\left(\pi \thetc^2\right)^2} \int_{\mathbb{R}^2}
  \diff^2{\bm \theta}_1 \diff^2{\bm \theta_2} \ Q_1Q_2\gamma_{{\rm t},1} \gamma_{{\rm t},2} \ . \label{eq:est3}
\end{align}
In the first step we took the joint PDF of spatial positions to be
simply the product of the independent 1-point PDFs for a uniform
random distribution. In the second step, on noting that $\gamma_{{\rm t},i}
= \gamma_{{\rm t},i}({\bm \theta}_i; {\bm\theta}_0)$, where
$\mathbf{\theta}_0$ is the same for all the galaxies, we used the fact
that the spatial integral will yield the same result no matter of the
indices - hence the change $(i,j)\rightarrow(1,2)$. In the last step,
we integrated out the remaining PDFs and rewrote the domain. Finally,
we perform the expectation over the cosmic fields:
\begin{align} E\left\{P\left[A\left(\widehat{{\mathcal M}_{\rm ap}^2(\thetc)}\right)\right]\right\}
& = \frac{(\pi\thetc^2)^2}{\sum_{i\neq j} w_i w_j}
\frac{\sum_{i\neq j} w_iw_j}{\left(\pi \thetc^2\right)^2} \nn\\
& \times \int_{\mathbb{R}^2}
\diff^2{\bm \theta_1} \diff^2{\bm \theta}_2 \ Q_1Q_2 \left<\gamma_{{\rm t},1} \gamma_{{\rm t},2}\right> \nn \\
& = \MapSq  \ . \label{eq:est4} \end{align}

In Appendix~\ref{app:var} we calculate the variance of the estimator
\Eqn{eq:est1} and, for a moment supressing  $\vartheta$, we find that it can be written as:
\begin{align}
	\!\!{\rm Var}\left[\widehat{{\mathcal M}_{\rm ap}^2}\right]
	&= \frac{\sum_{l\ne k\ne j\ne i}w_i w_j w_k w_l}{\left(\sum_{j\ne i}w_i w_j\right)^2}
        \left<{\mathcal M}_{\rm ap}^4 \right> \nn \\
        & + 4\frac{\sum_{k\ne j\ne i}w_i^2 w_j w_k}{\left(\sum_{j\ne i} w_i w_j\right)^2}
        \left<{\mathcal M}_{\rm ap}^2{\mathcal M}_{\rm s,2}\right>  \nn \\
        & + 2\frac{\sum_{j\ne i} w_i^2 w_j^2 }{\left(\sum_{j\ne i}w_iw_j\right)^2}
        \left<{\mathcal M}_{\rm s,2}^2\right> \nn \\
        & + \frac{2 \sum_{k\ne i \ne j} w_i w_j^2 w_k}{\left(\sum_{j\ne i} w_i w_j\right)^2}        
        \sigma^2_{\epsilon}G\left<{\mathcal M}_{\rm ap}^2\right> \nn \\
        & +\frac{2 \sum_{j\ne i}  w_i^2 w_j^2}{\left(\sum_{j\ne i}w_iw_j\right)^2}
        \sigma^2_{\epsilon}G\left<{\mathcal M}_{s,2}\right>\nn \\
        & + \frac{\sum_{j\ne i} w_i^2 w_j^2}{2\left(\sum_{j\ne i} w_i w_j\right)^2}\sigma^4_{\epsilon}G^2
        -\left<{\mathcal M}_{\rm ap}^2\right>^2 , \label{eq:varest15}
\end{align}
where $G$ and ${\mathcal M}_{s,2}$ are as defined as:
\begin{align}
\langle{\mathcal M}_{\rm
    s,2}\rangle(\vartheta) &\equiv \pi\thetc^2\int \diff^2{\bm \theta} \ Q^2(|\bthet|; \vartheta) \langle\gamma^2_{\rm t}\rangle(\bthet)
\nn \ ;\\
G(\vartheta) &\equiv \pi\thetc^2\int
\diff^2\btheta \ Q^2(|\bthet|; \vartheta) \ .
\end{align}
Importantly, in the limit where all of the source galaxy
weights are equal we recover the expression derived in
\citet{Schneideretal1998}.

It is interesting to obtain an approximate form for the above
variance. Firstly, let us consider the case where the number of
galaxies per aperture is large such that $N\gg 1$, whereupon we see
that all of the partial sums are approximately equivalent to the full
sums, e.g. $\sum_i\sum_{j\ne i}w_iw_j\approx (\sum_i
w_i)^2$. Consequently, all of the prefactors involving the weights
can be dramatically simplified to give:
\begin{align}
	\lim_{N\rightarrow \infty}{\rm Var}\left[\widehat{{\mathcal M}_{\rm ap}^2}\right]
	& \approx 
        \left<{\mathcal M}_{\rm ap}^4 \right> 
        + 4S_2\left<{\mathcal M}_{\rm ap}^2{\mathcal M}_{\rm s,2}\right>  
        + 2S_2^2\left<{\mathcal M}_{\rm s,2}^2\right> \nn \\
        & + 2S_2        
        \sigma^2_{\epsilon}G\left<{\mathcal M}_{\rm ap}^2\right> 
         +2S_2^2 
        \sigma^2_{\epsilon}G\left<{\mathcal M}_{s,2}\right> \nn \\
        & + \frac{1}{2}S_2^2\sigma^4_{\epsilon}G^2
        -\left<{\mathcal M}_{\rm ap}^2\right>^2 \ , \label{eq:varest16}
\end{align}
where we defined $S_2\equiv \sum_{i} w_i^2/(\sum_{i} w_i)^2$. Let us
inspect the quantity $S_2$ in more detail: the Cauchy-Schwarz inequality
tells us that $\left(\sum_{i}^N u_iv_i\right)^2\le \sum_i^N
u_i^2\sum_j^Nv_j^2$, where the elements of the sets $\{u_i\}$ and
$\{v_i\}$ are drawn from the reals. If we take $v_i=v$ for any $i$,
then we see that $\left(\sum_{i}^N u_i\right)^2\le \sum_i^N u_i^2 N$,
which in turn implies $\left< u\right>^2\le \left<u^2\right>$. On
applying this to our ratio $S_2$ we see that:
\be S_2\equiv \frac{1}{N} \frac{\overline{w^2}}{\overline{w}^2} \ge \frac{1}{N} \ee
where $\overline{w^2}=\sum_i w_i^2/N$ and $\overline{w}=\sum_i w_i/N$.
This insight leads us to make our next approximation, since $\mathcal
  {M}_{s,2}\sim{\mathcal M}^2_{\rm ap}$ we see that $S_2\mathcal{M}_{s,2}\ll
{\mathcal M}^2_{\rm ap}$, since $S_2\propto 1/N$. This, then, leads us to
write:
\begin{align}
	\lim_{N\rightarrow \infty}{\rm Var}\left[\widehat{{\mathcal M}_{\rm ap}^2}\right]
	& \approx 
        \left<{\mathcal M}_{\rm ap}^4 \right> 
        + 2S_2        
        \sigma^2_{\epsilon}G\left<{\mathcal M}_{\rm ap}^2\right> \nn \\
        & + \frac{1}{2}S_2^2\sigma^4_{\epsilon}G^2
        -\left<{\mathcal M}_{\rm ap}^2\right>^2 \ . \label{eq:varest17}
\end{align}
Thirdly, let us further assume that the underlying shear field is
Gaussian and hence $\left<{\mathcal M}_{\rm ap}^4 \right> =3\left<{\mathcal
  M}_{\rm ap}^2 \right>^2$. Under these circumstances, which will be fulfilled for large apertures, the variance can
be written as:
\begin{align}
	{\rm Var}\left[\widehat{{\mathcal M}_{\rm ap}^2}\right]
	&\approx  2\left<{\mathcal M}_{\rm ap}^2 \right>^2
        + 2S_2 \sigma^2_{\epsilon}G\left<{\mathcal M}_{\rm ap}^2\right>
        + \frac{1}{2}S_2^2\sigma^4_{\epsilon}G^2 \nn \\
        &= \left(\sqrt{2}\left<{\mathcal M}_{\rm ap}^2\right> +\frac{1}{\sqrt{2}}S_2 G \sigma^2_{\epsilon}\right)^2
        \ . \label{eq:varest18}
\end{align}
The first term in the bracket is cosmic variance and the last term denotes the shape noise contribution.

\begin{figure*}
  \centering {
    \includegraphics[angle=0,width=8.5cm]{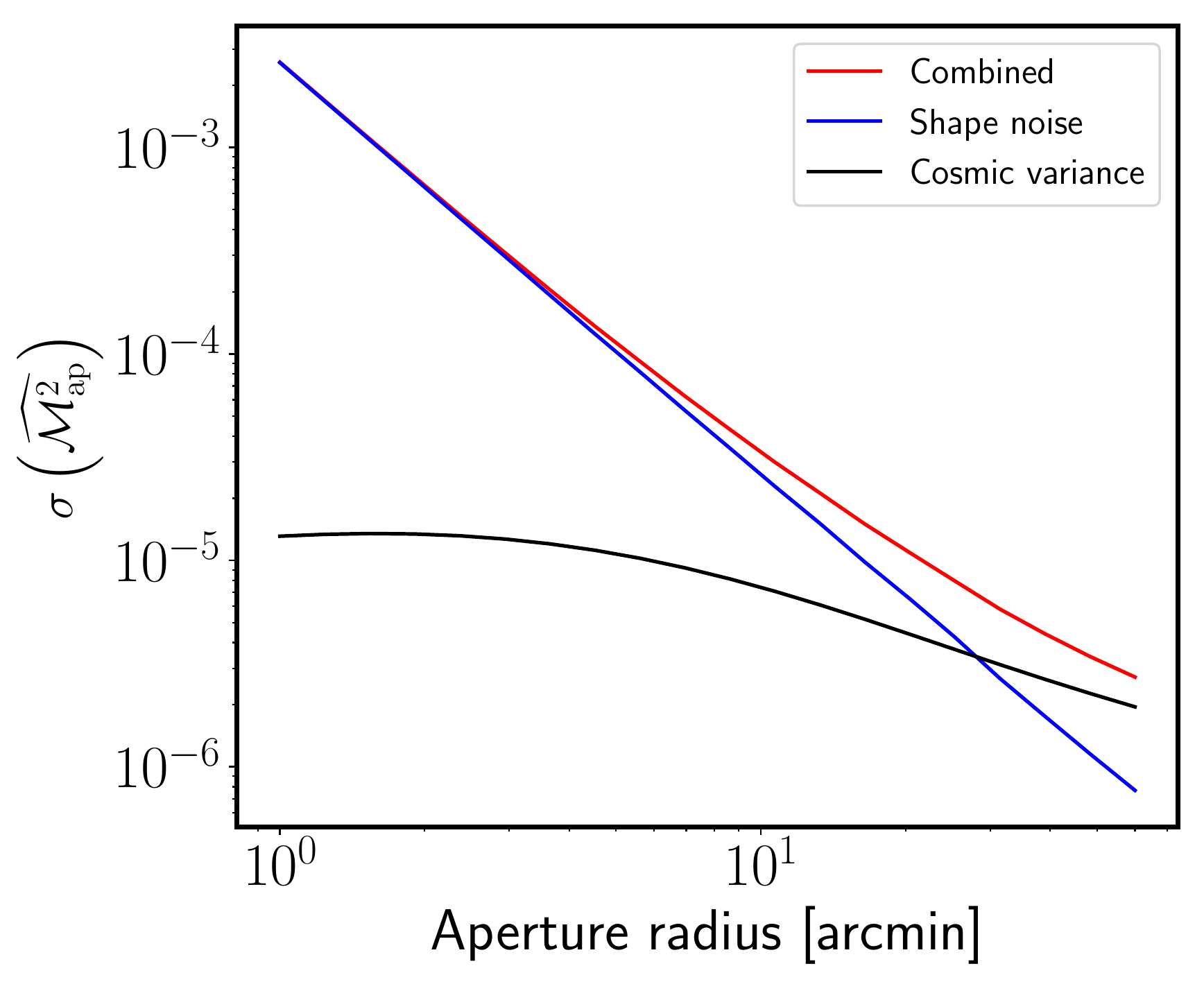}\hspace{0.2cm}
    \includegraphics[angle=0,width=8.5cm]{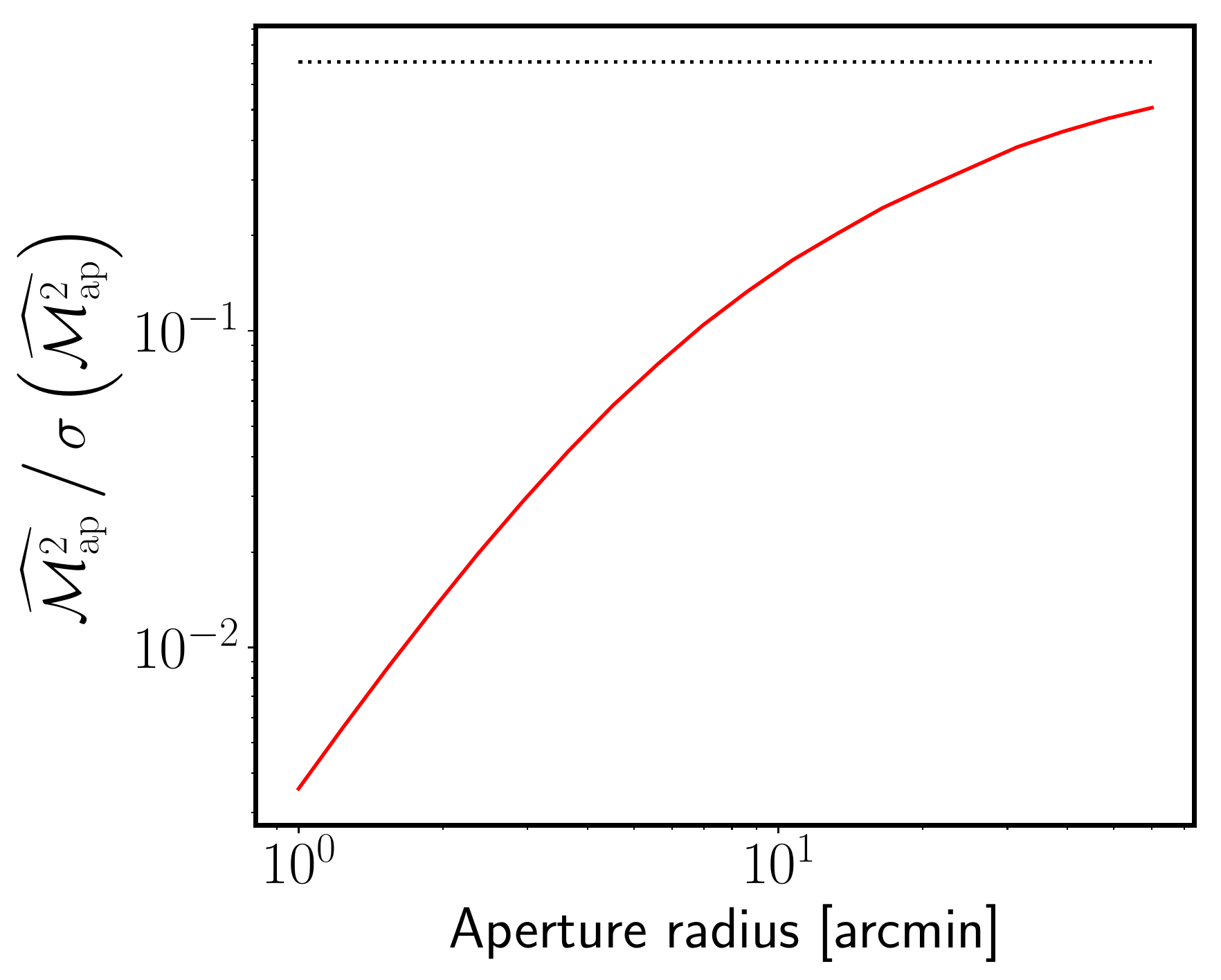}}
  \caption{\small{{\bf Left panel}:  Contributions to the variance of the estimate of the aperture mass dispersion in the \cfhtls data. We show the two contributions to the variance as predicted from \eqref{eq:varest18} as well as the combined result. Note that in adopting \eqref{eq:varest18} we have assumed that the convergence is Gaussianly distributed. {\bf Right panel}: The theoretical signal-to-noise per aperture as a function of the aperture size $\vartheta$ in the \cfhtls data. The black dotted line shows the cosmic variance limit.}
    \label{fig:SN}}
\end{figure*}

The left panel of Figure~\ref{fig:SN} shows the error on a given estimate
from a single aperture, using \Eqn{eq:varest18}. The right panel shows the corresponding prediction of the signal-to-noise on the aperture mass variance, per aperture, again using \Eqn{eq:varest18}. In order to generate these prediction we have used \eqref{eq:MapVar} as a model for the cosmic variance contribution.


\subsection{Acceleration of the direct estimator}\label{ssec:accel}

If we were to naively implement the direct estimator approach as given
by \Eqn{eq:est1} then we see that in order to compute the estimate of
the variance for a single aperture we need to compute the sum from
$N(N-1)$ galaxies. Thus one might conclude that the method scales as
typical $N^2$ pair counting approach for galaxies inside the
aperture. However, we now show that the method can be made to scale
linearly with the number of galaxies.  Let us consider again the
estimator from \Eqn{eq:est1}, and we notice that if we put back the
term that has $i\ne j$ and explicitly subtract it then we have:
\begin{align}
  \widehat{{\mathcal M}_{\rm ap}^2(\thetc|\bthet_{0,k})}
  & = (\pi\thetc^2)^2
  \frac{\sum_{i, j} w_i \ w_j \ Q_i \ Q_j \ \epsilon_{{\rm t},i}
    \epsilon_{{\rm t},j}}{\sum_{i\neq j} w_i w_j} \nn \\
  & - (\pi\thetc^2)^2
  \ \frac{\sum_{i} w_i^2\ Q_i^2 \epsilon^2_{t,i} }{\sum_{i\neq j} w_i w_j} \ .\label{eq:est5}
\end{align}
If we now introduce the estimators for aperture mass and $\mathcal{M}_{s,2}$ as discretised versions of their definition, 
\begin{align}
  \widehat{{\mathcal M}_{\rm ap}(\thetc|\bthet_{0,k})}
  & \equiv (\pi\thetc^2)
  \frac{\sum_{i} w_i \ Q_i \epsilon_{{\rm t},i}}{\sum_{i} w_i}
  \label{eq:est6} \\
  \widehat{{\mathcal M}_{\rm s,2}(\thetc|\bthet_{0,k})}
  & \equiv (\pi\thetc^2)
  \frac{\sum_{i} w_i^2 \ Q_i^2 \epsilon_{{\rm t},i}^2}{\sum_{i} w_i^2}
  \ , \label{eq:est6b}
\end{align}
we see that \Eqn{eq:est5} can be rewritten as: 
\begin{align}
  \widehat{{\mathcal M}_{\rm ap}^2(\thetc|\bthet_{0,k})}
  & =  \frac{1}{1-S_2}  \left(\frac{\pi\thetc^2 \sum_i w_i \ Q_i \epsilon_{{\rm t},i}}{\sum_i w_i}\right)^2 
  \nn \\ 
  &-\frac{\pi\thetc^2}{\frac{1}{S_2}-1} \ \pi\thetc^2 \ \frac{\sum_i w_i^2 \ Q_i^2 \epsilon_{{\rm t},i}^2}{\sum_i w_i^2}
  \nn \\ &=
  \frac{1}{1-S_2} \left[ (\widehat{{\mathcal M}_{\rm ap}})_k^2 -S_2 \pi\thetc^2 (\widehat{{\mathcal M}_{\rm s,2}})_k \right]
  \ ,\label{eq:est7}
\end{align}
where for brevity we used the notation $(\widehat{{\mathcal M}_{\rm
    ap}})_k\equiv \widehat{{\mathcal M}_{\rm
    ap}(\thetc|\bthet_{0,k})}$ and $(\widehat{{\mathcal M}_{\rm
    s,2}})_k\equiv \widehat{{\mathcal M}_{\rm
    s,2}(\thetc|\bthet_{0,k})}$. 
Note that both terms in the brackets receive an identical contribution of shape noise, hence the second term should not be neglected.

In general, the estimator \Eqn{eq:est7} is mathematically identical to
that of \Eqn{eq:est1} and therefore is also an unbiased estimator for
the variance of the aperture mass. However, algorithmically it has a
significant advantage in that it is linear in the number of galaxies.
This owes to the fact that all of the terms on the right-hand-side of
\Eqn{eq:est7} are linear in $N$. For example, the estimate of
$(\widehat{{\mathcal M}_{\rm ap}})_k$ is linear, so too are the
correction factors $(\widehat{{\mathcal M}_{\rm s,2}})_k$ and $S$.
As we will show in the second paper in this series (Porth~et~al., in
prep.), it can be shown that this can be naturally extended to higher
order aperture mass statistics. This acceleration of the method to
linear order opens the door to a significant advantage in speed for
estimation of aperture mass statistics at all order.


\subsection{Extending the estimate to an ensemble of fields}\label{ssec:ensemble}


The estimator \Eqn{eq:est7} is for a single aperture $k$ and as such will
provide a single low-signal-to-noise, albeit unbiased, estimate. We
now wish to make use of the full area of the survey available to us.
We are therefore confronted as to how to best achieve this. As
proposed by \citet{Schneideretal1998}, one simple approach would be to
sample well seperated apertures such that the shear in one field is
statistically independent from another. This would yield the estimator:
\begin{align}
  \widehat{{\mathcal M}_{\rm ap}^2(\thetc)} = \frac{\sum_k {\mathcal W}_k
  \widehat{{\mathcal M}_{\rm ap}^2(\thetc|\btheta_{0,k})}}{\sum_k{\mathcal W}_k} \ , \label{eq:Sest1}
\end{align}
where ${\mathcal W}_k$ are weights and the sum extends over the
$N_{\rm ap}$ apertures.  Since the estimates can be considered to be
statistically independent then the noise can be minimised by choosing
the weights to be given by:
\be {\mathcal W}_k\propto\frac{1}{ {\rm Var}\left[\widehat{
      {\mathcal M}_{\rm ap}^2(\thetc|\btheta_{0,k})}\right]} \ .\label{eq:Sest2} \ee
However, this approach would be suboptimal in that it does not take
advantage of the full area of the survey. In this case, the
signal-to-noise on the estimate for the full field can be achieved by
multiplying the estimates for the aperture mass variance per single
aperture by the square root of the number of independent apertures.

A much better approach, which makes better use of the full survey
area, is to oversample the apertures. Since the estimate for the
survey is still given by \Eqn{eq:Sest1} and since it is a linear
combination of the estimates for the single field, it too is unbiased:
\begin{align}
  \left<\widehat{{\mathcal M}_{\rm ap}^2(\thetc)}\right>
  & = \frac{\sum_k {\mathcal W}_k\left<\widehat{{\mathcal M}_{\rm ap}^2(\thetc|\btheta_{0,k})}\right>}{\sum_k{\mathcal W}_k}
  \nn \\
  & =\left<\widehat{{\mathcal M}_{\rm ap}^2(\thetc)}\right> \frac{\sum_k {\mathcal W}_k}{\sum_k{\mathcal W}_k} 
  = \left<{\mathcal M}_{\rm ap}^2(\thetc)\right> \ . \label{eq:Sest3}
\end{align}
However, the variance of the estimate for the survey is no longer
trivial to determine. This in turn means that the weights ${\mathcal
  W}_k$ from \Eqn{eq:Sest2} are no longer optimal. Computing the
optimal weights will be further complicated if we include incomplete
apertures in the estimate -- which we discuss next.


\subsection{Allowing incomplete aperture coverage to increase estimator efficiency}\label{ssec:coverage}

We next turn to the problem of aperture completeness. In real surveys
there are regions of the survey that are masked out due to bright
stars, chip gaps and the survey boundaries. The question now arises:
{\em what do we do if an aperture has some fraction of its area
  overlapping with the mask?} The simple answer would be that we
exclude all such apertures from the estimator. The problem with this
approach is that depending on the size of the aperture this may
significantly impact the available survey area from which to compute
the estimate and thus make the approach sub-optimal. Here we will
explore the idea of effectively including all apertures that fit
within the survey boundary, but apply weights to each of the form:
\be {\mathcal W}_k=f(c_k,{\rm Var},\dots) \ ,\ee 
where $c_k\equiv A_k/A$ is the completeness factor for the $k$th aperture,
where $A_k$ is the available area of the aperture and $A$ is the unmasked
area of the aperture, such that we have $c_k\le1$. $\rm Var$ is
related to the variance of the estimate in the aperture. The ellipsis
denotes that in general the weights could depend on other factors. In
this work we will explore three distinct choices:
\begin{align}
{\mathcal W}^{(1)}_k & = {\mathcal H}\left(c_k-\alpha\right) ; \label{eq:Weight1}\\
{\mathcal W}^{(2)}_k & = {\mathcal H}\left(c_k-0\right){\rm Var}\left[
 \left<\widehat{{\mathcal M}_{\rm ap}^2(\thetc|\bm\theta_{0,k})}\right>\right]^{-1} ; \label{eq:Weight2}\\
{\mathcal W}^{(3)}_k & = {\mathcal H}\left(c_k-\alpha\right){\rm Var}\left[
 \left<\widehat{{\mathcal M}_{\rm ap}^2(\thetc|\bm\theta_{0,k})}\right>\right]^{-1} , \label{eq:Weight3}
\end{align}
The first case corresponds to accepting all apertures whose completeness
factor $c_k\ge\alpha$ and for those that do we combine them in a
simple average with equal weights to arrive at our estimate for
$\widehat{{\mathcal M}_{\rm ap}^2}$. The second case corresponds to
accepting all apertures, irrespective of the completeness factor, but
combining all of the estimates using an inverse variance weighted
estimate, where the variance is approximated by \Eqn{eq:varest18}.
The third case is simply the product of the first and second case.

It is important to note that unless $\alpha=1$ our estimator given by
\Eqn{eq:Sest1} will formally become biased. This means that we will
expect some leakage of E/B modes. Postponing a thorough analytical and numerical analysis of incomplete aperture coverage and de-biasing strategies to a companion paper, we will
content our selves by investigating the degree of bias that is
introduced by computing the aperture-cross statistics. For reference,
these are defined in direct analogy with \Eqn{eq:est1}:
\begin{align}
  \widehat{{\mathcal M}_{\times}^2(\thetc|\btheta_{0,k})} & := (\pi\thetc^2)^2 \
  \frac{\sum_{i\neq j} w_i \ w_j \ Q_i \ Q_j \ \epsilon_{\times,i} \ \epsilon_{\times,j}}{\sum_{i\neq j} w_i w_j}
  \label{eq:CrossEst1}\ .
\end{align}
It can be proven that the expectation of this estimator vanishes; that
is provided we have no bias in the estimate we have
\mbox{$\left<\widehat{{\mathcal
      M}_{\times}^2(\thetc|\btheta_{0,k})}\right>=0$}. However, the
variance of the estimator does not vanish and it should be given by
the pure shape noise contribution to \Eqn{eq:varest15}. Under the
approximations of \Eqn{eq:varest18} this is: $ \frac{1}{2}{\rm
  Var}\left[\widehat{{\mathcal M}_{\times}^2(\thetc)}\right] \approx
S^2\sigma^4_{\epsilon}G^2 $ per aperture.

Finally, before moving on, we note that it is important to appreciate
that the weights ${\mathcal W}_k$ apply to how different fields are
combined, and that the weights $w_i$ from \Eqn{eq:est1} apply to how
the source galaxies are combined in arriving at an estimate for a
single field. We assume that these latter weights have been
pre-computed by the method for estimating galaxy ellipticities.


\subsection{Estimating computational complexity  for evaluation of the
direct estimator for $\widehat{{\mathcal M}_{\rm ap}^2}$}\label{ssec:complex}

Before moving on to the computation of the estimator with real data
let us estimate the computational cost for an evaluation of ${\mathcal
  M}_{\rm ap}^2$. As described above, the actual implementation is
built from a series of algorithmic blocks.
\begin{enumerate}
\item We first construct a KD-tree data structure for the galaxy
  catalogue.
\item The full survey is tiled with overlapping apertures, where the
  centres are separated by a distance $d$.
\item The aperture coverage map is computed to give the $c_k$ values
  for every aperture.
\item For apertures that pass the selection cut (${\mathcal
  H}(c_k-\alpha)$), a KD-tree range search locates all particles that
  lie inside the aperture radius $\thetc$.
\item Estimate the aperture mass statistics and its variance according to \Eqn{eq:varest15} for the $k$th aperture.
\item Combine the $N_{\rm ap}$ estimates through a weighted mean of
  the resulting estimates.
\end{enumerate}
In the above algorithm we shall assume that Steps~(i)--(iii), and Step
(vi) are performed once and therefore are not the limiting factors for
the execution of the code. We do note, however, that the construction
of the KD-tree may have a large memory footprint and will take some
non-negligible time for the initial construction. The parts of the
method that require some consideration are Steps~(iv) and (v).

Step (iv) is a range search routine and Step (v) is a routine that
evaluates the sums in \Eqn{eq:est7}. To compute the complexity for
these steps we first identify some parameters: let $p$ specify the
order of the statistics; $\thetc$ describe the aperture radius; and
$\zeta$ be a parameter that determines the spacing $d$ between
apertures: $d \equiv \thetc/\zeta$. We note that for a non-overlapping
field of apertures whose circumferences just touch each other, we
would set $\zeta=1/2$. Further, the number of apertures is thus a
function of $\thetc$ and $\zeta$, $N_{\rm ap}(\thetc,\zeta)$. The
order for the complexity can thus be computed as follows:
\begin{align}
  \mathcal{O}(\widehat{{\mathcal M}_{\rm ap}^p}|p, \thetc, \zeta) 
  &= \sum_{k=1}^{N_{\rm ap}(\thetc,\zeta)}
  \bigg[\mathcal{O}({\rm range \ search}|N_{k},N_{\rm nodes},\thetc) \nn \\
    & \ \ +  \mathcal{O}({\rm compute\ statistic}|N_{k},p)\bigg] \nn\\
  &\approx
  N_{\rm ap}(\thetc,\zeta) \bigg[\mathcal{O}({\rm range \ search}|\overline{N},N_{\rm Tot},\thetc) \nn \\
    & \ \ +  \overline{N}\mathcal{O}({\rm compute\ statistic}|N=1,p)\bigg] \ .
\end{align}
The first thing to notice is that the number of apertures scales all
of the computations, so if we fix the parameter $\zeta$, then the
total number of apertures will scale as $N_{\rm ap}\propto\Omega_{\rm
  s}/\thetc^2$, where $\Omega_{\rm s}$ is the survey area. The first
term in the square brackets gives the computational time for a range
search to deliver back the $N_k$ galaxies in the aperture. If the
distribution of source galaxies is roughly randomly distributed on the
sky, then we make the approximation $N_k\approx \overline{N}=N_{\rm
  Tot} A/\Omega_{\rm s}$. Each such range search operation then takes
of the order $\mathcal{O}(\log N_{\rm Tot})$ time to execute, but this
factor will also scale with the aperture radius and how clustered the
source galaxy data are, and also the depth we need to go in the tree
walk.

Considering the second term in square brackets, this is the required
time for computation of the estimate for the $p$th order aperture mass
statistic. As was described earlier for $\widehat{{\mathcal M}_{\rm
    ap}^2}$, the estimator scales linearly with the number of galaxies
in the aperture, thus in the second line we simply scale up the
complexity to estimate the statistic for a single galaxy by the number
of galaxies in the aperture. As we will explore in our companion
paper, owing to this linear scaling, there is no additional overhead
in extending the method to compute higher order statistics, beyond the
variance, such as the skewness $p=3$ and kurtosis $p=4$.


\section{Application to the \cfhtls data}\label{sec:cfhtlens}

We now turn to estimating the aperture mass variance in the \cfhtls 
data and in a large series of mocks generated by ray-tracing
through $N$-body simulations.

\subsection{\cfhtls shear data}\label{ssec:cfhtdata}

The Canada-France-Hawaii Telescope Lensing Survey (hereafter
\cfhtl) is a weak lensing survey that was completed around 2010.
It covers 154~${\rm deg}^2$ of the sky in five optical bands
$\{u^*,\ g',\ r',\ i',\ z'\}$ with a $\sim5\sigma$ point source
limiting magnitude in the $i'$ band of $i'_{\rm AB}\sim25.5$. The
survey measures galaxy ellipticities for use in weak lensing analysis
from multicolour data obtained as part of the CFHT Legacy
Survey\footnote{www.cfht.hawaii.edu/Science/CFHLS/}. The survey data
is distributed into four well spaced fields, three of which (W1, W2 \&
W4) lie close to the equatorial plane, and the third (W3) lies at high
declination. Full details of the survey can be found in
\citet{Heymansetal2012}.


\begin{figure*}
  \centering {
    \includegraphics[width=8.5cm]{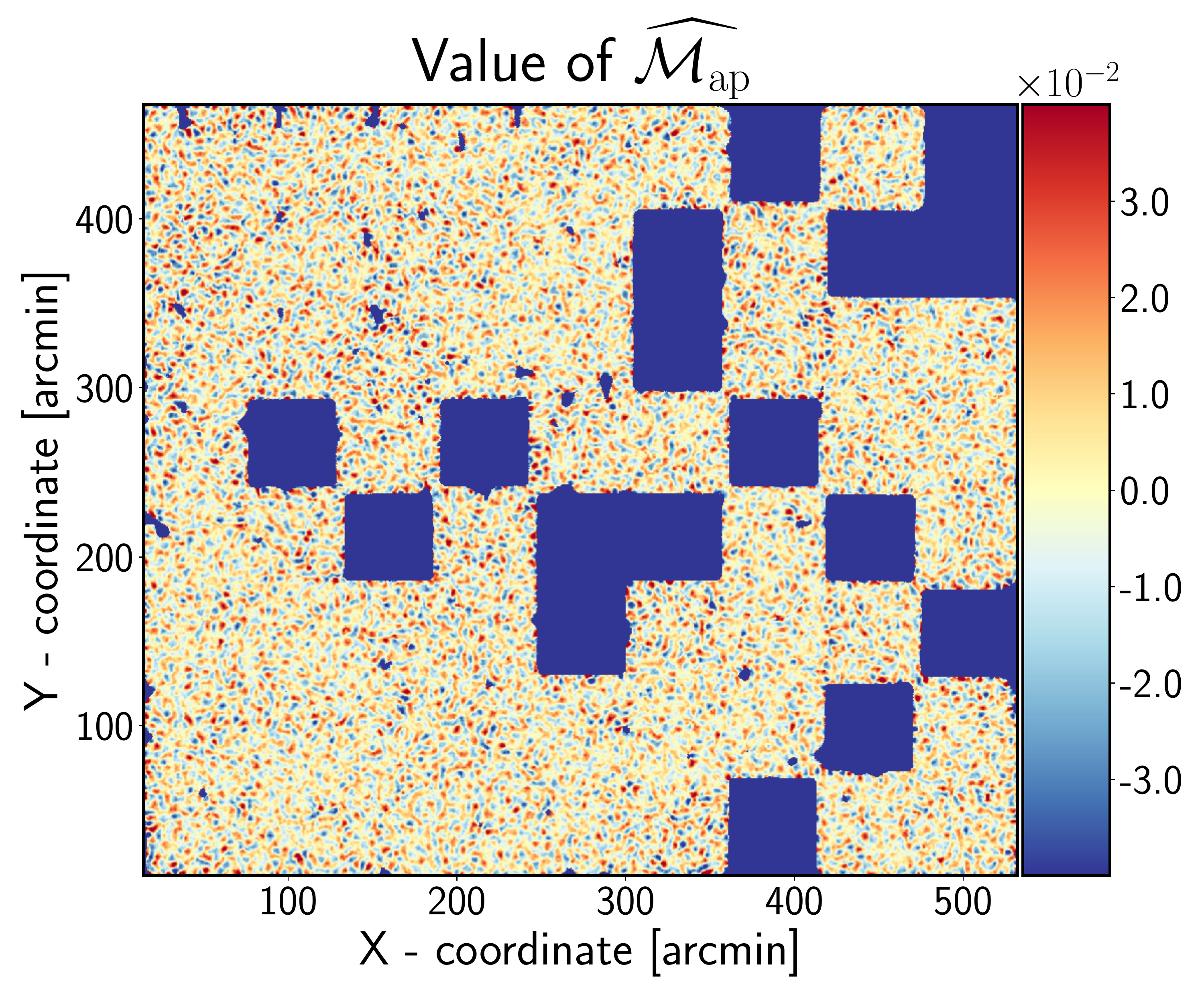}\hspace{0.2cm}
    \includegraphics[width=8.5cm]{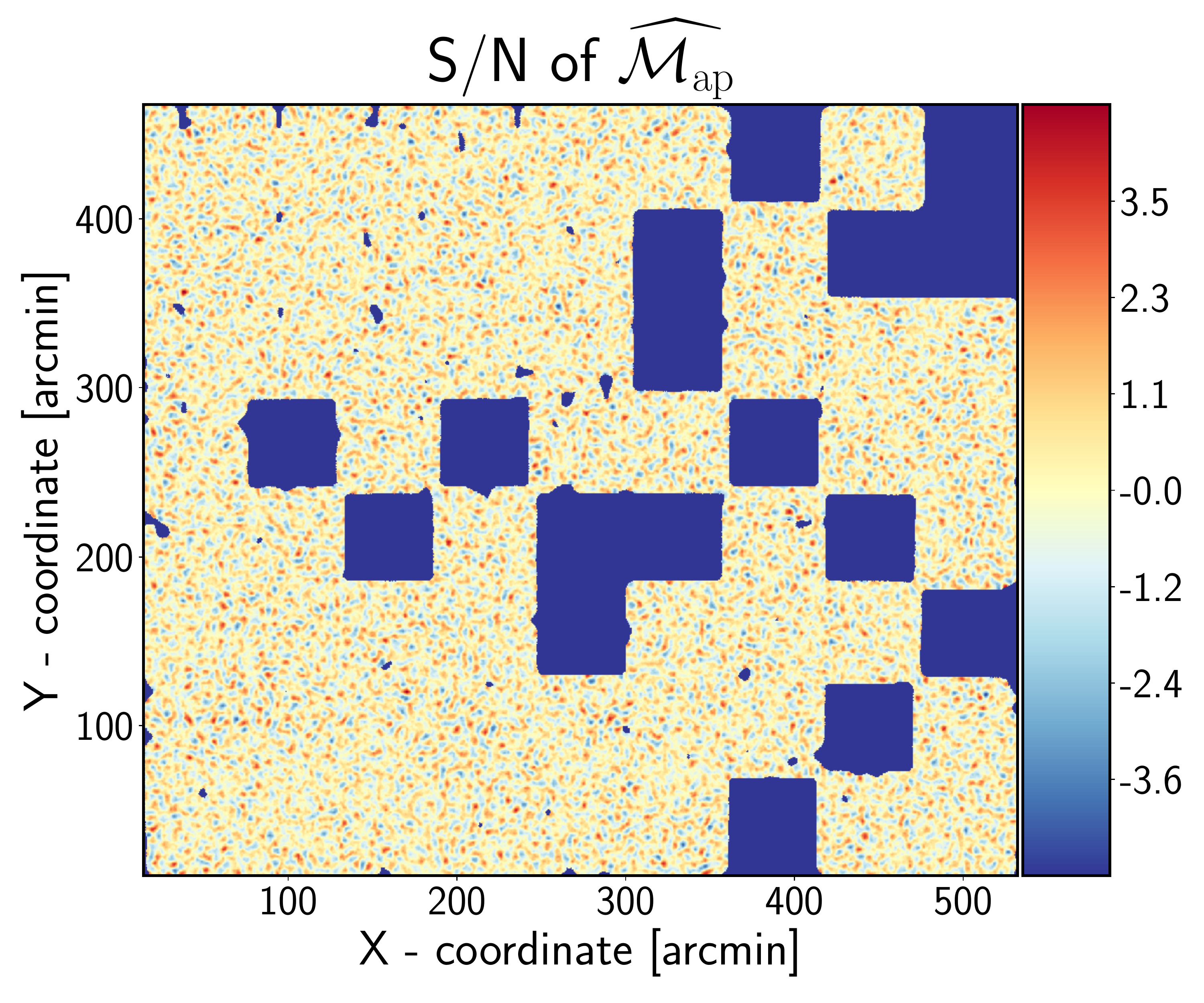}}
  \centering {
    \includegraphics[width=8.5cm]{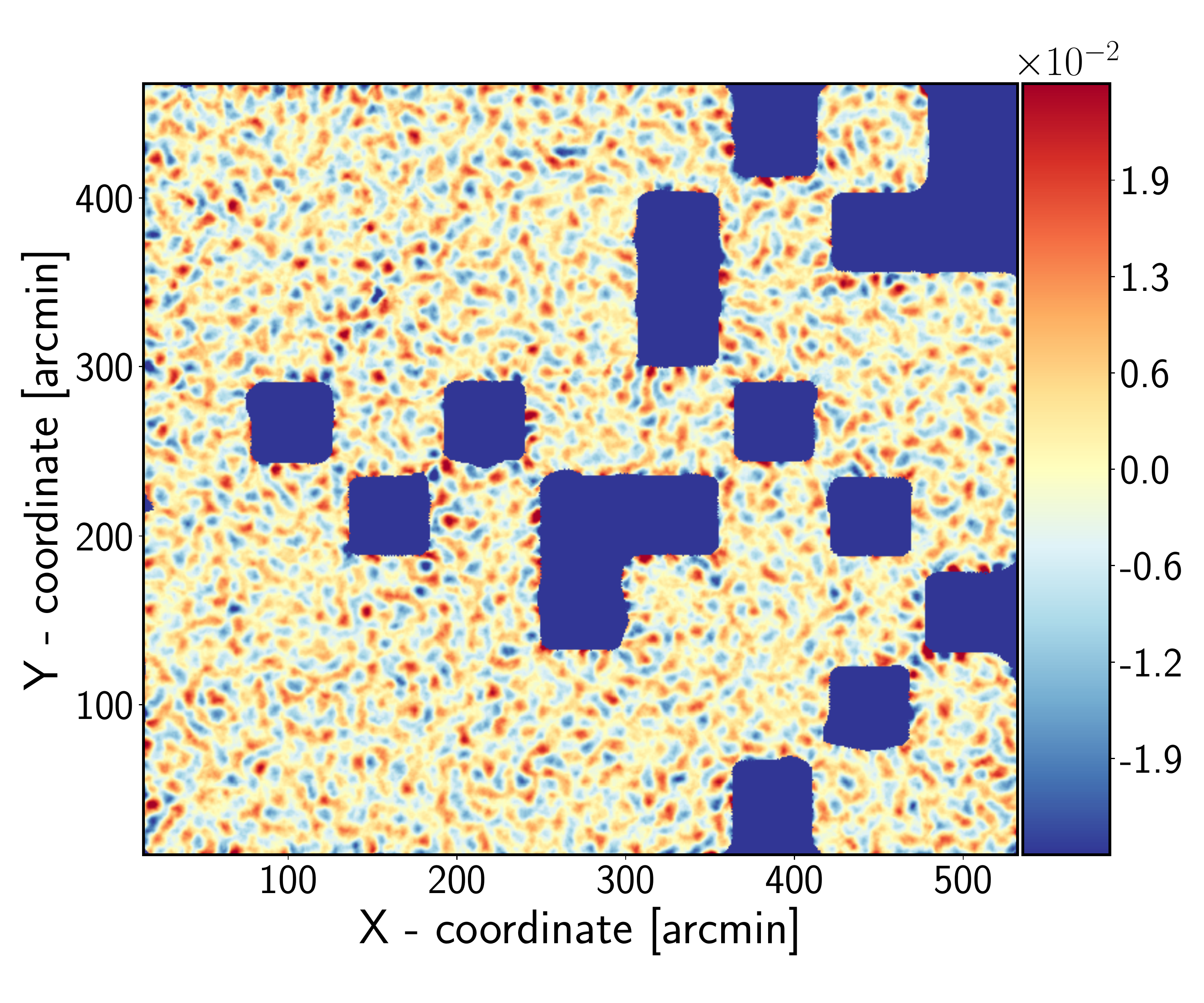}\hspace{0.2cm}
    \includegraphics[width=8.5cm]{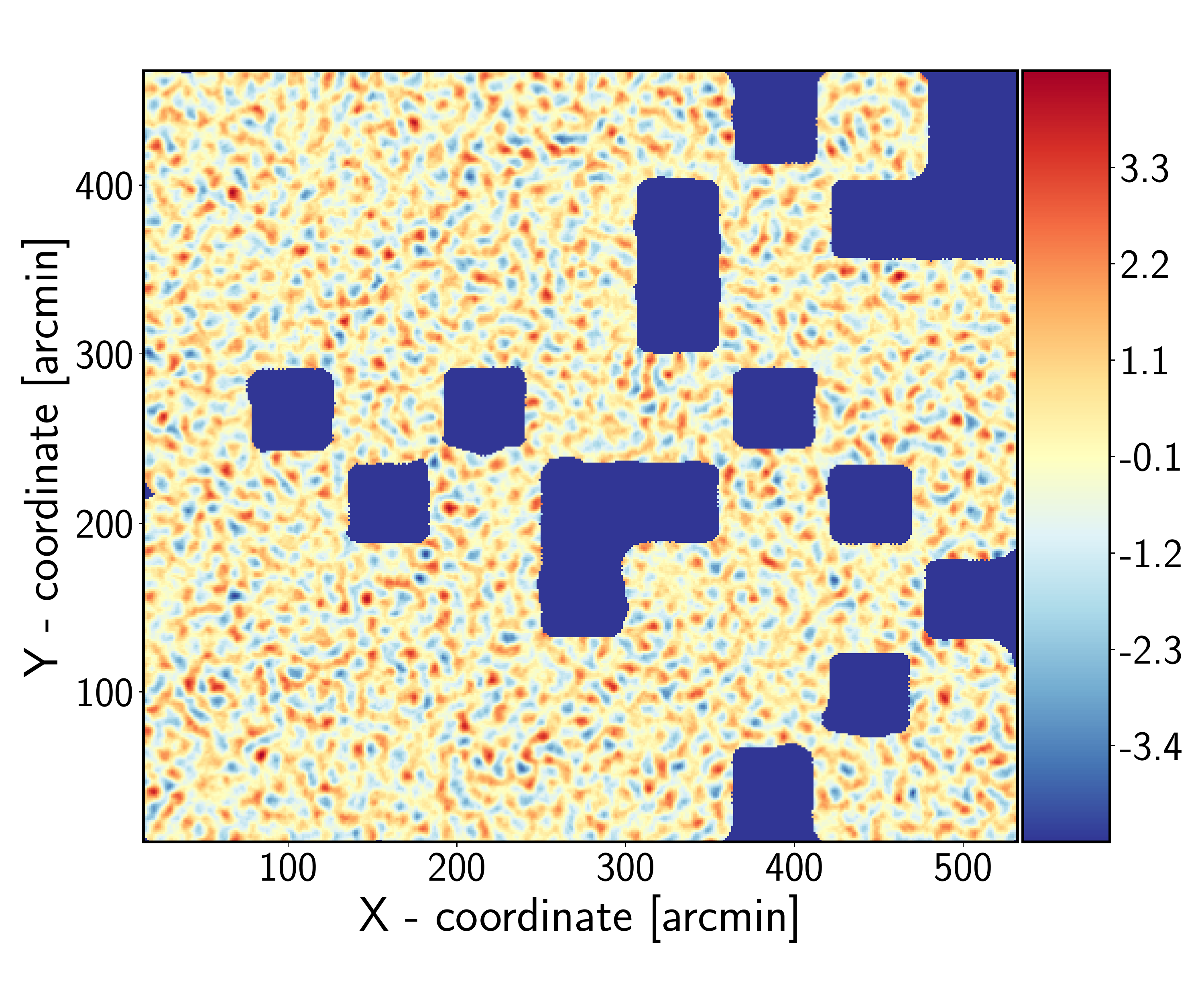}}
  \centering {
    \includegraphics[width=8.5cm]{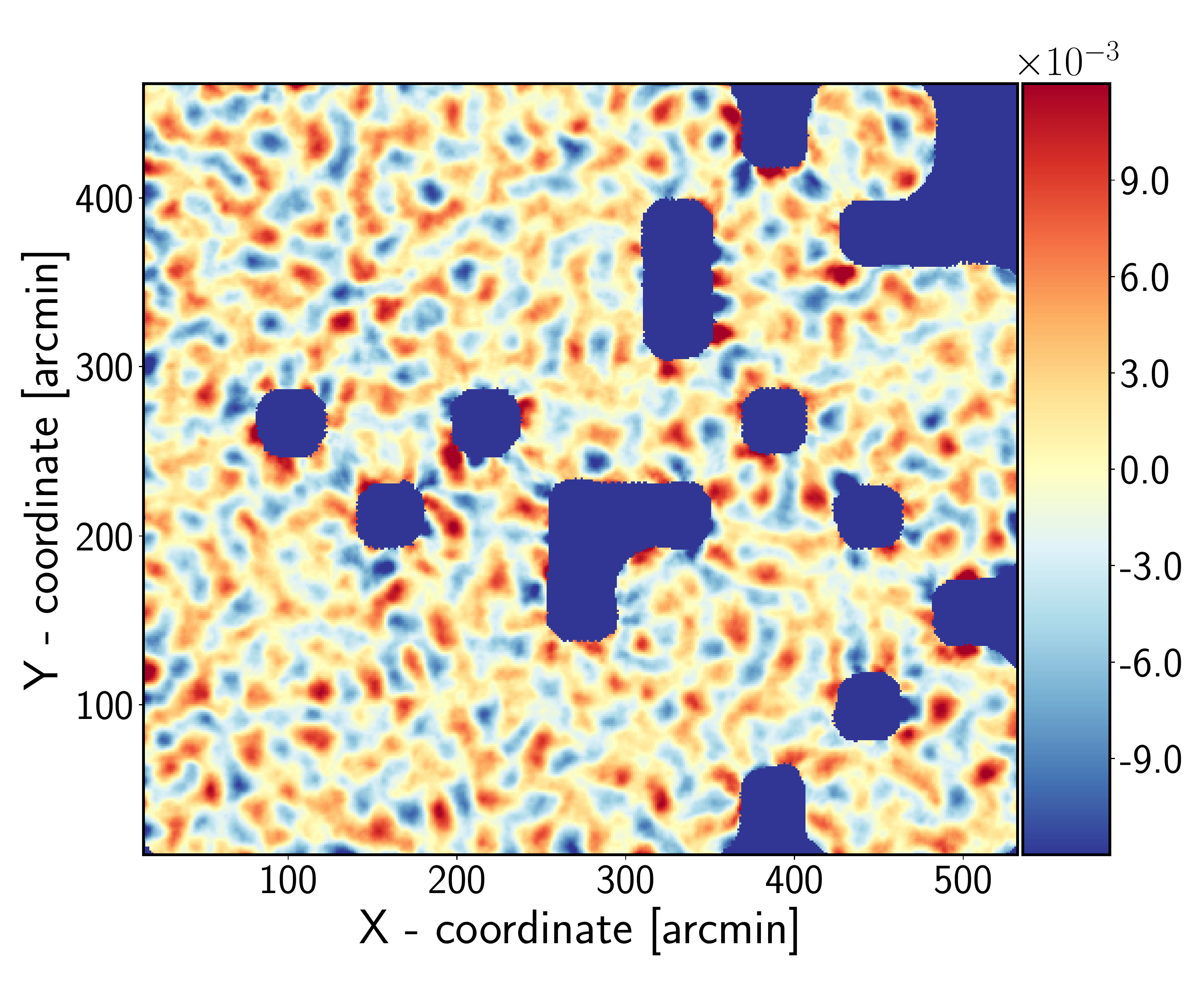}\hspace{0.2cm}
    \includegraphics[width=8.5cm]{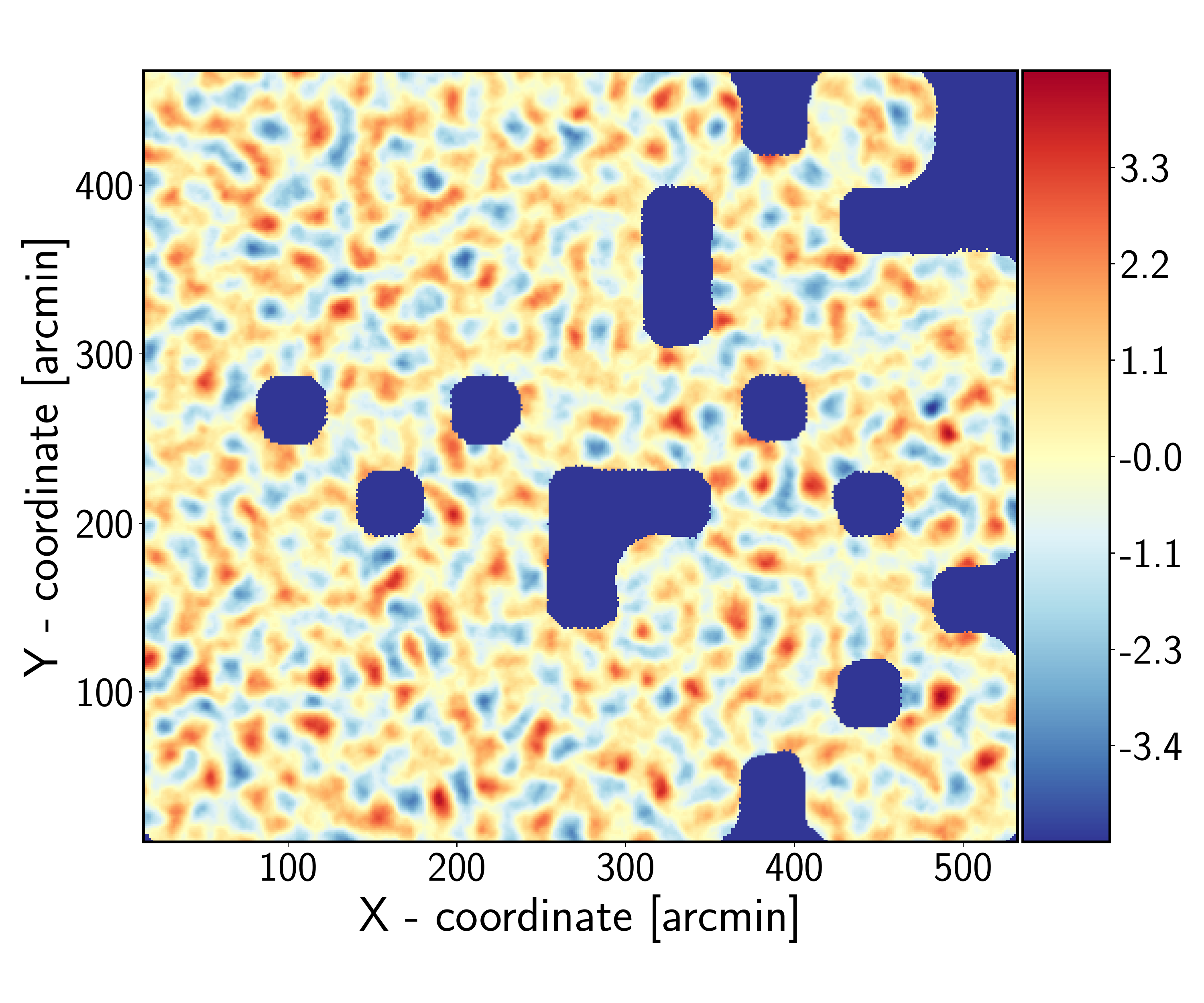}}
      \caption{\small{Mass related maps measured from the W1 field of the \cfhtls
          data. The x- and y-axes show the right-ascension and
          declination of the survey, in arcminutes. The left and right
          columns show the results for the aperture mass (${\mathcal M_{\rm
          ap}}(\bthet_0|\thetc)$) and the corresponding signal-to-noise ratio. The top, middle and bottom rows show the results for aperture radii of $5'$, $10'$, and $20'$, respectively.
          The value of the field at each location is indicated by the
          colour bar on the right.}\label{fig:massmap1}}
\end{figure*}

\begin{figure*}
  \centering {
    \includegraphics[width=8.5cm]{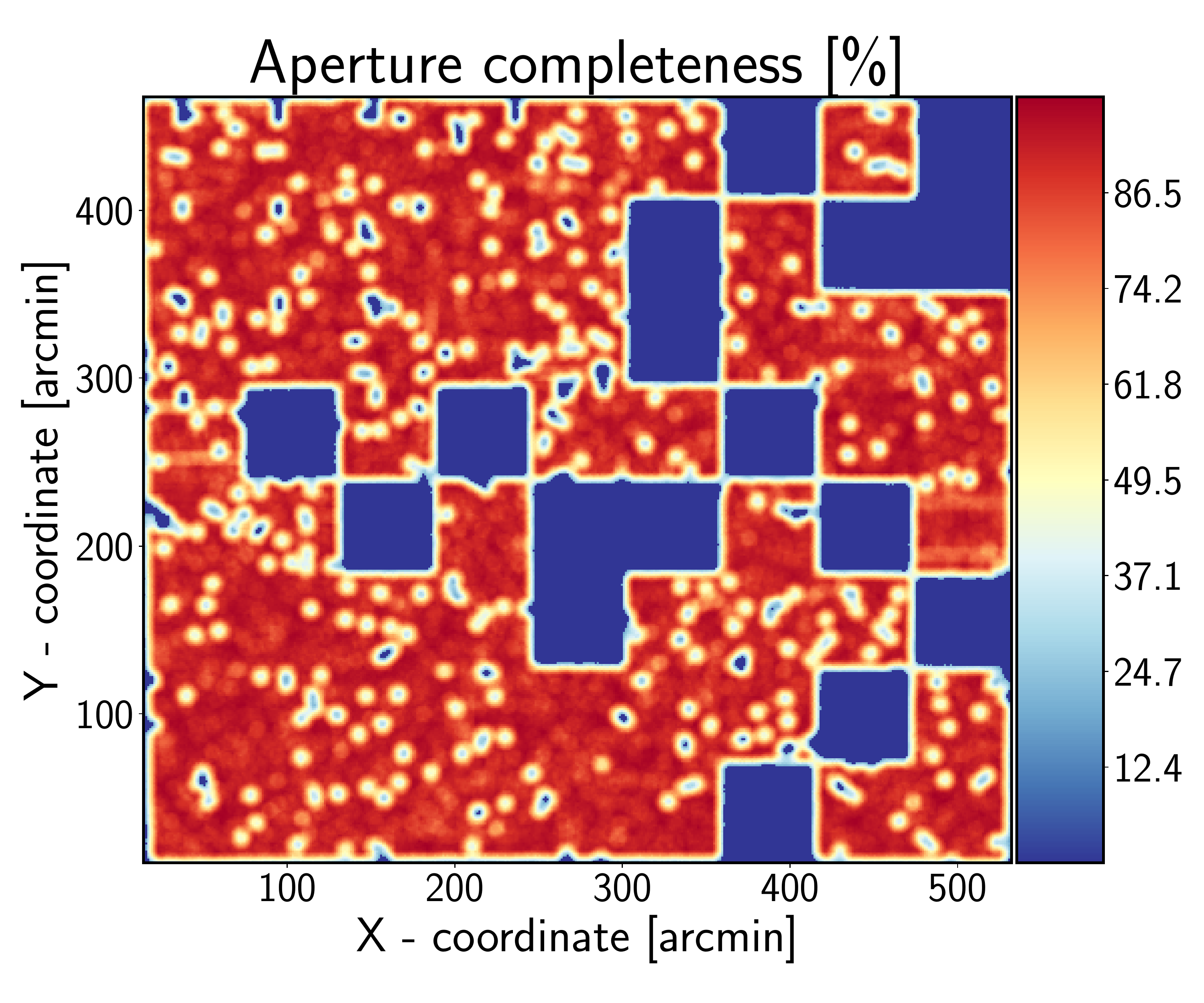}\hspace{0.2cm}
    \includegraphics[width=8.5cm]{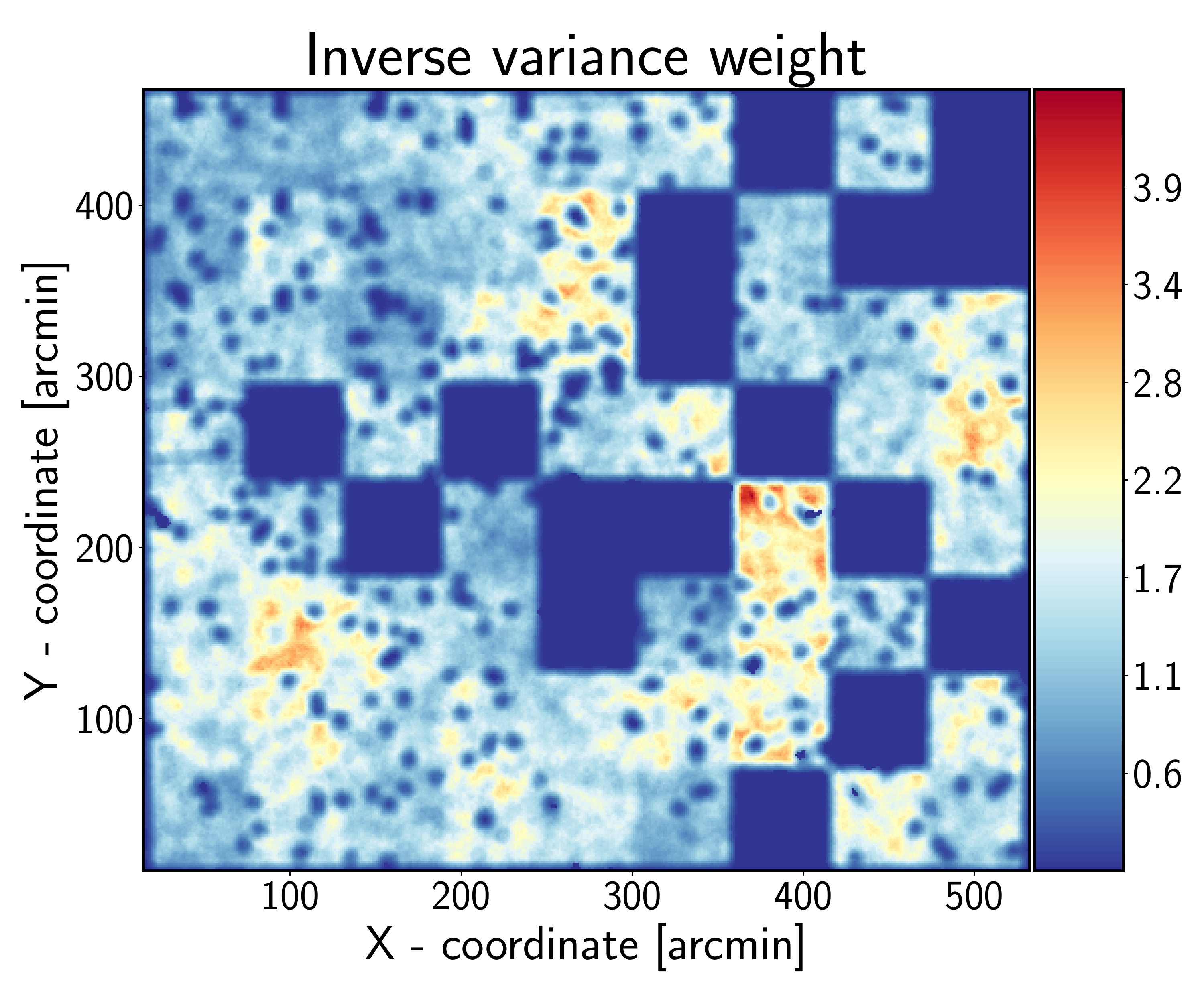}}
  \centering {
    \includegraphics[width=8.5cm]{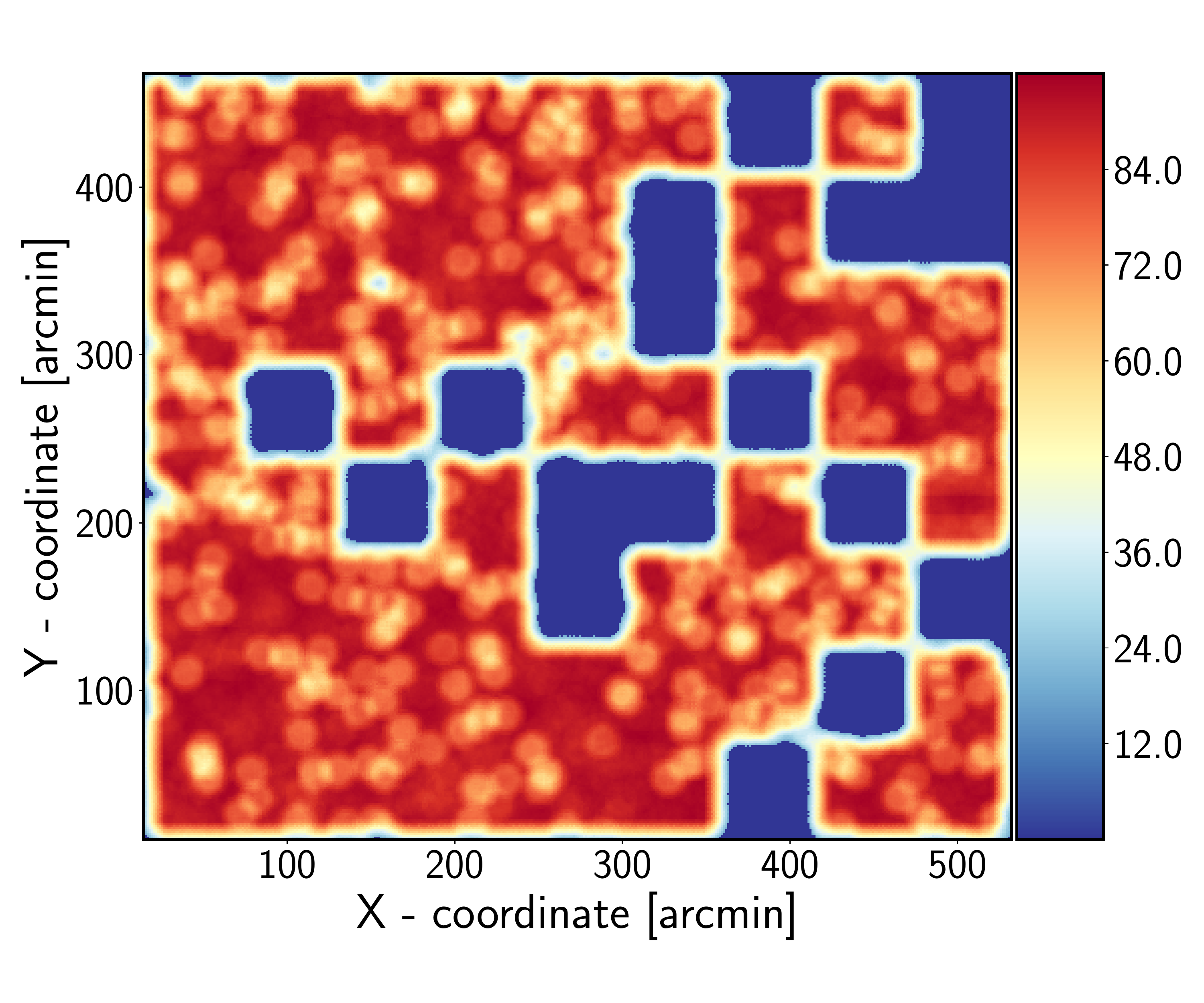}\hspace{0.2cm}
    \includegraphics[width=8.5cm]{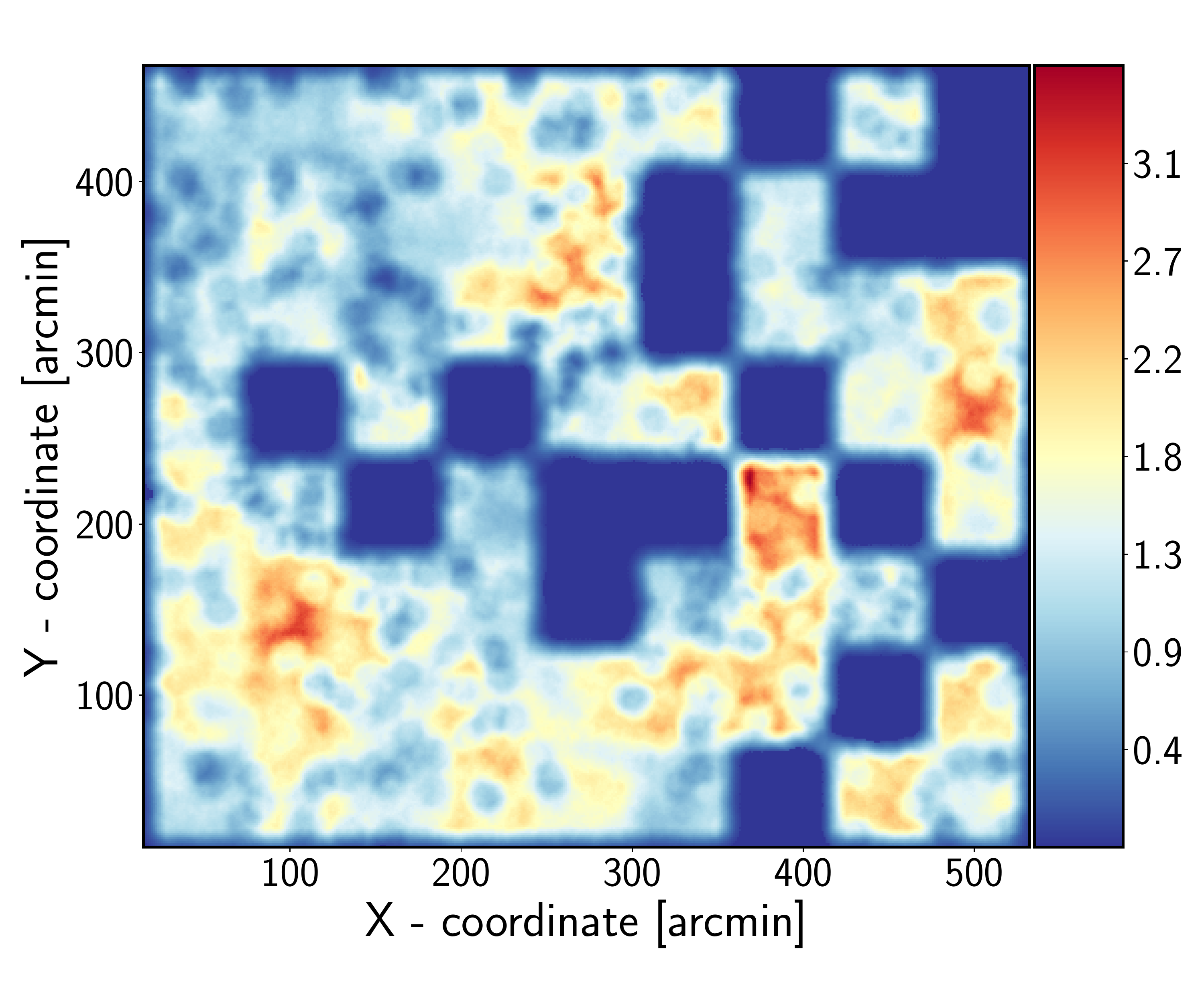}}
  \centering {
    \includegraphics[width=8.5cm]{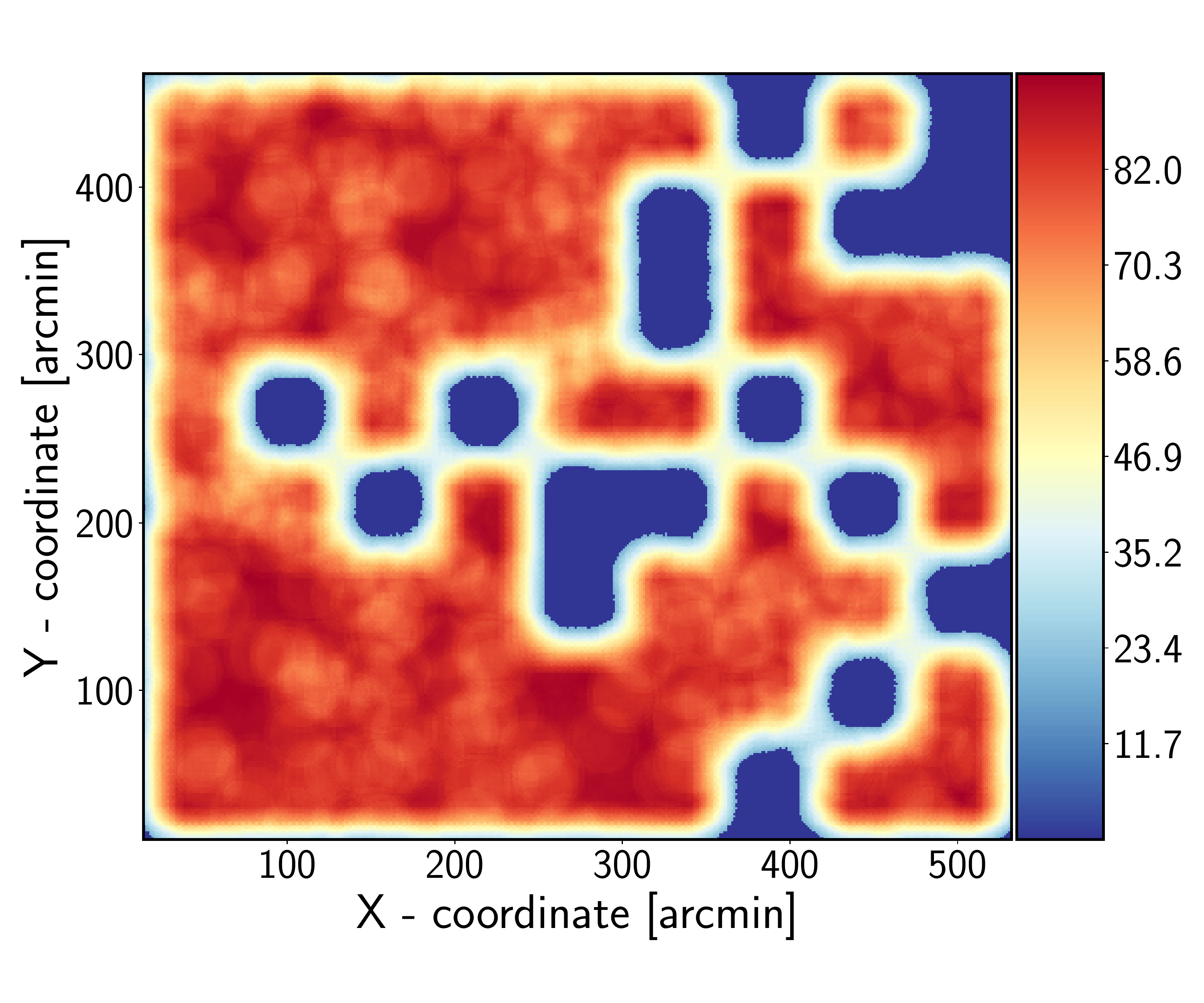}\hspace{0.2cm}
    \includegraphics[width=8.5cm]{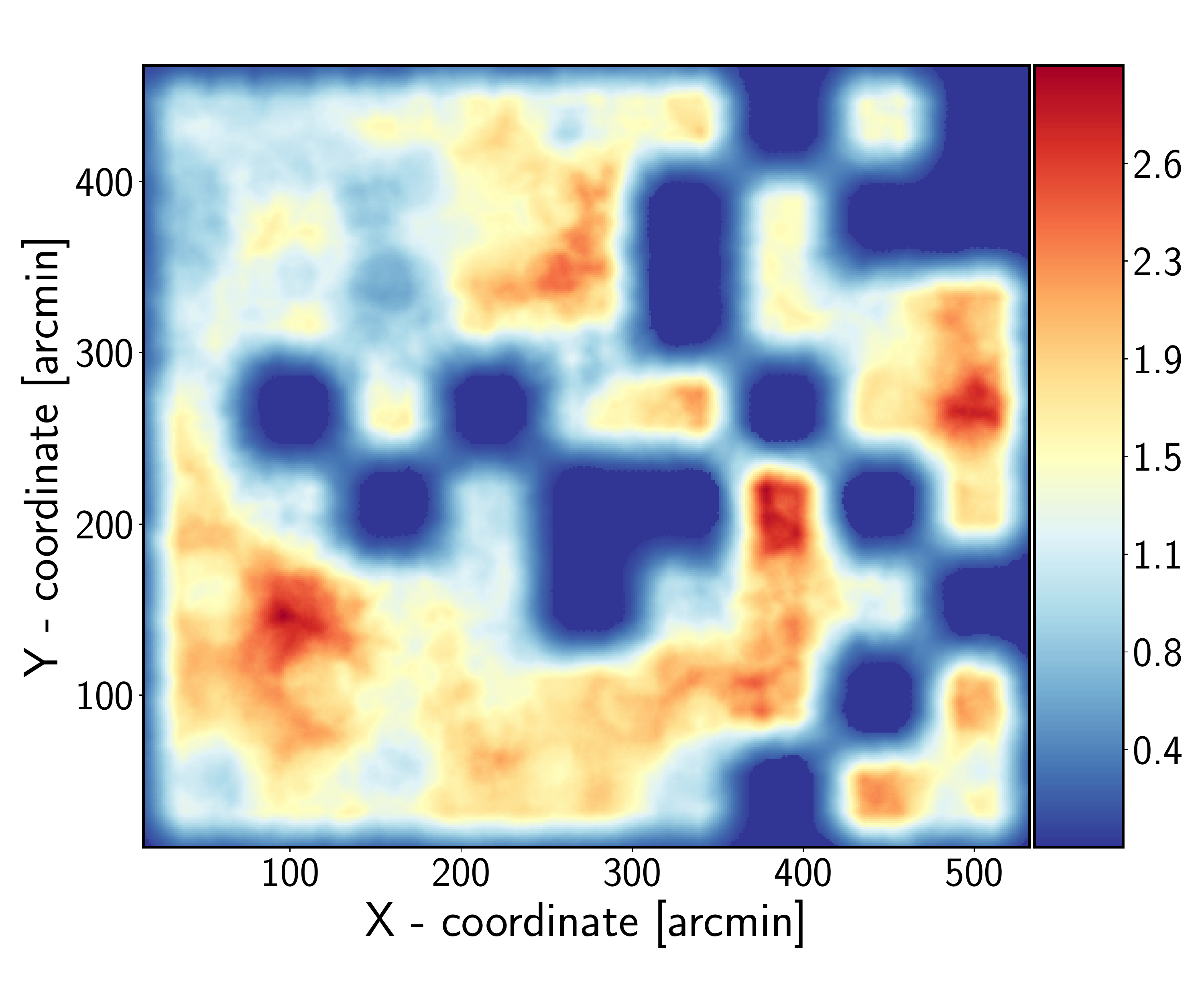}}
      \caption{\small{Same as Figure~\ref{fig:massmap1}, but this time
          columns indicate aperture completeness (left) and the aperture weights derived from \Eqn{eq:Weight3}, using the shot noise dominated limit of \Eqn{eq:varest18} (right). The weight maps shown are rescaled by their inverse mean. As expected, the weight depends on the local galaxy number density and on the aperture completeness. \label{fig:massmap2}}}
\end{figure*}

%
%
\begin{table}
\caption{Overview of the \cfhtls data. 
\label{tab:cfhtlens}}
\vspace{0.2cm}
\centering{
  \begin{tabular}{|l|c|c|c|}
    \hline
    Field & $\#$ of galaxies &  Angular Area $\Omega_s \ [{\rm deg}]^{2}$ & $\nbar \ [{\rm arcmin}]^{-2}$
    \\
    \hline
    W1    & 1871709          &  42.36    &  12.27 \\
    W2    & 499372           &  11.72    &  11.84 \\
    W3    & 1192084          &  25.23    &  13.12 \\
    W4    & 558515           &  12.55    &  12.36 \\
    \hline
  \end{tabular}}
\end{table}
%


In this work we make use of the final public data release. The
combined data set of W1, W2, W3 and W4 contains ellipticity
measurements for 4,121,680 galaxies. In Table~\ref{tab:cfhtlens} we
provide a summary overview of the data. Associated with each galaxy
are: the angular positions RA and DEC, in radians; the $x$- and
$y$-pixel coordinates in the projected tangent map; the ellipticity
estimates $\epsilon_1$ and $\epsilon_2$, the lens weights $w$; the
shear bias correction $c$ and the multiplicative bias correction $m$;
photometric redshift estimate $z_{\rm phot}$. Figure \ref{fig:reddist}
shows the redshift distribution and lensing efficiency for the sources
in each of the four \cfhtls fields.


\begin{table*}
\caption{{\zhorizon} cosmological parameters. Columns are (from left to right): density
  parameters for matter, dark energy and baryons; the amplitude of the
  power spectrum; the dark energy equation of state parameters; the
  spectral index of the primordial power spectrum; the Hubble
  parameter.
\label{tab:zHORIZONcospar}}
\vspace{0.2cm}
\centering{
\begin{tabular}{|c|c|c|c|c|c|c|c|c|}
\hline 
Cosmological model & $\Omega_{\rm m,0}$ & $\Omega_{\rm DE,0}$ & $\Omega_{\rm b,0}$ & 
$\sigma_8$ &  $w_0$  & $w_a$ &  $n_s$ &  $h$\\
\hline
{\tt Fiducial}    & 0.25 & 0.75 & 0.04 & 0.8 & -1  & 0 &  1 & 0.7\\
{\tt $\Omega_{\rm m,0}^{\pm}$}  & 0.2/0.3 & 0.8/0.7 & 0.04 & 0.8 & -1 & 0 & 1 & 0.7\\
{\tt $\Omega_{\rm b,0}^{\pm}$}  & 0.25 & 0.75 & 0.035/0.45 & 0.8 & -1 & 0 & 1 & 0.7\\
{\tt $\sigma_8^{\pm}$} & 0.25 & 0.75 & 0.04 & 0.7/0.9 & -1 & 0 & 1 & 0.7\\
{\tt $w_0^{\pm}$}  & 0.25 &  0.75 & 0.04 &  0.8 & -1.2/-0.8  & 0 & 1 & 0.7\\
{\tt $w_a^{\pm}$}  & 0.25 &  0.75 & 0.04 &  0.8 & -1 & -0.2/0.2 & 1 & 0.7\\
{\tt $n_s^{\pm}$}  & 0.25 &  0.75 & 0.04 &  0.8 & -1 & 0 & 0.95/1.05 & 0.7\\
{\tt $h^{\pm}$}  & 0.25 &  0.75 & 0.04 &  0.8 & -1 & 0 & 1 & 0.65/0.75\\
\hline 
\end{tabular}}
\end{table*}


\subsection{Simulating the \cfhtls data}\label{ssec:mocks}


In order to understand the statistical properties of the data we have
generated a large set of simulated \cfhtls skies.  These mock data
were generated from ray-tracing through $N$-body simulations. We used
the \zhorizon simulations, performed on the {\tt zBOX-2} and {\tt
  zBOX-3} supercomputers at the University of Z\"{u}rich, described in
detail in \cite{Smith2009}.  Each of the \zhorizon simulations
was performed using the publicly available {\tt Gadget-2} code
\citep{Springel2005}, and followed the nonlinear evolution under
gravity of $N=750^3$ equal-mass particles in a comoving cube of length
$L_{\rm sim}=1500\Mpc$; the softening length was $l_{\rm soft}=60\,
\kpc$. For all realizations 11 snapshots were output between redshifts
$z=[0,2]$; further snapshots were at redshifts $z=\{3,4,5\}$. The
transfer function for the simulations was generated using the publicly
available {\tt cmbfast} code \citep{SeljakZaldarriaga1996}, with high
sampling of the spatial frequencies on large scales. Initial
conditions were set at redshift $z=50$ using the serial version of the
publicly available {\tt 2LPT} code
\citep{Scoccimarro1998,Crocceetal2006}. The simulations correspond to
several cosmological models, with parameters varying around a fiducial
model. The latter closely matched the results of the WMAP experiment
\citep{Komatsuetal2009}. There are 8 simulations of the fiducial
model, and 4 of each variational model, matching the random
realization of the initial Gaussian field with the corresponding one
from the fiducial model.  Table~\ref{tab:zHORIZONcospar} summarizes
the cosmological parameters that we simulated.

From each \zhorizon simulation, 16 large fields of view were
generated by choosing different observer positions within the
simulation box. These large fields have side lengths of $12\,\degt$
and are covered by a regular mesh of $4096^2$ pixels. For each pixel,
a light ray was traced back through the simulation by a
multiple-lens-plane algorithm
\citep[see][]{Hilbertetal2007,Hilbertetal2009}, and its distortion due
to gravitational lensing was recorded for a set of $45$ source
redshifts between 0 and 4.

Each of the large fields was used to create four simulated
mock-\cfhtls wide field source galaxy catalogues for each of the
different \cfhtls fields W1 to W4 (i.e., 64 mock catalogues per
\cfhtls field and \zhorizon simulation). The basis for the
simulated source galaxy catalogues are the actual \cfhtls source
catalogues, from which the angular positions and redshifts were
taken. The reduced shear $g \equiv \gamma / (1-\kappa)$ for each galaxy in the mock catalogues was
computed by multi-linear interpolation of the simulated lensing
distortions onto the source galaxy's angular position and redshift
(using a different angular offset within the $12\times12\,\degt^2$
simulated fields for each mock catalogue). The `observed' source galaxy
ellipticities in the simulated catalogues were then computed by
combining the reduced shear from the ray-tracing and the randomly
rotated observed ellipticities from the actual source galaxy catalogue.


\subsection{Maps from the \cfhtls data}\label{ssec:maps}

As a first step in our analysis of the \cfhtls data, we compute
several aperture based maps for the four fields of the \cfhtl. To generate these maps the survey area was pixelated and the corresponding maps were computed for apertures located at the pixel centers. We furthermore only include apertures that are at least $20$ per cent complete, the map values for all less complete apertures are set to the minimum value and therefore appear as blue pixels. Images of the official \cfhtls masks are presented in Appendix~\ref{app:masks}.

In Figure~\ref{fig:massmap1} we show the aperture mass map and its corresponding signal-to-noise map for the W1 field. The aperture masses are estimated using \Eqn{eq:est6} while the noise for each aperture was estimated as \citep{Hetterscheidtetal2005}:
\be
\widehat{\sigma^2 \left( \mathcal{M}_{\mathrm ap} (\bthet_0|\thetc) \right)}
 = 
\frac{(\pi\thetc^2)^2\sum_i w_i^2Q_i^2\epsilon_{{\rm t}, i}^2}{2\left(\sum_iw_i\right)^2} \ .
\ee
The results are shown for the Schneider filter \eqn{eq:schneiderQ} with aperture extent in the
set $\thetc\ \in\ [5',\ 10',\ 20']$. It is interesting to note that near the survey mask boundaries
the value of the aperture mass obtains large positive and negative
values. This is due to the fact that for incomplete apertures the ring
averaged tangential shear becomes biased as the mask induces
non-vanishing $B$-modes (cross-shear). How to deal with the effects of
masked apertures will be discussed in detail below and in a companion paper. 

In Figure~\ref{fig:massmap2} we show the aperture completeness map as well as a map of the aperture weights derived from \Eqn{eq:Weight3}, using the shot noise dominated limit of \Eqn{eq:varest18}. We rescaled the latter map by its inverse mean such that the mean weight becomes unity. From there we can explicitly see that such aperture weights depend on the aperture completeness, cosmic structures and on the local depth of the survey. 
Analogous maps for the other three fields are presented in Appendix~\ref{app:CFHTLOtherMaps}.


\begin{figure*}
  \centering {
    \includegraphics[width=8.7cm]{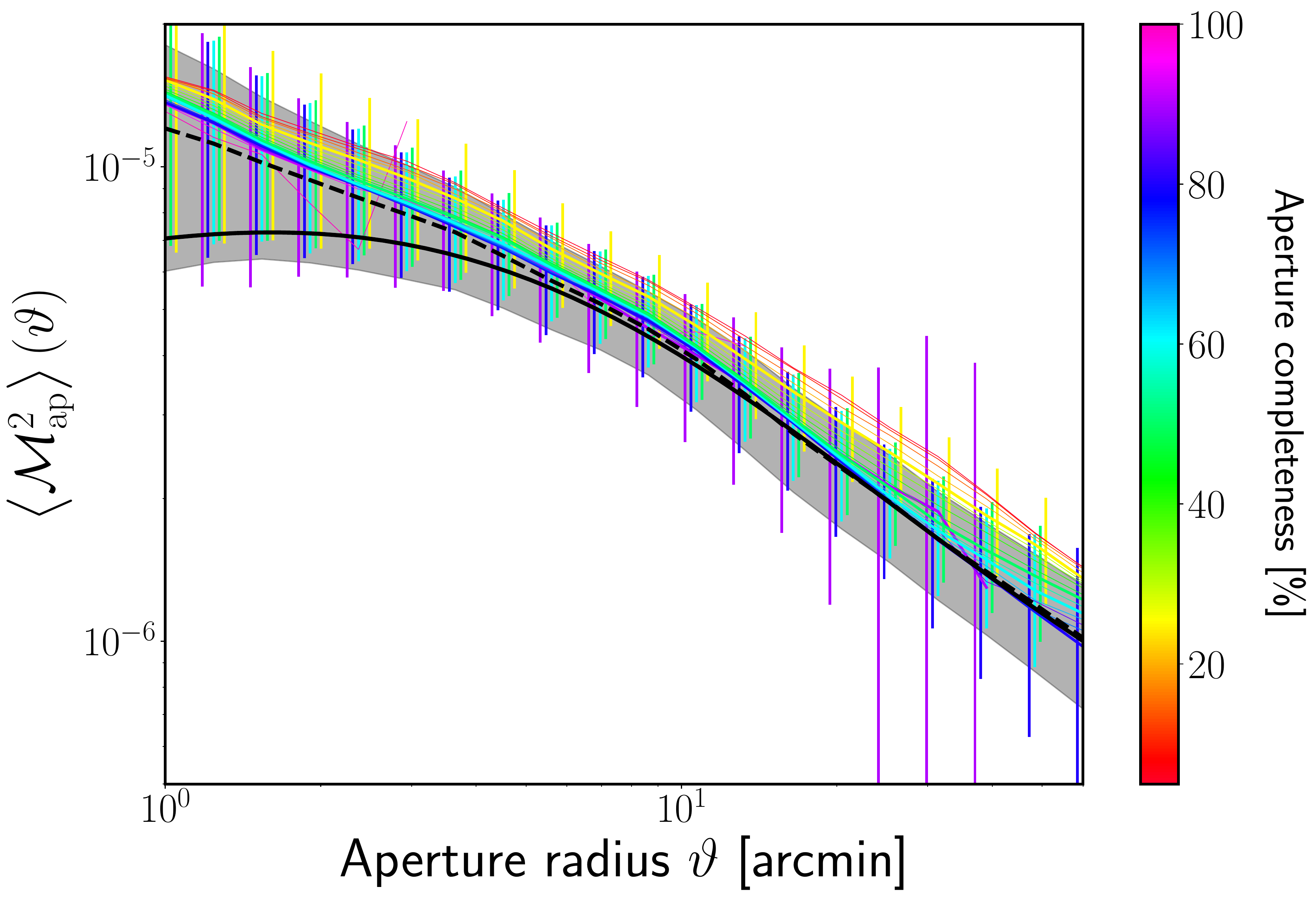}
    \includegraphics[width=8.7cm]{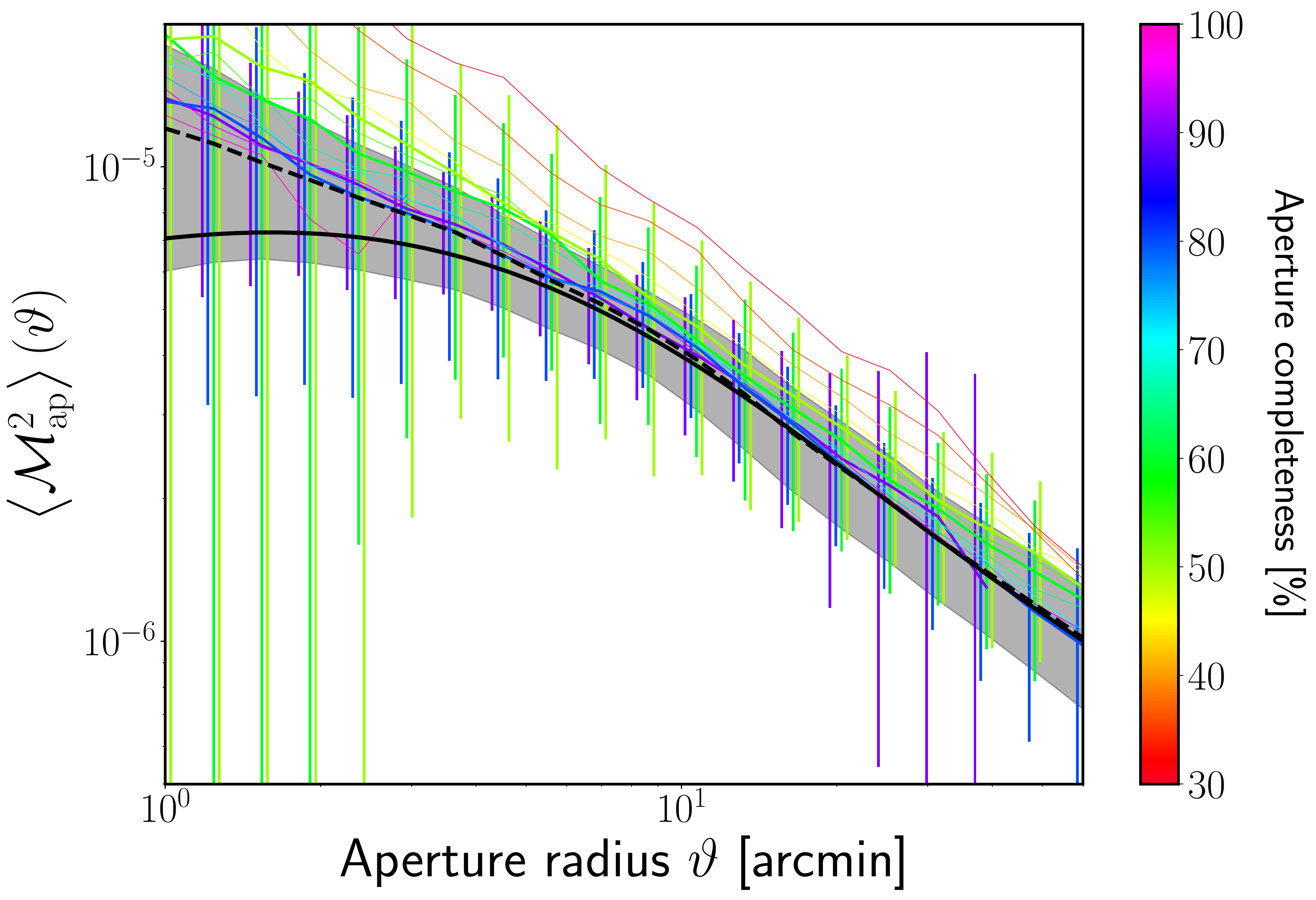}}
  \caption{\small{{\bf Left panel}: Aperture mass variance as a
      function of aperture radius computed from the 512 full-ray
      traced \cfhtls mock catalogues of the fiducial runs of the W1
      field for the different weighting schemes described in the
      text. The thick solid line gives the theoretical prediction evaluated via \eqn{eq:MapVar}. The thin solid lines show results from the direct approach
      where the estimates from each apertures are combined with equal
      weight (this is \Eqn{eq:Weight1} in the text). The colour of the
      lines indicates the value of the aperture completeness parameter
      $c_k=A_k/A$ -- only apertures with a completeness greater than
      this value are used in the estimate. The thick dashed line gives
      the result from the correlation function approach as obtained
      using the \treecorr routine \citep{Jarvisetal2004}. The
      error bars show the error for a single realisation of the
      \cfhtls W1 field. The grey shaded region are for \treecorr. Note that the errors from the direct estimator
      approach have been slightly offset for clarity. {\bf Right
        panel}: shows the same as the left panel, except this time
      the estimates from the direct estimator have been computed in
      bins of aperture completeness. The diffenence between the simulations and the theoretical predictions for small aperture radii can be attributed to shot noise and the line-of-sight discretization from which the simulated data suffers \citep{Hilbertetal2020}.}\label{fig:map2mockW1a}}
\end{figure*}


\begin{figure*}
  \centering {
    \includegraphics[width=8.7cm]{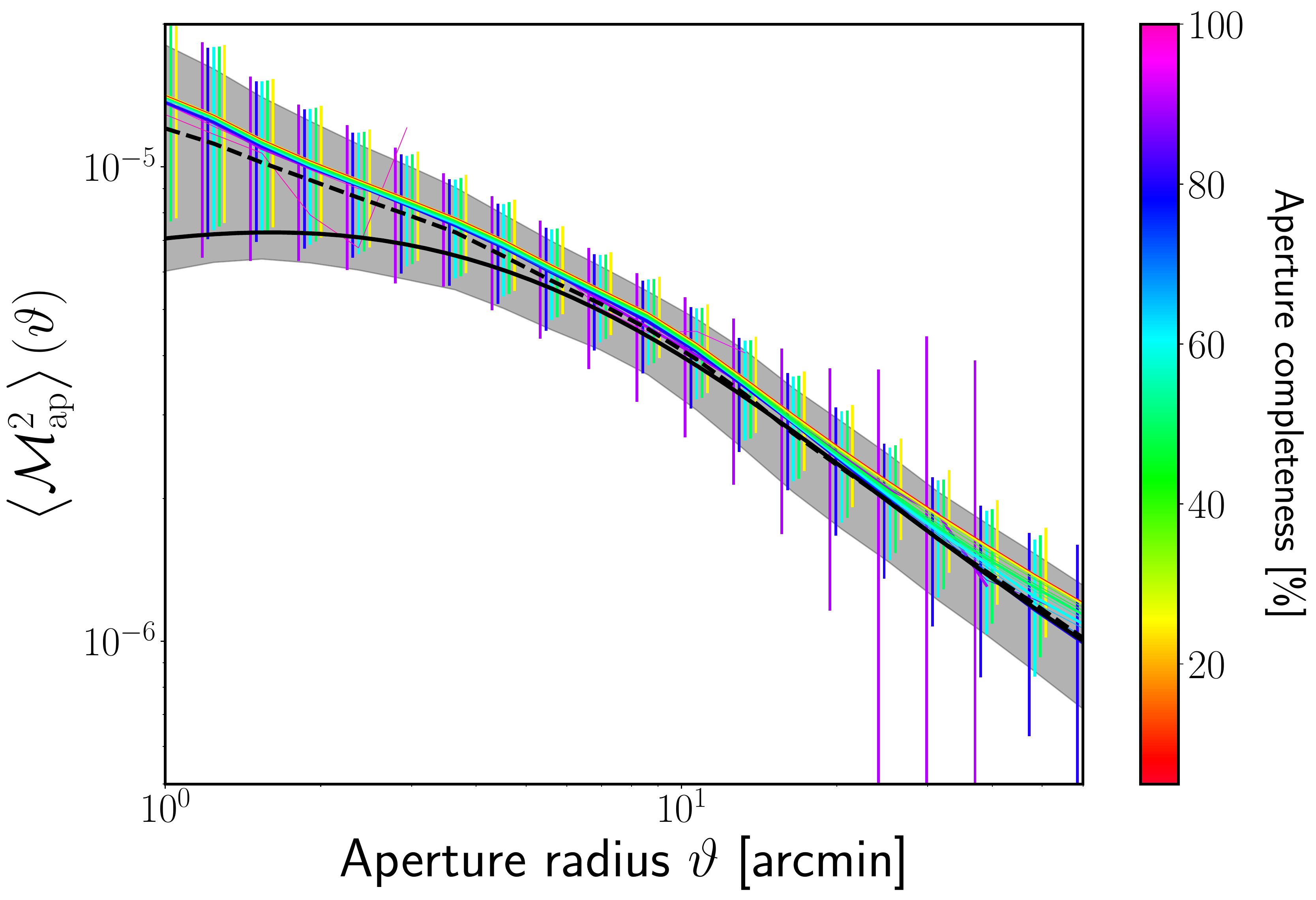}
    \includegraphics[width=8.7cm]{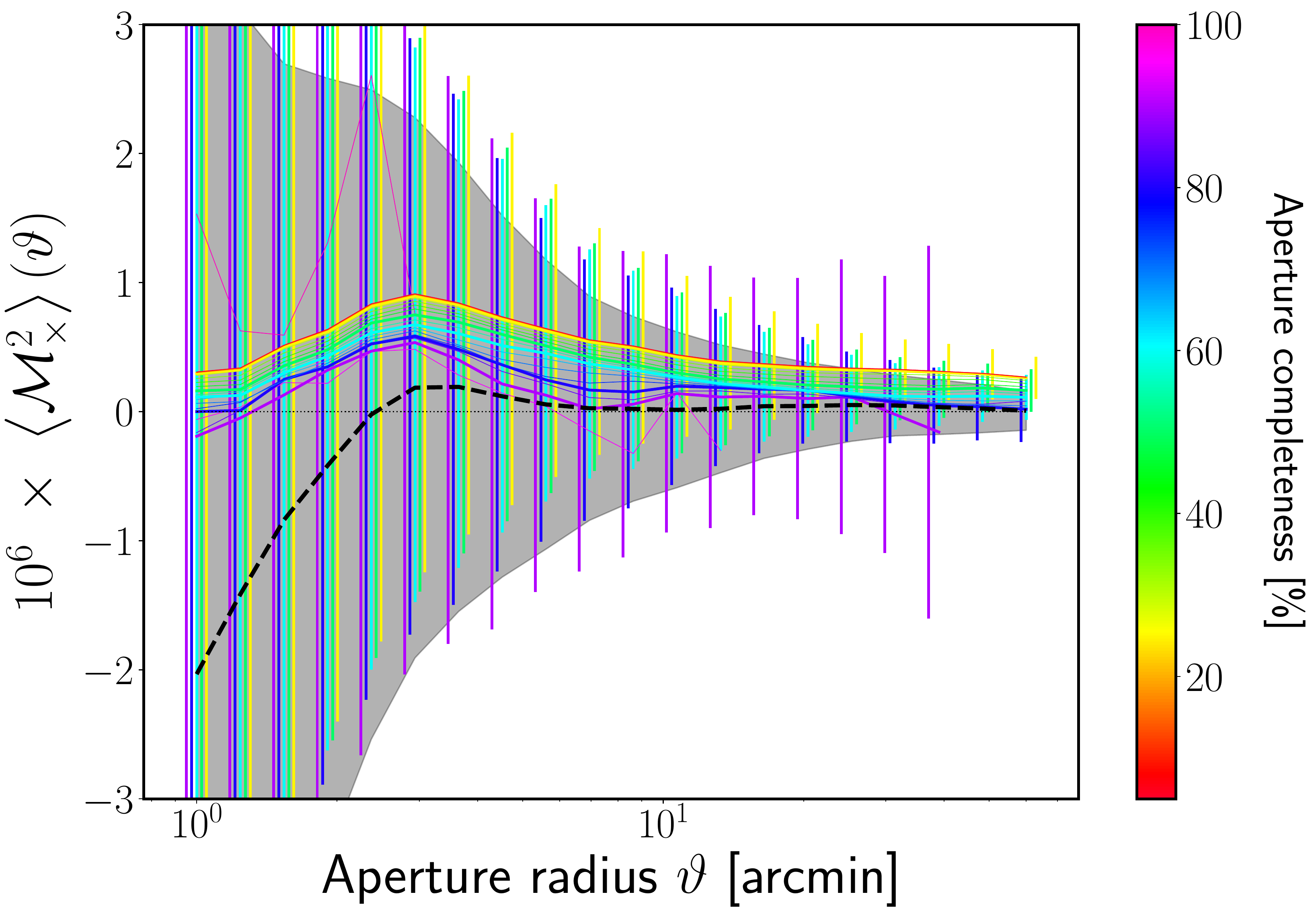}}
  \caption{\small{{\bf Left panel}: Same as
      Figure~\ref{fig:map2mockW1a}, however this time the estimates for
      each aperture are combined using an inverse variance weighting
      scheme (this is \Eqn{eq:Weight3} in the text). Note that when
      $c_k\lesssim10\%$ is equivalent to the case where all apertures
      are considered irrespective of coverage and their results are
      combined using an inverse variance weight.  {\bf Right Panel}:
      Same as the left panel, but this time showing the variance of
      the cross component of aperture mass. As shown in
      \Eqn{eq:CrossEst1} and in the absence of systematic errors,
      $\left<{\mathcal
        M}_{\times}^2(\thetc)\right>=0$. }\label{fig:map2mockW1b}}
\end{figure*}


\begin{figure*}
  \centering{
    \includegraphics[width=8.7cm]{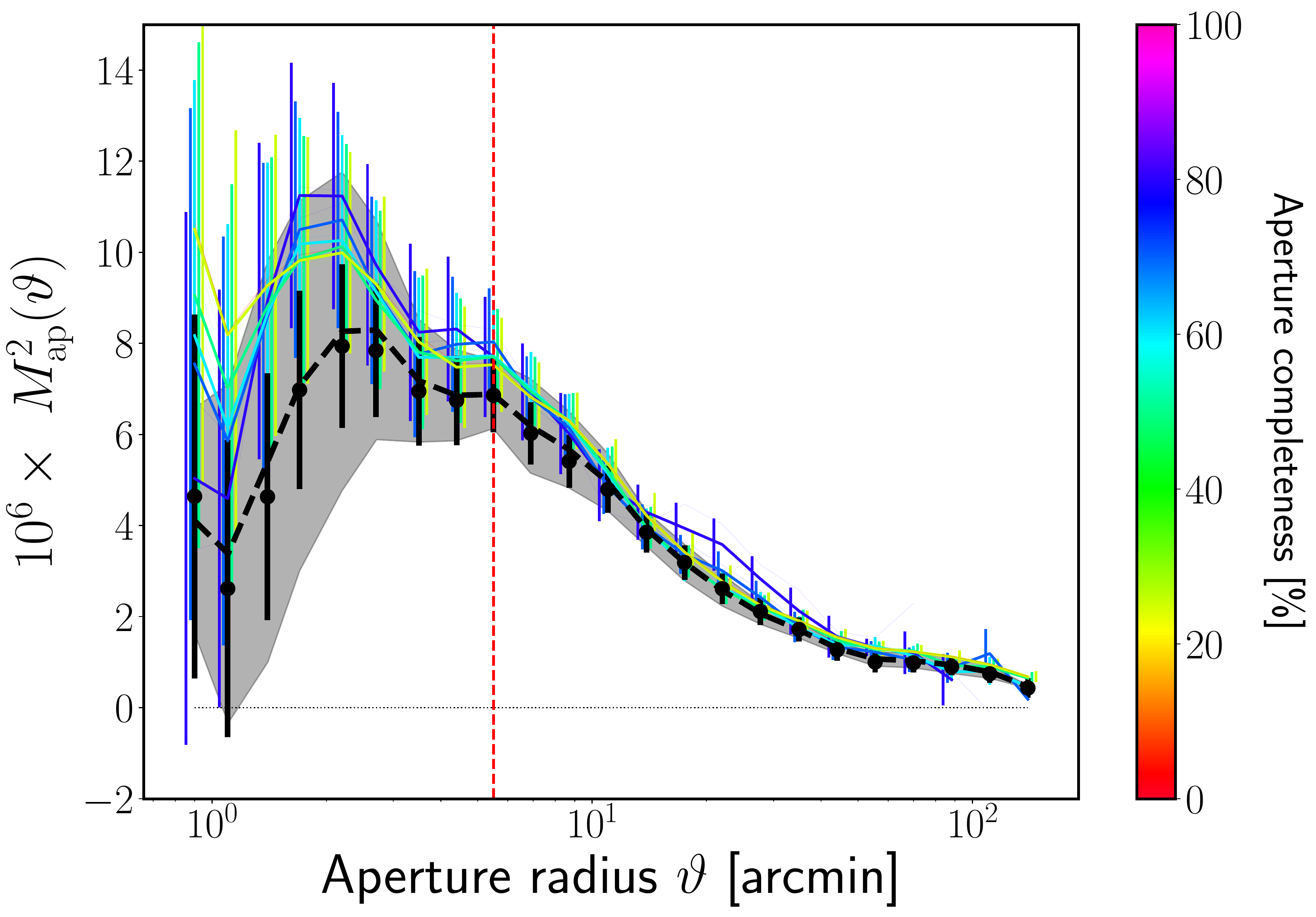}\hspace{0.2cm}
    \includegraphics[width=8.7cm]{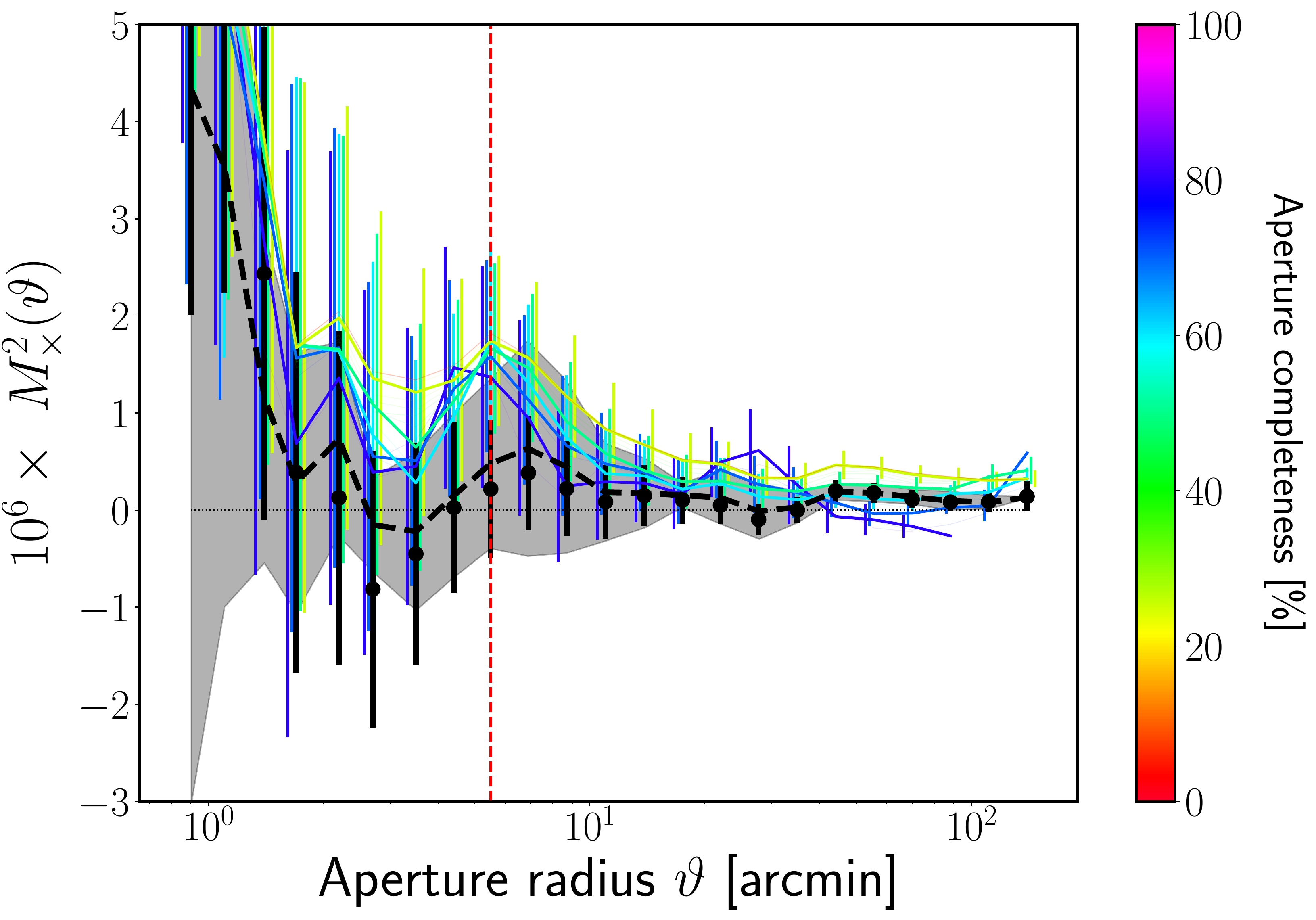}}
  \caption{\small{{\bf Left panel}: Aperture mass variance as a
      function of aperture cut-off scale as measured in the \cfhtls
      data. The results for the direct estimator are indicated as
      coloured lines, where the line colour indicates the value of
      aperture completeness that was used. The thick dashed line gives
      the result from the correlation function approach as obtained
      using the \treecorr routine \citep{Jarvisetal2004}. The
      error bars show the error for the full \cfhtl. {\bf
        Right panel}: Same as the left panel, but for the variance of
      the cross component of the aperture mass.}\label{fig:map2cfht_raw}}
\end{figure*}



\section{Measurements of the aperture mass variance}\label{sec:results}

We now turn to the estimation of the aperture mass variance from the
\cfhtls data using the direct and correlation function estimators.


\subsection{Analysis of the \cfhtls mock skies}\label{ssec:resmocks}

Before performing our statistical analysis of the real \cfhtls data
we first make a study of the direct and correlation function
estimators as applied to the mocks. From this we will be able to
determine whether the methods are consistent with one another to
within the errors and also which of the three weighting schemes given
by \Eqns{eq:Weight1}{eq:Weight3} provides the better method for
estimating $\left<{\mathcal M}_{\rm ap}^2\right>(\thetc)$.

The left panel of Figure~\ref{fig:map2mockW1a} shows the variance of
the aperture mass estimated from the mocks, as a function of the
aperture radius. For the results presented in this section we used a spacing of $d=0.25 \ \vartheta$ between the apertures. The thin coloured lines show the results from
the direct estimator approach where the estimates from individual
apertures are combined with equal weight, but where a completeness
thresholds $c_k$ has been adopted -- this is equivalent to weight
scheme ${\mathcal W}_1$ c.f. \Eqn{eq:Weight1}.  For example the
magenta lines show a conservative case, where only apertures with
completeness $c_k\gtrsim 90\%$ are taken, whereas the red lines are
the most relaxed where all apertures with $c_k\gtrsim 10\%$ are
allowed. The thick dashed line shows the results obtained from the
correlation function method, where the pair counts have been measured
using the \treecorr code of \citet{Jarvisetal2004}\footnote{We computed the shear correlation functions down to $6''$, used $100$ logarithmically spaced bins per decade and set the {\tt binslop} parameter to $0.1$.}. The grey
shaded regions shows the standard error for the correlation function
method and the error bars are the errors on direct estimator. In all
cases these were determined from the ensemble of 512 mocks of our
fiducial model. The right panel of Figure~\ref{fig:map2mockW1a} shows
the same as the left panel, however the results from the direct
estimator have been binned in completeness. Both panels also show the theoretical prediction evaluated via \eqn{eq:MapVar}. 

The figure shows that there is a clear bias in the direct estimator
approach when the apertures have a low completeness. This is manifest
as an increased amplitude of the signal on all scales. However, for
apertures with $c_k\gtrsim75\%$ completeness (blue line) we find that
the results are in good agreement with the case of $c_k\gtrsim90\%$
(magenta line) and that these are fully consistent with the
correlation function results on large scales, to within the errors.
We note that on scales smaller than $\sim5'$ the results from \treecorr appear to be biased slightly low to those from the direct estimator approach. We furthermore note a difference between the theoretical prediction and the measurements for small aperture radii which is due to shot noise and the line-of-sight discretization from which the simulated data suffers \citep{Hilbertetal2020}.

The left panel of Figure~\ref{fig:map2mockW1b} again shows the
aperture mass variance estimated from the mocks for the two methods,
but this time the estimates from individual apertures are combined
using the inverse variance weighting schemes with a completeness
threshold -- that is we now employ ${\mathcal W}_2$ and ${\mathcal
  W}_3$, c.f. \Eqns{eq:Weight2}{eq:Weight3}. The right panel of
Figure~\ref{fig:map2mockW1b} shows the variance of the cross component
of the aperture mass $\left<{\mathcal M}_{\times}^2\right>(\thetc)$
from \Eqn{eq:CrossEst1}, which for B-mode free fields should
vanish.

There are a number of interesting points to note from this
analysis. First, we see that all of the estimates from the direct
estimator are consistent with one another and that they all lie within
the error bars of the \treecorr result. Nevertheless, for scales
$\thetc>20'$ we still observe that the direct estimator approach
appears to be slightly biased high on large scales for aperture
completeness values $c_k\gtrsim 40\%$ compared to the correlation
function method. However, for aperture completeness levels
$c_k\gtrsim75\%$ we see excellent agreement between the two methods.
On small scales, $\thetc<5'$, the correlation function method gives slightly lower results than the direct
estimator.  Here we believe that the direct estimator is correct,
since as was noted in \citet{Kilbingeretal2006}, the correlation
function method is biased low on scales of the order $\thetc\sim1'$
due to the absence of correlation function bins on small scales. In
the \treecorr code the small-scale cutoff in the pair counts is
set at $\thetc>6''$. Note that in the mock data there is no image
blending and so the direct estimator should not suffer from this
suppression. For the observed \cfhtls data this is not necessarily
the case.

We also note that, as shown in the right panel of
Figure~\ref{fig:map2mockW1b}, the cross component of the aperture mass
variance is consistent with zero to within the error bars for all
completeness fractions used. However, the $c_k\gtrsim40\%$ shows a
small positive offset from zero at the level of below $5 \times
10^{-7}$. On large and small scales the bias is very small for
$c_k\gtrsim75\%$. This gives us further confidence that the
discrepancy between the direct estimator and \treecorr on small
scales is due to the bias in the correlation function approach.

Another point of interest, is that we see the error bars on the most
conservative completeness cuts, $c_k\gtrsim95\%$, are significantly
larger than those obtained from the correlation function method. This
makes sense, since, for the case of the most conservative cuts, one can
find only a few apertures that meet the criterion. On the other hand,
for completeness fractions of the order $c_k\gtrsim 70\%$, the error
bars between the two methods are comparable.

Based on the discussion above, we will be using the weighting scheme ${\mathcal W}_3$ for the analysis of the observed \cfhtls data.


\subsection{Calibration of the estimators for ellipticity bias}\label{ssec:bias}

We now turn to the analysis of the observed data. As discussed in
\S\ref{sec:est} the aperture mass variance for a set of apertures can
be directly estimated using \Eqns{eq:Sest1}{eq:est1}.  However, owing
to calibration errors in the \lensfit shape estimation algorithm
\citep{Milleretal2013}, each galaxies ellipticity has a corresponding
multiplicative $m$ and additive $c$ bias. Hence, the observed and true
ellipticity components of the $i$th galaxy are related through:
\begin{align}
\epsilon_{1,i}^{\rm obs} & = (1+m_i)\epsilon_{1,i}^{\rm true} +c_{1,i} \ ; \label{eq:epsi1Obs}\\
\epsilon_{2,i}^{\rm obs} & = (1+m_i)\epsilon_{2,i}^{\rm true} +c_{2,i} \ . \label{eq:epsi2Obs}
\end{align}
For the \cfhtls $c_1$ was found to be consistent with zero, however
$c_2$ was found to have a $S/N$ and size dependent bias that was
subtracted from each galaxy. On average this correction was of the
order $2\times10^{-3}$.

\vspace{0.1cm}
\noindent {\bf Correction for shear correlation functions:} following
\citet{Milleretal2013}, the shear correlation functions can be
corrected for multiplicative bias through the following procedure. The
'raw' shear correlation functions are first estimated from the
`observed' shears using the estimator:
\be \widehat{\xi^{\rm raw}_{\pm}}(\theta_{\alpha}) =
\frac{\sum_i\sum_{j\ne i} w_i w_j \left[\epsilon^{\rm obs}_{{\rm t},i} \epsilon^{\rm obs}_{{\rm t},j} \pm \epsilon^{\rm obs}_{\times,i}
    \epsilon^{\rm obs}_{\times,j}\right]\Pi(\theta_{ij}|\theta_{\alpha})}
{\sum_{i} \sum_{j\ne i} w_iw_j\Pi(\theta_{ij}|\theta_{\alpha})}\ , \ee
where $\theta_{ij}\equiv |\bthet_i-\bthet_j|$ and where
$\Pi(\theta_{ij}|\theta_{\alpha})$ is the pair binning function which
is unity if $\theta_{ij}$ lies in the range
$[\theta_{\alpha}-\Delta\theta/2,\theta_{\alpha}+\Delta\theta/2)$.  If
  we now insert \Eqns{eq:epsi1Obs}{eq:epsi2Obs} into the above
  estimator (taking $c_{1,i}=0$ and $c_{2,i}=0$) we find:
\be \widehat{\xi^{\rm raw}_{\pm}(\theta_{\alpha})} =
\frac{\sum_{j\ne i} w_i' w_j' 
  \left[\epsilon^{\rm true}_{{\rm t},i} \epsilon^{\rm true}_{{\rm t},j} \pm \epsilon^{\rm true}_{\times,i}
    \epsilon^{\rm true}_{\times,j}\right]\Pi(\theta_{ij}|\theta_{\alpha})}
{\sum_{j\ne i} w_iw_j\Pi(\theta_{ij}|\theta_{\alpha})}\ , \ee
where in the above we introduced the new weights $w_i'\equiv
w_i(1+m_i)$, then we see that the above equation can be rewritten:
\begin{align}
  \widehat{\xi^{\rm raw}_{\pm}(\theta_{\alpha})} & = \left(\frac{\sum_{j\ne i} w'_i w'_j 
  \left[\epsilon^{\rm true}_{{\rm t},i} \epsilon^{\rm true}_{{\rm t},j} \pm \epsilon^{\rm true}_{\times,i}
    \epsilon^{\rm true}_{\times,j}\right]\Pi(\theta_{ij}|\theta_{\alpha})}
    {\sum_{j\ne i} w_i'w_j'\Pi(\theta_{ij}|\theta_{\alpha})}\right)\nn \\
    & \times \left(\frac{\sum_{j\ne i} w_i'w_j'\Pi(\theta_{ij}|\theta_{\alpha})}
         {\sum_{j\ne i} w_iw_j\Pi(\theta_{ij}|\theta_{\alpha})}\right)\ ,
\end{align}
The first term on the right in parenthesis is the `calibrated' true
shear correlation which we can write as $\xi^{\rm cal}_{\pm}$, hence we
can write:
\be
\widehat{\xi^{\rm cal}_{\pm}(\theta_{\alpha})}  = \frac{\xi^{\rm raw}_{\pm}(\theta_{\alpha})}{1+{\mathcal K}(\theta_{\alpha})}
\ \ ; \ \  
1+{\mathcal K}(\theta_{\alpha})  \equiv
\frac{\sum_{j\ne i} w_i' w_j' \Pi(\theta_{ij}|\theta_{\alpha})}
     {\sum_{j\ne i} w_iw_j\Pi(\theta_{ij}|\theta_{\alpha})} \ .
\ee

\vspace{0.1cm}
\noindent {\bf Correction for direct aperture mass variance estimator:} following in
the footsteps of the shear correlation function approach, we define the raw uncorrected
aperture mass variance estimate as:
\begin{align}
  \left[\widehat{{\mathcal M}_{\rm ap}^2(\thetc|\btheta_{0,k})}\right]^{\rm raw} & = (\pi\thetc^2)^2 \
  \frac{\sum_{j\ne i} w_i \ w_j \ Q_i \ Q_j \ \epsilon^{\rm obs}_{{\rm t},i} \ \epsilon^{\rm obs}_{{\rm t},j}}{\sum_{j\ne i}
    w_i w_j}
  \label{eq:_est1}\ ,
\end{align}
recalling that the sums only extend over the galaxies in the aperture.
If we now insert \Eqns{eq:epsi1Obs}{eq:epsi2Obs} into the above
estimator, as before, then we find:
\begin{align}
  \left[\widehat{{\mathcal M}_{\rm ap}^2(\thetc|\btheta_{0,k})}\right]^{\rm raw} & = (\pi\thetc^2)^2 \
  \frac{\sum_{j\ne i} w_i' \ w_j' \ Q_i \ Q_j \ \epsilon^{\rm true}_{{\rm t},i} \
    \epsilon^{\rm true}_{t,j}}{\sum_{j\ne i} w_i w_j}
  \label{eq:_est1}\ .
\end{align}
Again, on redefining the \lensfit weights this leads us to write:
\begin{align}
  \left[\widehat{{\mathcal M}_{\rm ap}^2(\thetc|\btheta_{0,k})}\right]^{\rm raw} & = 
  \left((\pi\thetc^2)^2\frac{\sum_{j\ne i} w_i' \ w_j' \ Q_i \ Q_j \ \epsilon^{\rm true}_{{\rm t},i} \ \epsilon^{\rm true}_{{\rm t},j}}
       {\sum_{j\ne i} w_i' w_j'} \right)\nn \\
       & \times \left(\frac{\sum_{j\ne i} w_i' w_j'}{\sum_{j\ne i} w_i w_j}
       \right)
  \label{eq:_est1}\ .
\end{align}
Thus we see that the calibrated estimate of the aperture mass variance
can be written:
\be
  \left[\widehat{{\mathcal M}_{\rm ap}^2(\thetc|\btheta_{0,k})}\right]^{\rm cal} =
  \frac{\left[\widehat{{\mathcal M}_{\rm ap}^2(\thetc|\btheta_{0,k})}\right]^{\rm raw}}{1+{\mathcal K}^{\rm ap}(\thetc|\btheta_{0,k})}
  \label{eq:est1a}\ ,
\ee
where the normalisation factor is
\be  1+{\mathcal K}^{\rm ap}(\thetc|\btheta_{0,k})
\equiv
\frac{\sum_{j\ne i} w_i w_j (1+m_i)(1+m_j)}
     {\sum_{j\ne i} w_iw_j} \ .
  \label{eq:est1b}
\ee

\subsection{Analysis of the \cfhtls data}\label{ssec:rescfht}


In left panel of Figure~\ref{fig:map2cfht_raw} we show the aperture mass
variance and in the right the variance of the cross component
variance, estimated from the \cfhtl.  The results shown are
for the full combination of the W1, W2, W3 and W4 survey areas. For
the direct aperture mass variance approach the estimate was obtained using the weighting scheme ${\mathcal W}_3$, c.f. \Eqn{eq:Weight3}, and the error was obtained using a weighted jackknife resampling of the data in patches having an area of roughly $(1 \ \mathrm{deg})^2$. For the method of measuring the
aperture mass variance from the shear correlation functions (thick,
black dashed line), when combining the results from each field, we
have simply weighted the estimates in proportion to the field area.

In \Fig{fig:map2cfht_raw} the vertical dashed line indicates the scale
$\thetc=5.5'$ identified in \citet{Kilbingeretal2013}, above which the
E/B-mode leakage is claimed to be $\lesssim 1.5\%$. The E/B mode
leakage on smaller scales originates from the $9''$ cut-off in the
shear-correlation functions, which originates from the blending of
galaxy images and a shear bias for close pairs (see \Sect{ssec:closepairs}).

The important point to note from the figure is that on scales
$\thetc>5.5'$ the measurements from the direct and correlation
function estimators are in very good agreement to within the errors
for a wide range of aperture completeness thresholds. In addition,
with the exception of all but the lowest $c_k$ thresholds, the
variance of the cross component of the aperture mass variance is
consistent with zero on the same scales. On scales $\thetc\le 5.5'$,
both estimators appear to have non-zero B-modes, which rise sharply on
scales $\thetc\lesssim 2'$.


\subsection{Impact of close pair image blending on the variance}\label{ssec:closepairs}

As noted in \citet{Milleretal2013}, \lensfit galaxies with
separations closer than $9''$ tend to have a bias in their
ellipticities which is in the direction of the line connecting the
centres of the galaxy images. For the shear correlation functions,
\citet{Kilbingeretal2013} stated that this bias can be removed by
only computing the shear correlation functions down to $9''$. Their justification was that the alignment direction
of the pair is, to a very good approximation, randomly orientated.  We
now investigate how such a bias may contaminate the direct estimator
approach for aperture mass variance by comparing the measurements of the full catalogue to a reduced one in which clustered galaxies below the blending scale are removed.

We proceed by describing our algorithm for removing close pairs of
galaxies to generate a reduced catalogue.
\begin{enumerate}
\item We begin by initializing a Boolean mask value for each
  galaxy. These values are all initially set to unity.
\item We then spatially organize the data using a hierarchical
  KD-tree.
\item We next set the pair-cut-off scale $\theta_{\rm pc}$, within
  which galaxy pairs are to be expunged from the catalogue.
\item We now loop over all galaxies and perform a range search.  If
  any of the galaxy positions lie within the sphere of radius
  $\theta_{\rm pc}$ and have their Boolean flag set to unity, then the
  Boolean mask value associated with the current galaxy is set to zero
  and the galaxy will henceforth be excluded when estimating
  statistics from the data.
\end{enumerate}
We apply this method to the \cfhtls data and set the pair-cut-off
scale $\theta_{\rm pc}=9''$. This yields a reduced catalogue that
contains $\sim$70\% of the original galaxies.

\begin{figure}
  \centering {
    \includegraphics[width=8cm,angle=0]{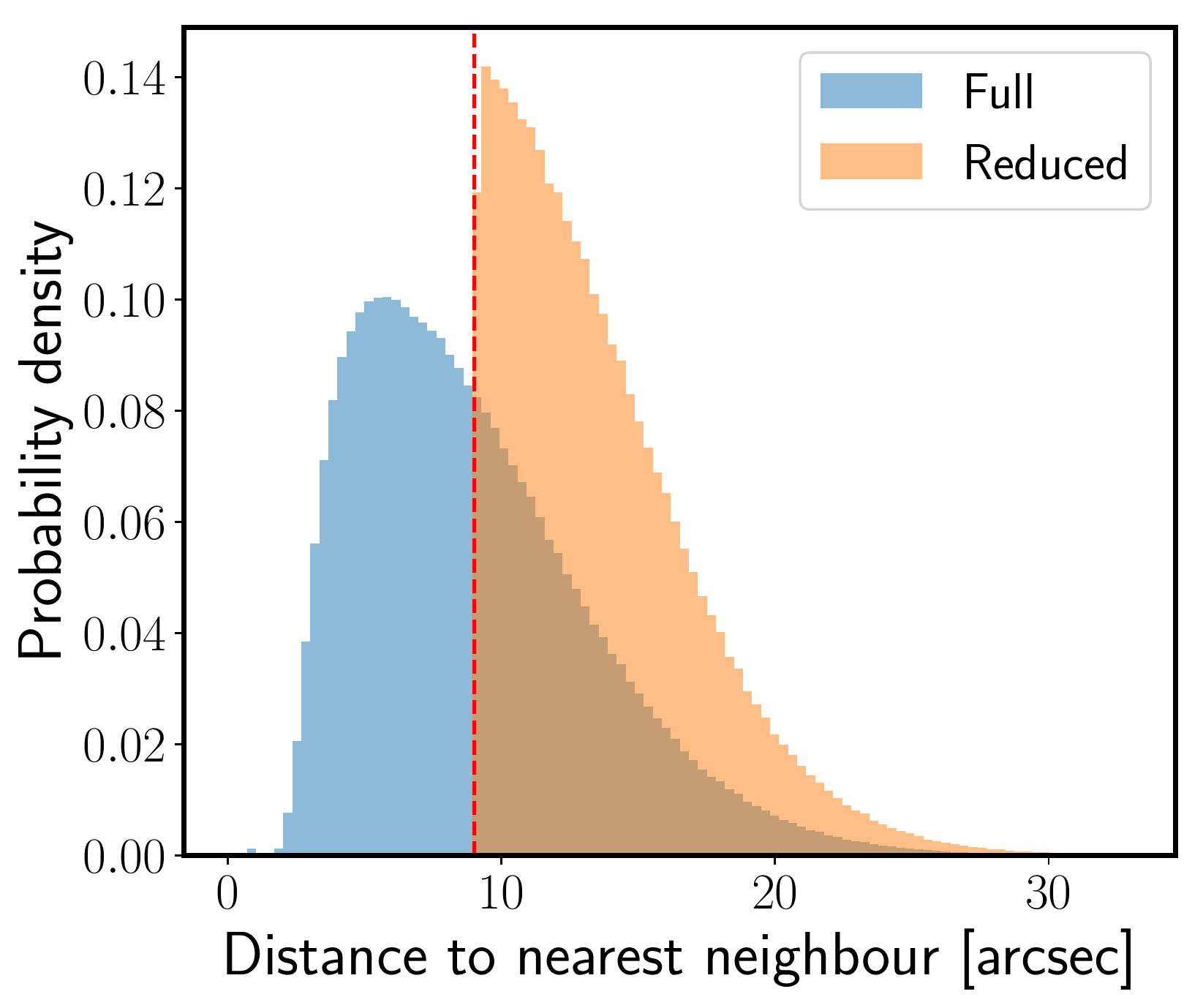}}
  \caption{\small{The probability density functions of the nearest
      neighbour distances in the \cfhtl. The blue histogram shows
      the distribution before applying our exclusion algorithm and the
      orange one shows the result after setting the pair-cut-off scale
      $\theta_{\rm pc}=9''$. The vertical dashed red line indicates
      $\theta_{\rm pc}$. }\label{fig:NNpdf}}
\end{figure}

Figure~\ref{fig:NNpdf} shows the probability density function of the
distance to the nearest-neighbour galaxy. Note that these curves have
been normalised so that the area under the graph gives unity.
%
%
%

\begin{figure*}
  \centering{
    \includegraphics[width=8.5cm]{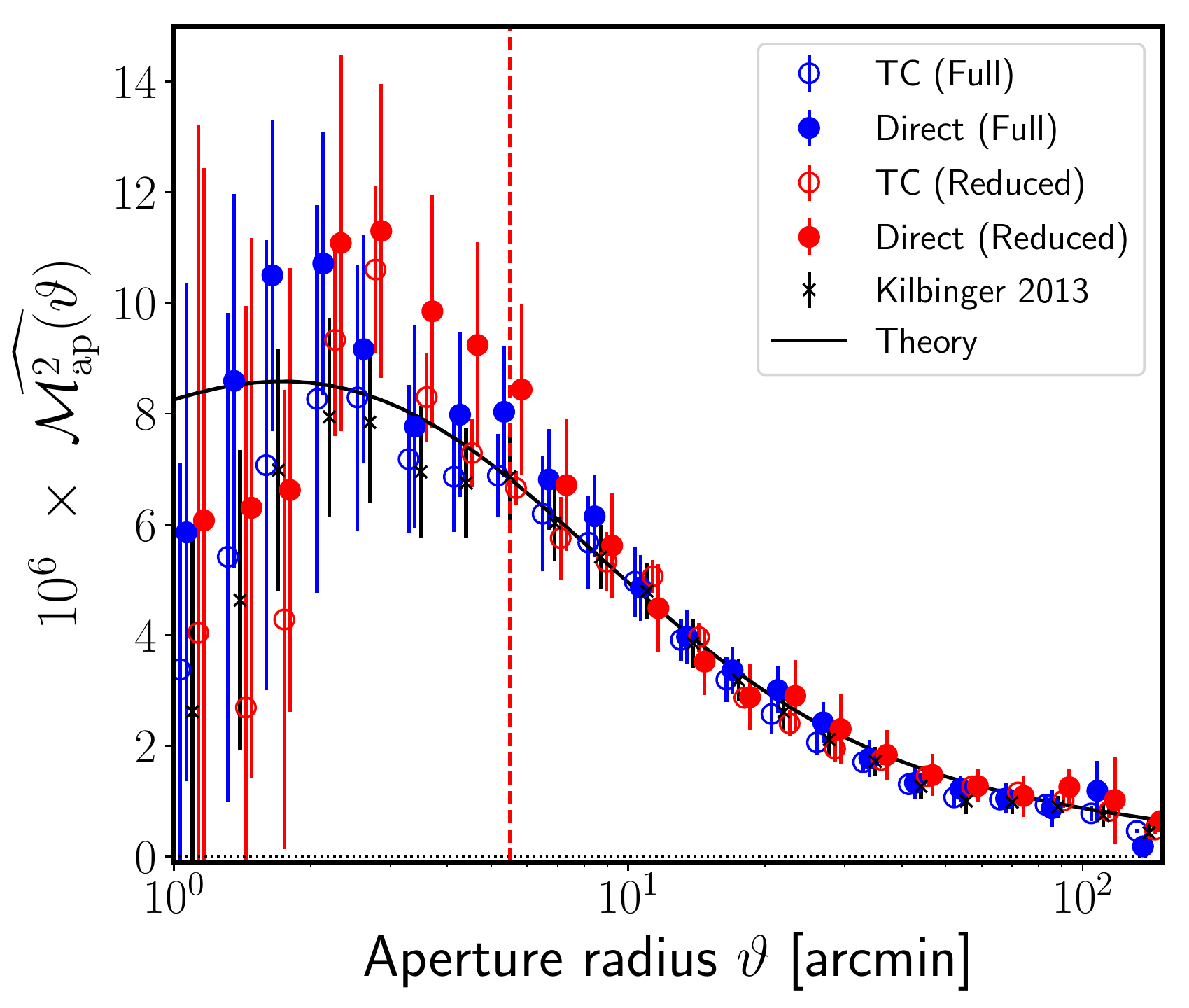}\hspace{0.2cm}
    \includegraphics[width=8.5cm]{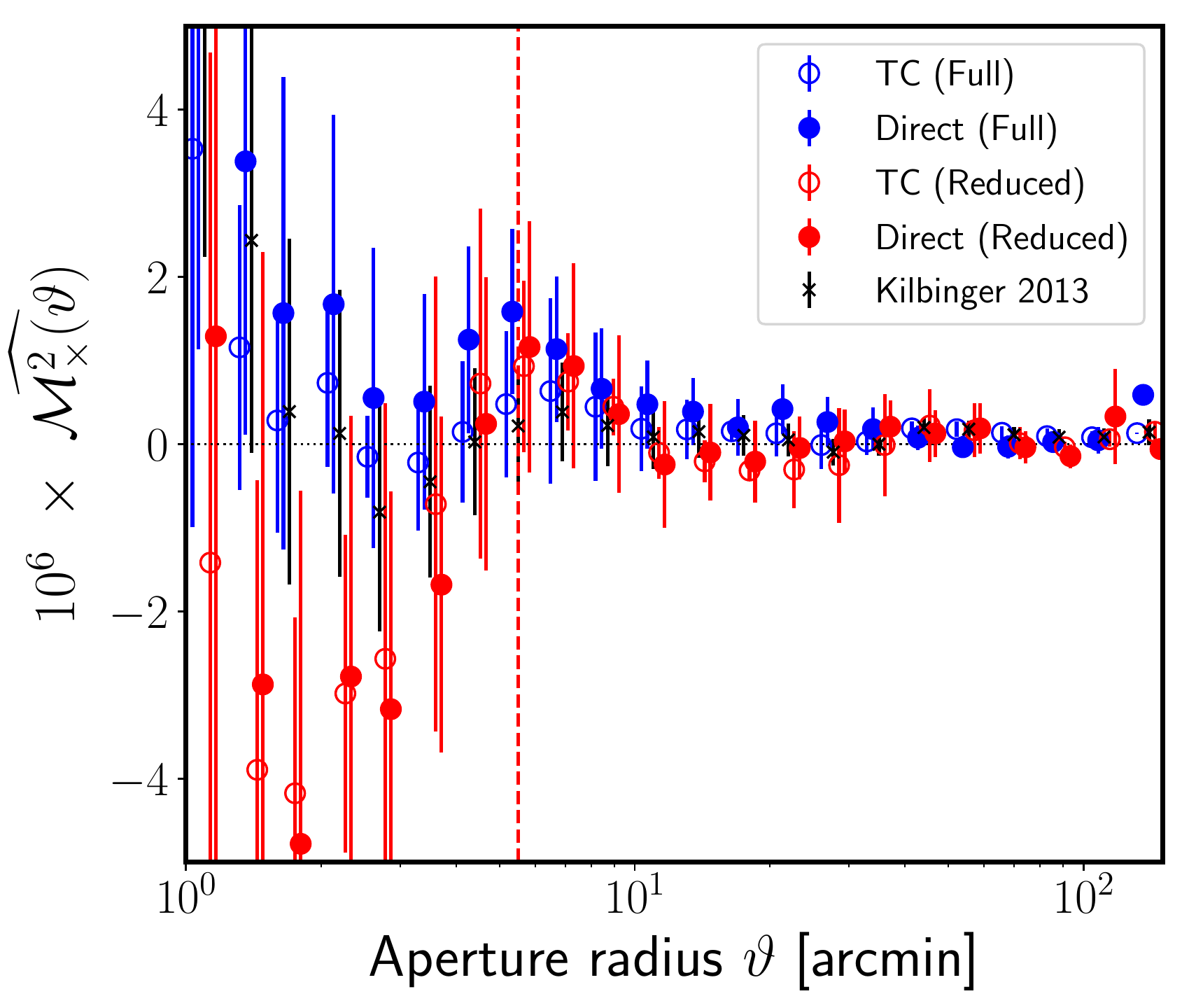}}
  \caption{\small{({\bf Left panel}): Comparison of the aperture mass
      variance from the full and small scale-pair reduced \cfhtls
      catalogues. The open circles show the results from \treecorr and the solid points the results from the direct
      estimator when including all apertures with $c_k>0.7$. The red and blue colours indicate the full and reduced
      catalogues, respectively. The black crosses and the black line show the published measurements from \citet{Kilbingeretal2013}  and the corresponding theoretical prediction for their best-fit cosmology evaluated via \eqn{eq:MapVar}. The vertical red
      dashed line indicates the scale above which the E/B mode leakage
      should be less than 1\% for the correlation function
      method. ({\bf Right panel}): same as left panel, except this
      time for the cross component of the aperture mass
      variance.\label{fig:map2cfht}}}
\end{figure*}


Figure~\ref{fig:map2cfht} shows our measurements of the aperture mass
variance and the variance of the cross component of aperture mass, in
the full and the close-pair reduced \cfhtls shear catalogues.  We
compare the results from both the direct and the correlation function
estimators. For the direct estimator approach we have employed the
weighting function ${\mathcal W}_3$, with the completeness threshold
set to $c_k=0.7$. Some important points can be noted: looking at the
left panel of \Fig{fig:map2cfht} the result of excluding pairs of
galaxies that are closer then 9'' lowers the amplitude of
${\left<{\mathcal M}_{\rm ap}^2\right>}$ by less than $15\%$ for small
aperture scales ($\sim 10'$). For larger apertures the changes are
smaller. Interestingly, the results from the \treecorr analysis
of the full and reduced catalogue also show similar differences: for
small aperture scales, the aperture mass variance appears to be lower
in the reduced catalogue. In addition, we see that all of the
estimators agree to within the errors on all scales. However, the
agreement between the direct estimator and the \treecorr result
is exceptionally good for the measurements from the reduced catalogue.

It is also interesting to compare these results with the published
measurement from \citet{Kilbingeretal2013} (denoted as the black
crosses in the plot). Here we see that our measurements are fully
consistent with the \citet{Kilbingeretal2013} measurements, to within
the errors. The right panel of Figure~\ref{fig:map2cfht} shows that
the variance of the cross-component of the aperture mass is
reassuringly consistent with zero for both estimators applied to both
the full and reduced catalogues.


\begin{figure*}
  \centering {
    \includegraphics[width=8.5cm]{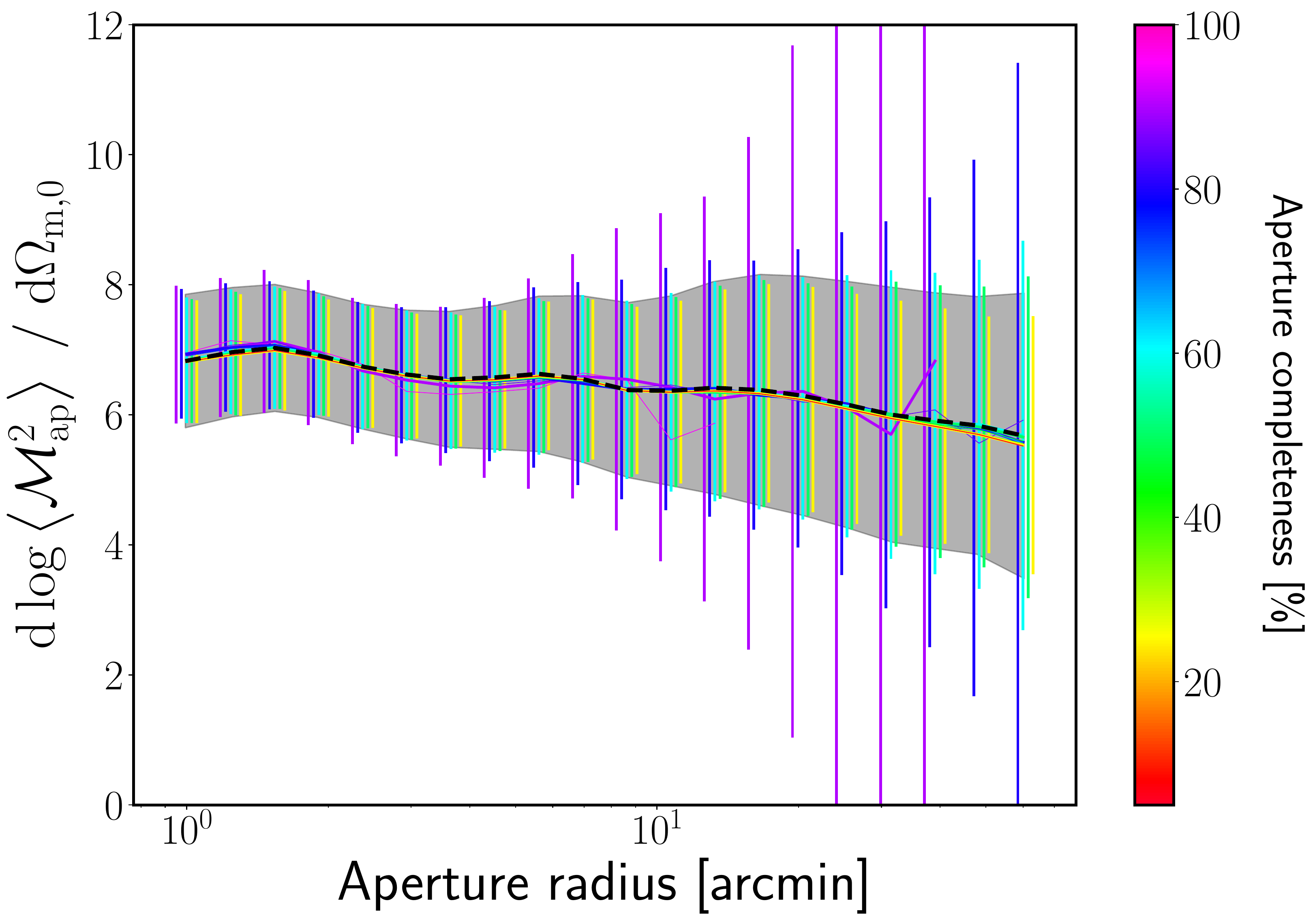}\hspace{0.2cm}
    \includegraphics[width=8.5cm]{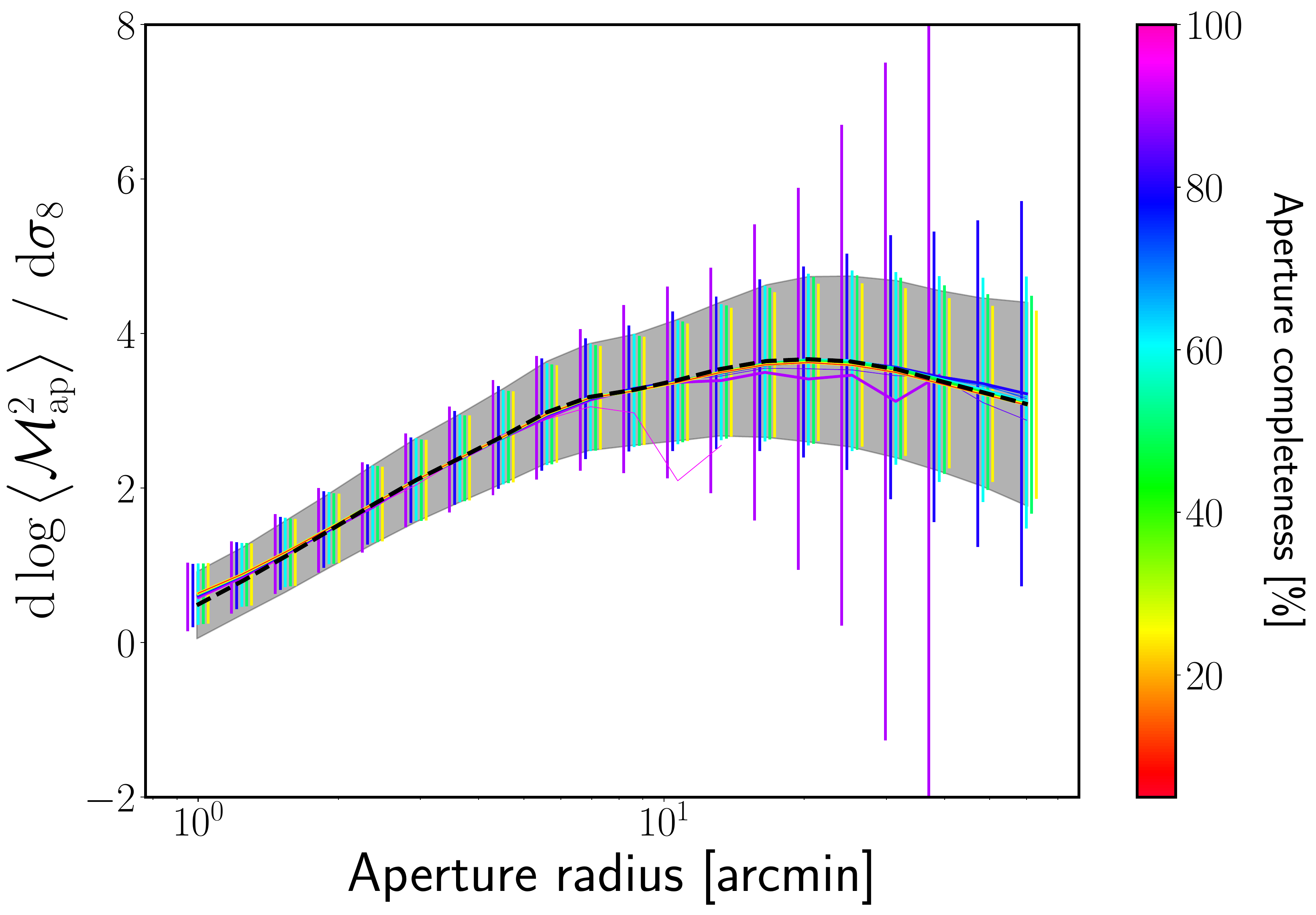}}
  \caption{\small{Estimates of the derivatives of the aperture mass
      variance with respect to the cosmological parameters. The left
      and right panels show the results for the variations with
      respect to the matter density parameter $\Omega_{\rm m,0}$ and
      the variance of matter fluctuations $\sigma_8$,
      respectively. The solid lines all show the results measured from
      the direct estimator approach, where the colour of the line
      indicates the value for the aperture completeness parameter that
      was adopted. The error bars indicate the errors in the
      ensemble. The black dashed line shows the results from the
      correlation function approach as measured using \treecorr. The shaded region shows the standard error on the
      \treecorr estimates.}\label{fig:map2Fish1}}
\end{figure*}


\section{Information content of the estimators}\label{sec:fish}

We now turn our attention to the question of addressing the possible
information loss in using the direct estimator approach. We will do
this using the Fisher matrix formalism. If we assume that the
likelihood function for measuring the aperture mass variance for a set
of $N_d$ aperture scales is Gaussian, and that the priors on the
cosmological parameters are flat, then the Fisher information matrix
can be written \citep{Tegmarketal1997}:
\be {\mathcal F}_{\alpha\beta} = \frac{1}{2}{\rm Tr}
\left[
  \mathbfss{C}^{-1}\frac{\partial \mathbfss{C}}{\partial p_\alpha}
  \mathbfss{C}^{-1}\frac{\partial \mathbfss{C}}{\partial p_\beta}
  \right]
+\frac{\partial\bm\mu^{\rm T}}{\partial p_\alpha}
\mathbfss{C}^{-1}\frac{\partial\bm\mu}{\partial p_\beta}\ , \label{eq:Fish}
\ee
where $\bm\mu^{\rm T}=\left(\widehat{\mathcal M^2_{\rm ap}(\thetc_1)},\dots,\widehat{\mathcal{
  M}^2_{\rm ap}(\thetc_{N_d})}\right)$ is the set of model means
measured at the bin centres and $\mathbfss{C}$ is the model covariance
matrix.  The vector $\vec{p}$ is the set of cosmological
parameters. In this study we will restrict our attention to the
cosmological parameters $\sigma_8$ and $\Omega_{\rm m,0}$, as these
are the most readily constrained from lensing data. The minimum
variance bounds on a given cosmological parameter, after marginalising
over all other parameters, can be obtained as:
\be \sigma^2_{p_{\alpha}} = \left[{\mathcal F}^{-1}\right]_{\alpha\alpha} \ .\ee

In order to simplify the calculation, we will assume that the first
term on the right-hand-side of \Eqn{eq:Fish} is significantly smaller
than the other term. As was noted in \S6 of \citet{SmithMarian2015}
this can be justified in the high-$k$ limit through mode counting
arguments, however, it is in general an incorrect
assumption\footnote{On the other hand, as was discussed in
  \citet{Takahashietal2011} the likelihood is not Gaussian, since the
  power spectrum estimator is $\chi^2$ distributed. Thus one should
  not quote over-optimistic errors based on the wrong form for the
  likelihood function.}.

In order to compute the second term of \Eqn{eq:Fish} we follow the
approach lain down in \citet{Smithetal2014} and use the mocks
to evaluate all quantities. That is we measure the derivatives of the
model mean with respect to the cosmological parameters and we also
estimate the precision matrix $\mathbfss{C}^{-1}$. To do these tasks
we make use of the ray traced mock \cfhtls data. The
derivatives are estimated using:
\be \widehat{\frac{\partial \mu_i}{\partial p_\alpha}} =
\sum_{j=1}^{N_{\rm ens}}\frac{
  \widehat{{\mathcal M}_{\rm ap, {(j)}}^2}(\thetc_i|p_\alpha+\Delta p_\alpha)-
  \widehat{{\mathcal M}_{\rm ap, {(j)}}^2}(\thetc_i|p_\alpha-\Delta p_\alpha)} {2N_{\rm ens}\Delta p_{\alpha}} \ ,
\label{eq:diff}\ee
where $N_{\rm ens}$ is the number of realisation of the ensemble,
and ${\mathcal M}_{\rm ap, {(j)}}^2(\theta_i|p_\alpha+\Delta
p_\alpha)$ is the estimate of the aperture mass variance on scale
$\theta_i$ in the $j$th realisation of the mocks, for the simulation
with cosmological parameters $p_\alpha+\Delta p_\alpha$. In computing
\Eqn{eq:diff} we use the 256 mock ray-tracing simulations of each
cosmological variation of the \cfhtls data. Note that the above
estimator will reduce the cosmic variance, since when running the
modified cosmology simulations we used Fourier phase realisations that
were identical to the fiducial model runs. In addition, when
estimating the precision matrix we take account of the bias in the
estimator using the method described in \citet{Hartlapetal2007}:
\be \widehat{\mathbfss{C}^{-1}} = \frac{N_{\rm ens}-N_\mathrm{d}-2}{N_{\rm
    ens}-1} \left[\widehat{\mathbfss{C}}\right]^{-1} \ ,\ee
where $N_\mathrm{d}$ is the dimension of the data vector and
$\widehat{\mathbfss{C}}$ is the standard maximum likelihood estimator
of the covariance matrix of the data.

Figure~\ref{fig:map2Fish1} shows the derivatives of the aperture mass
variance with respect to the cosmological parameters, estimated from
the 256 ray-traced mock simulations. The agreement between the results
from the direct estimator approach is excellent, for all of the $c_k$
values that we considered, where we used the inverse variance
weighting approach of ${\mathcal W}_3$ to combine estimates from
individual apertures. In addition, these also agree with the results
from the correlation function approach to a high degree of accuracy.
The only issue to note is that for high completeness fractions
$c_k\sim1$ the estimates become very noisy\footnote{In making the
  Fisher forecasts we need to take into account the error on the model
  means and the precision matrix, since not doing so could lead to
  over-optimistic forecasts. This could be done by generalising the
  approach described in \citet{Tayloretal2013}.}.


\begin{figure}
  \centering {
    \includegraphics[width=8.5cm]{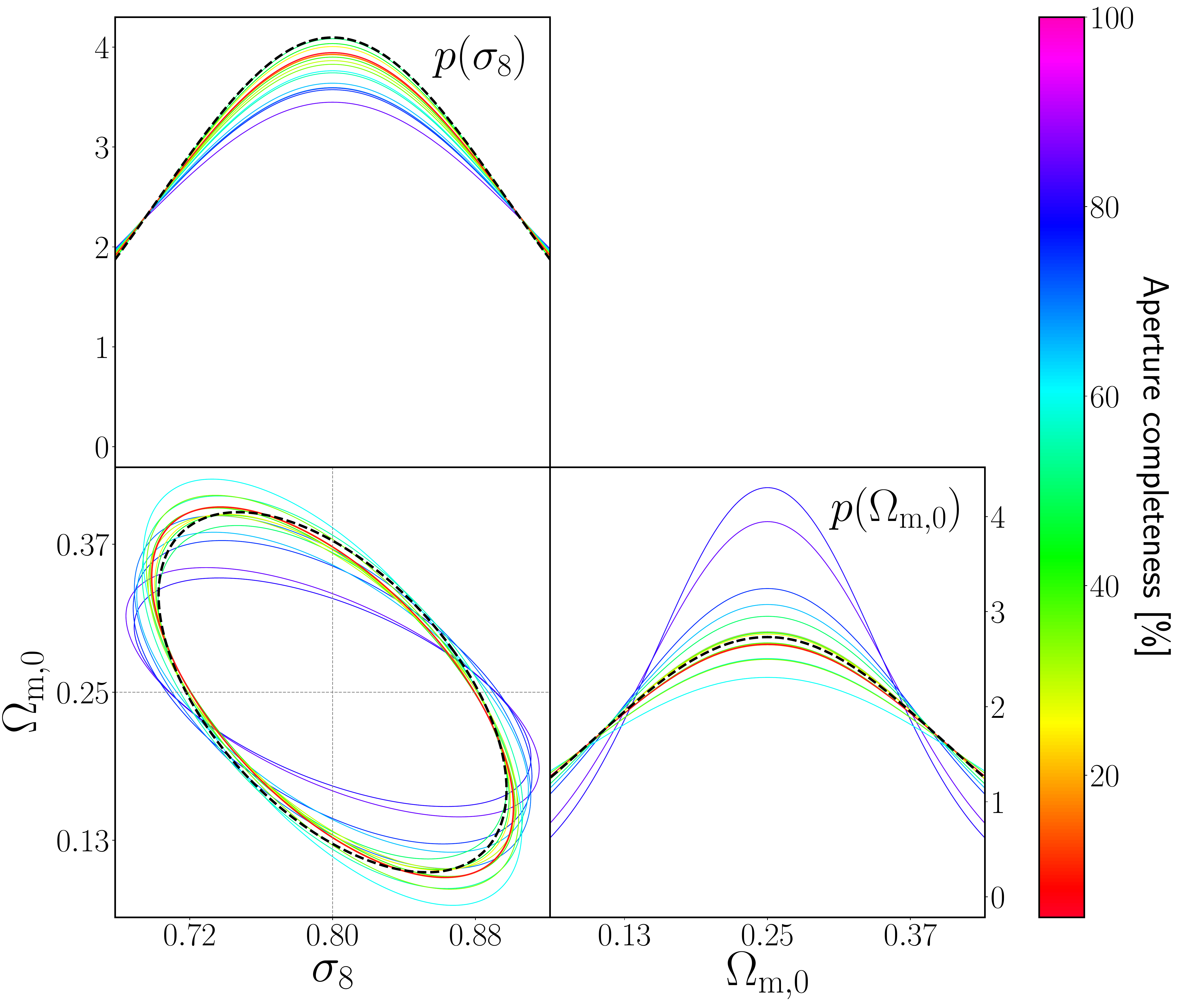}}
  \caption{\small{Top left and bottom right panels show the 1D
      posterior distributions, marginalised over all other parameters,
      for $\Omega_{\rm m,0}$ and $\sigma_8$, respectively. The bottom
      left panel shows the 2D confidence interval ellipse for the
      joint posterior distribution of $\Omega_{\rm m,0}$ and
      $\sigma_8$. The coloured lines again show the results from the
      Fisherm matrix forecast using the direct estimator approach for
      various aperture completeness thresholds $c_k$. The dashed line
      shows the results from \treecorr.}\label{fig:map2Fish2}}
\end{figure}


Figure~\ref{fig:map2Fish2} shows the 2D confidence contour for
$\Omega_{\rm m,0}$ and $\sigma_8$, and also the 1D posterior
distributions, marginalised over all other parameters, for the same
two parameters. Focusing on the 2D contours, we notice a number of
interesting points: first, considering the results for the direct
estimator, we see that as we decrease the threshold completeness value
$c_k$, the ellipses rotate clockwise by a small amount. This difference in orientation can be explained by noting that for a \cfhtls like survey the off-diagonal elements of the covariance matrix are getting more and more noisy for increasing values of $c_k$. The reason for this is that the effective survey footprint changes for different aperture radii and that this difference is most prominent for conservative aperture completeness cuts. When choosing a $c_k$ of around $0.7$ or lower, the covariance matrices become stable and so does the orientation and area of the ellipse.

Similar observations can be made for the 1D marginalized posterior distributions which lets us conclude that, within the Fisher Matrix formalism, the information content of the direct estimator is comparable with the correlation function method.

There are some caveats that must be mentioned to the interpretation of
these results. Firstly, both the precision matrix and the model means
are estimated from the mock simulations of the \cfhtl, and
are therefore subject to errors. This means that the forecasted
constraining power should also come with an error. Since we are only
interested in the relative information, we have not taken this into
account. A more detailed study is required in order to make a more
precise statement, and this is beyond the scope of this work.


\section{Conclusions and discussion}\label{sec:conclusions}

In this paper we have explored an alternative method for estimating
the aperture mass statistics in weak lensing cosmic shear surveys. Our
method is a direct estimator of the variance. With the use of a
hierarchical KD-tree algorithm for ordering the data we found that
the computational time for execution of this estimator was linear in
the number of galaxies per aperture and the number of apertures used in
the estimate. This paper, the first in a series, focused on the
two-point statistics, and in particular the aperture mass variance. The
summary is:

In \S\ref{sec:theory} and \S\ref{sec:aperturemass} we reviewed the
background theory of weak lensing in the cosmological context and the
aperture mass variance and its connection to the matter power
spectrum. Here we also discussed the standard approach for estimating
this quantity, which relies on measurements of the shear correlation
functions.

In \S\ref{sec:aperturemass} we introduce the direct estimator for the
aperture mass variance that we employ. We show that when including
ellipticity weights the estimator is unbiased. We also compute the
variance of the estimator and show that in the limit of Gaussian shear
signal, no source clustering and a large number of galaxies per
aperture that the variance reduces to a simple expression. We then
show how the original estimator introduced by
\citet{Schneideretal1998} can be accelerated to linear order in the
number of galaxies per aperture. We also discuss various weighting
schemes for combining the estimates from different apertures. Finally,
we illuminate the computational complexity of our estimator.

In \S\ref{sec:est} we give an overview of the \cfhtls data and we
also describe our method for generating mocks of the survey using full
gravitational ray-tracing simulations through $N$-body simulations. As
a first test, we measured the aperture mass maps from the survey data.

In \S\ref{sec:results} we computed the aperture mass variance from our
mock surveys, using both the direct estimator and correlation function
method. We found that if we included incomplete apertures in the
direct estimate method, and if we combined all apertures equally that
there was a significant bias in the final result. This bias vanished
if only complete apertures were used. However, the errors on the
estimates increased significantly. We then explored an alternative
weighting scheme, where the apertures were combined using an inverse
variance weighting approach, where the variance was assumed to be
dominated by shape noise. The results in this case were found to be in
excellent agreement with the alternative method, even in the case
where incomplete apertures were included in the estimate. We found
that an aperture completeness threshold of $\sim0.7$ gave very good
results and contained only a small residual of B-modes, that were
contained well within the error tolerance.

We then turned to the application of the method to the \cfhtls
data. It was necessary to account for two additional observational
biases: first, we derived the correction factors required to account
for the ellipticity bias in our estimator; second, we created a
modified source galaxy catalogue that removed pairs of galaxies whose
images were in close projection on the sky whose ellipticities are
biased by an artefact in the ellipticity estimator algorithm \emph{lens}fit. On taking account of these we found that our direct
estimator approach and our estimates using the shear correlation
function were in excellent agreement with the published data from
\citet{Kilbingeretal2013}.

In \S\ref{sec:fish} we explored the information content of the direct
estimator approach and compared it with that from the correlation
function method. We found that the 1D marginalised posterior
distributions for $\sigma_8$ were less constraining for high aperture
completeness than the shear correlation function method, but that as
some incomplete apertures were included in the estimate the
distributions became very similar. This trend was mirrored for
$\Omega_{\rm m,0}$, except that for high completeness the distribution
for the direct estimator was the most constraining. This may be due to
errors in the forecast due to uncertainties in the precision matrix
and derivatives. The 2D confidence contours for $\Omega_{\rm m,0}$ and
$\sigma_8$ were roughly the same size for all aperture completeness
thresholds, however they rotated to be in the same direction as those
of the correlation function approach as lower thresholds were taken.
This leads us to conclude, at this stage, that the information content
of the two estimators is comparable.

The main advantage of this development in not to replace the
correlation function approach as {\em the way} to measure aperture
mass statistics and other associated statistics, but to show that it
is a credible method. As mentioned earlier, the real advantage of this
approach is that it can be easily generalised to enable the
measurement of higher order aperture mass statistics such as the
skewness and kurtosis with very little extra effort. While these can
of course also be estimated using the shear three-point and four-point
correlation functions, the task of measuring these correlation
functions and all of their configurations becomes increasingly onerous
and time consuming. The method that we have developed will scale
linearly with the number of galaxies in the aperture. This will be the
subject of our upcoming study. Besides this, for application to more
recent large-scale lensing surveys like {\tt DES} and \kids\ the
analysis will need to be extended to include curved sky effects and we
see no obvious issues with this.


\section*{Acknowledgements}
LP acknowledges support from a STFC Research Training Grant (grant number ST/R505146/1). RES acknowledges support from the STFC (grant number ST/P000525/1,  ST/T000473/1). This work used the DiRAC@Durham facility managed by the Institute for Computational Cosmology on behalf of the STFC DiRAC HPC Facility (www.dirac.ac.uk). The equipment was funded by BEIS capital funding via STFC capital grants ST/K00042X/1, ST/P002293/1, ST/R002371/1 and ST/S002502/1, Durham University and STFC operations grant ST/R000832/1. DiRAC is part of the National e-Infrastructure.



\bibliographystyle{mnras}
\bibliography{ms}

\vspace{5mm}
 

\onecolumn

\appendix


\section{Estimating the variance of the estimator -- including the source weights}\label{app:var}

We can also calculate the variance of the estimator:
\begin{align}
	{\rm Var}\left[\widehat{{\mathcal M}_{\rm ap}^2(\thetc)}\right]
	&= \left< (\pi\thetc^2)^2 \
        \frac{\sum_{i\neq j} w_i \ w_j \ Q_i \ Q_j \ \epsilon_{{\rm t},i} \ \epsilon_{{\rm t},j}}{\sum_{i\neq j} w_i w_j}
        (\pi\thetc^2)^2 \
  \frac{\sum_{k\neq l} w_k \ w_l \ Q_k \ Q_l \ \epsilon_{{\rm t},k} \ \epsilon_{{\rm t},l}}{\sum_{k\neq l} w_k w_l}\right>
  -\MapSq^2 \nn \\
  & =  \frac{(\pi\thetc^2)^4}{\left(\sum_{i\neq j} w_i w_j\right)^2}
  \sum_{i\neq j}\sum_{k\neq l} w_i w_j w_k w_l Q_i Q_j Q_k Q_l
  \left<\epsilon_{{\rm t},i} \epsilon_{{\rm t},j}\epsilon_{{\rm t},k}\epsilon_{{\rm t},l}\right>
  -\MapSq^2 \label{eq:varest1}
\end{align}
Following the recipe as for the mean estimator, we first calculate the
average over the source galaxies and this yields
\citet{Schneideretal1998}:
\begin{align}
  A\left[\epsilon_{{\rm t},i} \epsilon_{{\rm t},j}\epsilon_{{\rm t},k}\epsilon_{{\rm t},l}\right]
& = \gamma_{{\rm t},i} \gamma_{{\rm t},j}\gamma_{{\rm t},k}\gamma_{{\rm t},l}
+A\left[\epsilon^{({\rm s})}_{{\rm t},i} \epsilon^{({\rm s})}_{{\rm t},j}\epsilon^{({\rm s})}_{{\rm t},k}\epsilon^{({\rm s})}_{t,l}\right]
+\frac{\sigma^2_{\epsilon}}{2}
\left[
 \gamma_{{\rm t},i} \gamma_{{\rm t},j}\delta^{K}_{kl}
+\gamma_{{\rm t},i} \gamma_{{\rm t},k}\delta^{K}_{jl}
+\gamma_{{\rm t},i} \gamma_{{\rm t},l}\delta^{K}_{jk}\right. \nn \\
& \left.
+\gamma_{{\rm t},j} \gamma_{{\rm t},k}\delta^{K}_{il}
+\gamma_{{\rm t},j} \gamma_{{\rm t},l}\delta^{K}_{ik}
+\gamma_{{\rm t},k} \gamma_{{\rm t},l}\delta^{K}_{ij}
  \right] \ .\label{eq:varest2}
\end{align} 

Let us now work out the first term on the right-hand-side of the above
expression. We see that we have the following possibilities: all
indices different $(i\ne j\ne k\ne l) \rightarrow \circled{1a}$; two
indices equal and three not $(i\ne j\ne k=l) + \rm {perms}\rightarrow
\circled{1b}$; $(i\ne j\ne k=l) + \rm {perms}\rightarrow
\circled{1b}$ two sets of indices equal; three indices equal and
one not; all indices equal.

\begin{itemize}

\item $\circled{1a}\rightarrow (i; j\ne i; k\ne i\ne j; l \ne i\ne j\ne k)$. Averaging
  over the source galaxy positions gives:
  \begin{align}
    \circled{1a}_{\rm part} & = P\left(\sum_{i}\sum_{j\ne i}\sum_{k\ne j\ne i}\sum_{l\ne k\ne j\ne i}
    w_i w_j w_k w_l Q_i Q_j Q_k Q_l \gamma_{{\rm t},i} \gamma_{{\rm t},j}\gamma_{{\rm t},k}\gamma_{{\rm t},l}\right) \nn \\
    & = \prod_{\alpha=1}^N\left\{\int\frac{\diff^2\bthet_{\alpha}}{\pi\thetc^2}\right\}
    \sum_{i}\sum_{j\ne i}\sum_{k\ne j\ne i}\sum_{l\ne k\ne j\ne i}
    w_i w_j w_k w_l Q_i Q_j Q_k Q_l \gamma_{{\rm t},i} \gamma_{{\rm t},j}\gamma_{{\rm t},k}\gamma_{{\rm t},l} \nn \\
    & = \frac{1}{(\pi\thetc^2)^4}
    \sum_{i}\sum_{j\ne i}\sum_{k\ne j\ne i}\sum_{l\ne k\ne j\ne i}w_i w_j w_k w_l 
    \int \diff^2\theta_1 \dots \diff^2\theta_4 Q_1 \dots Q_4 \gamma_{{\rm t},1} \dots \gamma_{{\rm t},4} \nn \\ 
    & = \frac{1}{(\pi\thetc^2)^4}
    \sum_{i}\sum_{j\ne i}\sum_{k\ne j\ne i}\sum_{l\ne k\ne j\ne i}w_i w_j w_k w_l 
    {\mathcal M}_{\rm ap}^4 \label{eq:varest3}
  \end{align}
  Finally, on averaging over the shear-field realisations and putting back the factors we see that:
  \begin{align}
    {\rm E}\left[\circled{1a}_{\rm full}\right] & =
    \frac{(\pi\thetc^2)^4}{\left(\sum_{i}w_i\sum_{j\ne i}w_j\right)^2}\frac{1}{(\pi\thetc^2)^4}
    \sum_{i}\sum_{j\ne i}\sum_{k\ne j\ne i}\sum_{l\ne k\ne j\ne i}w_i w_j w_k w_l \left<{\mathcal M}_{\rm ap}^4\right> \nn \\
    & = \frac{\sum_{i}\sum_{j\ne i}\sum_{k\ne j\ne i}\sum_{l\ne k\ne j\ne i}w_i w_j w_k w_l}{\left(\sum_{i}w_i\sum_{j\ne i}w_j\right)^2}
     \left<{\mathcal M}_{\rm ap}^4 \right> \label{eq:varest4}
  \end{align}

\item $\circled{1b}\rightarrow (i; j\ne i; k\ne i\ne j; l = i \ne j \ne k)+{\rm perms}$. Averaging
  over the source galaxy positions gives:
  \begin{align}
    \circled{1b}_{\rm part} & = P\left(\sum_{i}\sum_{j\ne i}\sum_{k\ne j\ne i}
    w_i^2 w_j w_k Q_i^2 Q_j Q_k \gamma^2_{{\rm t},i} \gamma_{{\rm t},j}\gamma_{{\rm t},k}\right) \nn \\
    & = \prod_{\alpha=1}^N\left\{\int \frac{\diff^2\bthet_{\alpha}}{\pi\thetc^2}\right\}
    \sum_{i}\sum_{j\ne i}\sum_{k\ne j\ne i}
    w_i^2 w_j w_k Q_i^2 Q_j Q_k \gamma_{{\rm t},i}^2 \gamma_{{\rm t},j}\gamma_{{\rm t},k} \nn \\
    & = \frac{1}{(\pi\thetc^2)^3}
    \sum_{i}\sum_{j\ne i}\sum_{k\ne j\ne i}w_i^2 w_j w_k 
    \int \diff^2\theta_1 \dots \diff^2\theta_3 Q_1^2 \dots Q_3 \gamma_{{\rm t},1}^2 \dots \gamma_{{\rm t},3} \nn \\ 
    & = \frac{1}{(\pi\thetc^2)^3}
    \sum_{i}\sum_{j\ne i}\sum_{k\ne j\ne i}w_i^2 w_j w_k 
        {\mathcal M}_{\rm ap}^2\frac{{\mathcal M}_{\rm s,2}}{\pi\thetc^2} \ ,\label{eq:varest5}
  \end{align}
  where in the above we have defined the quantity: ${\mathcal M}_{\rm
    s,2}:=\pi\thetc^2\int \diff^2\theta_1 Q^2(\bthet_1) \gamma^2_{\rm
    t}(\bthet_1)$. We note that there are three identical contributions
  of this type that arise from the terms where $i=k$, $j=l$, and
  $j=k$. Hence, on averaging over the shear-field realisations and
  putting back the factors we see that the sum of these terms becomes:
  \begin{align}
    {\rm E}\left[\circled{1b}_{\rm full}\right]
    & = 4\frac{\sum_{i}\sum_{j\ne i}\sum_{k\ne j\ne i}w_i^2 w_j w_k}{\left(\sum_{i}w_i\sum_{j\ne i}w_j\right)^2}
    \left<{\mathcal M}_{\rm ap}^2{\mathcal M}_{\rm s,2}\right> \ .\label{eq:varest6}
  \end{align}
  
\item $\circled{1c}\rightarrow (i; j\ne i; k=i; l = j)+(i; j\ne i; k=j; l = i)$. Averaging
  over the source galaxy positions gives:
  \begin{align}
    \circled{1c}_{\rm part} & = P\left(\sum_{i}\sum_{j\ne i}
    w_i^2 w_j^2 Q_i^2 Q_j^2  \gamma^2_{{\rm t},i} \gamma^2_{{\rm t},j}\right) \nn \\
    & = \prod_{\alpha=1}^N\left\{\int \frac{\diff^2\bthet_{\alpha}}{\pi\thetc^2}\right\}
    \sum_{i}\sum_{j\ne i}
    w_i^2 w_j^2  Q_i^2 Q_j^2 \gamma_{{\rm t},i}^2 \gamma_{{\rm t},j}^2 \nn \\
    & = \frac{1}{(\pi\thetc^2)^2}
    \sum_{i}\sum_{j\ne i} w_i^2 w_j^2
    \int \diff^2 \theta_1  Q_1^2 \gamma_{{\rm t},1}^2 \int \diff^2\theta_2Q_2^2 \gamma^2_{{\rm t},2} \nn \\ 
    & = \frac{1}{(\pi\thetc^2)^2}
    \sum_{i}\sum_{j\ne i}w_i^2 w_j^2
        \frac{{\mathcal M}_{\rm s,2}^2}{(\pi\thetc^2)^2} \ .\label{eq:varest7}
  \end{align}
  We note that the second term $(i; j\ne i; k=j; l = i)$ will be
  identical to the first after summation, and thus gives us an extra factor of 2.
  Hence, on averaging over the shear-field realisations and putting
  back the factors we see that the sum of these terms becomes:
  \begin{align}
    {\rm E}\left[\circled{1c}_{\rm full}\right]
    & = 2\frac{\sum_{i} w_i^2 \sum_{j\ne i} w_j^2 }{\left(\sum_{i}w_i\sum_{j\ne i}w_j\right)^2}
    \left<{\mathcal M}_{\rm s,2}^2\right> \ . \label{eq:varest8}
  \end{align}
\end{itemize}
The terms where three or four indices are the same vanish due to the
constraints on the sums. 

Turning now to the second term on the right-hand-side of
\Eqn{eq:varest2}. Performing the average over the source galaxy
ellipticities we see that, since $i\ne j$ and $k\ne l$, we have two
possibilities:
\be A\left[\epsilon^{({\rm s})}_{{\rm t},i} \epsilon^{({\rm s})}_{{\rm t},j}\epsilon^{({\rm s})}_{{\rm t},k}\epsilon^{({\rm s})}_{{\rm t},l}\right]
= \frac{\sigma^4_{\epsilon}}{4}\left[\delta^{K}_{i,k}\delta^K_{j,l}+\delta^{K}_{i,l}\delta^K_{j,k}\right] \ , \label{eq:varest10}
\ee
Let us call these two terms $\circled{2a}$ and $\circled{2b}$.  On
performing the average over galaxy positions the first term becomes:
\begin{align}
  \circled{2a}_{\rm part}
  & \equiv {\rm P}\left( \sum_{i} \sum_{j\neq i} \sum_{k} \sum_{k\neq l}  w_i w_j w_k w_l Q_i Q_j Q_k Q_l 
  \frac{\sigma^4_{\epsilon}}{4}\delta^{K}_{i,k}\delta^K_{j,l} \right) \nn \\
  & =  {\rm P}\left(\frac{\sigma^4_{\epsilon}}{4} \sum_{i} \sum_{j\neq i}    w_i^2 w_j^2 Q_i^2 Q_j^2\right)\nn \\
  & = \frac{\sigma^4_{\epsilon}}{4} \prod_{\alpha=1}^N\left\{\int \frac{\diff^2\bthet_{\alpha}}{\pi\thetc^2}\right\}
  \sum_{i} \sum_{j\neq i} w_i^2 w_j^2 Q_i^2 Q_j^2\nn\\
  & = \frac{\sigma^4_{\epsilon}}{4} \sum_{i} \sum_{j\neq i} w_i^2 w_j^2
  \int \frac{\diff^2\bthet_{1}}{\pi\thetc^2} Q_1^2 \int \frac{\diff^2\bthet_{1}}{\pi\thetc^2} Q_2^2 \nn \\
  & = \frac{\sigma^4_{\epsilon}}{4} \sum_{i} \sum_{j\neq i} w_i^2 w_j^2 \frac{G^2}{(\pi\thetc^2)^4}\ ,\label{eq:varest11}
\end{align}
where in the above we have defined the quantity $G:= \pi\thetc^2\int
\diff^2\theta_1 Q^2(\bthet_1)$. The second term $\circled{2b}$ is
identical and so gives us a factor of 2. Finally the expectation over
the ensemble of realisations of the shear field, with the
normalisation factors restored, gives us:
\begin{align}
  {\rm E}\left[\circled{2a}_{\rm full}+\circled{2b}_{\rm full}\right]
  & = \frac{\sum_{i} \sum_{j\neq i} w_i^2 w_j^2}{2\left(\sum_{i}\sum_{j\ne i}w_iw_j\right)^2} \sigma^4_{\epsilon} G^2 \ .
  \label{eq:varest12}
\end{align}

Turning now to the third term on the right-hand-side of
\Eqn{eq:varest2} we see that when summing over allowed indices the
first and last terms in the square bracket will not contribute. Let us
label the remaining terms four terms $\circled{3a}$--$\circled{3d}$.
On averaging over the source galaxy positions the first of these
terms can be written as:
\begin{align}
  \circled{3a}_{\rm part} & =
  {\rm P}\left(\sum_{i} \sum_{j\ne i} \sum_k \sum_{l \ne k}
  w_i w_j w_k w_l Q_i Q_j Q_k Q_l \frac{\sigma^2_{\epsilon}}{2}\gamma_{{\rm t},i} \gamma_{{\rm t},k}\delta^{K}_{jl}\right) \nn \\
  & = \frac{\sigma^2_{\epsilon}}{2}
  {\rm P}\left(\sum_{i} \sum_{j\ne i} \sum_{k\ne i \ne j} 
  w_i w_j^2 w_k Q_i Q_j^2 Q_k \gamma_{{\rm t},i} \gamma_{{\rm t},k}
  +\sum_{i} \sum_{j\ne i}  
  w_i^2 w_j^2 Q_i^2 Q_j^2 \gamma_{{\rm t},i}^2 \right) \nn \\
  & = \frac{\sigma^2_{\epsilon}}{2}
  \prod_{\alpha=1}^N\left\{\int \frac{\diff^2\bthet_{\alpha}}{\pi\thetc^2}\right\}
  \left(\sum_{i} \sum_{j\ne i} \sum_{k\ne i \ne j} 
  w_i w_j^2 w_k Q_i Q_j^2 Q_k \gamma_{{\rm t},i} \gamma_{{\rm t},k}
  +\sum_{i} \sum_{j\ne i}  
  w_i^2 w_j^2 Q_i^2 Q_j^2 \gamma_{{\rm t},i}^2 \right)\nn \\
  & = \frac{\sigma^2_{\epsilon}}{2}
  \sum_{i} \sum_{j\ne i} \sum_{k\ne i \ne j} 
  w_i w_j^2 w_k \int \frac{\diff^2\bthet_{1}}{\pi\thetc^2}  Q_1 \gamma_{{\rm t},1} \int \frac{\diff^2\bthet_{2}}{\pi\thetc^2} Q_2^2
  \int \frac{\diff^2\bthet_{3}}{\pi\thetc^2} Q_3  \gamma_{{\rm t},2}
  + \frac{\sigma^2_{\epsilon}}{2}
  \sum_{i} \sum_{j\ne i}  w_i^2 w_j^2
  \int \frac{\diff^2\bthet_{1}}{\pi\thetc^2}  Q_1^2\gamma_{{\rm t},1}^2\int \frac{\diff^2\bthet_{2}}{\pi\thetc^2}  Q_2^2
  \nn \\
  & =
  \frac{\sigma^2_{\epsilon}}{2(\pi\thetc^2)^4}
  \sum_{i} \sum_{j\ne i} \sum_{k\ne i \ne j} 
  w_i w_j^2 w_k {\mathcal M}^2_{\rm ap}G
  +\frac{\sigma^2_{\epsilon}}{2(\pi\thetc^2)^4}
  \sum_{i} \sum_{j\ne i}  w_i^2 w_j^2 {\mathcal M}_{\rm s,2}G \ ,\label{eq:varest13}
\end{align}
If we repeat the above calculation for the other 3 remaining terms
$\circled{3b}$--$\circled{3d}$, then we see that they yield exactly
the same answer. This means that when summing these contributions we
simply multiply our answer by a factor of 4. Finally, the expectation
over the ensemble of realisations of the shear field, with the
normalisation factors restored, gives us:
\begin{align}
  {\rm E}\left[\circled{3a}_{\rm full}+\dots+\circled{3d}_{\rm full}\right]
  & = 
  \frac{2\sum_{i} \sum_{j\ne i} \sum_{k\ne i \ne j}
  w_i w_j^2 w_k}{\left(\sum_{i}\sum_{j\ne i}w_iw_j\right)^2} \sigma^2_{\epsilon} {\mathcal M}^2_{\rm ap}G
  +
  \frac{2\sum_{i} \sum_{j\ne i}  w_i^2 w_j^2}{\left(\sum_{i}\sum_{j\ne i}w_iw_j\right)^2}
  \sigma^2_{\epsilon}{\mathcal M}_{\rm s,2}G \ .\label{eq:varest14}
\end{align}

On summing all of the contributions to the variance that come from
terms \Eqn{eq:varest4}, \Eqn{eq:varest6}, \Eqn{eq:varest8},
\Eqn{eq:varest12} and \Eqn{eq:varest14} we see that the weighted variance
of the estimator for the aperture mass variance can be written:
\begin{align}
	{\rm Var}\left[\widehat{{\mathcal M}_{\rm ap}^2(\thetc)}\right]
	&= \frac{\sum_{i}\sum_{j\ne i}\sum_{k\ne j\ne i}\sum_{l\ne k\ne j\ne i}w_i w_j w_k w_l}{\left(\sum_{i}w_i\sum_{j\ne i}w_j\right)^2}
        \left<{\mathcal M}_{\rm ap}^4 \right> 
        + 4\frac{\sum_{i}\sum_{j\ne i}\sum_{k\ne j\ne i}w_i^2 w_j w_k}{\left(\sum_{i}w_i\sum_{j\ne i}w_j\right)^2}
        \left<{\mathcal M}_{\rm ap}^2{\mathcal M}_{\rm s,2}\right>  \nn \\
        & + 2\frac{\sum_{i} w_i^2 \sum_{j\ne i} w_j^2 }{\left(\sum_{i}w_i\sum_{j\ne i}w_j\right)^2}
        \left<{\mathcal M}_{\rm s,2}^2\right>
        + \frac{2\sum_{i} w_i \sum_{j\ne i} w_j^2 \sum_{k\ne i \ne j} w_k}{\left(\sum_{i}w_i\sum_{j\ne i} w_j\right)^2}        
        \sigma^2_{\epsilon}G\left<{\mathcal M}_{\rm ap}^2\right> \nn \\
        & +\frac{2\sum_{i} \sum_{j\ne i}  w_i^2 w_j^2}{\left(\sum_{i}\sum_{j\ne i}w_iw_j\right)^2}
        \sigma^2_{\epsilon}G\left<{\mathcal M}_s^2\right>
        + \frac{\sum_{i} w_i^2 \sum_{j\ne i} w_j^2}{2\left(\sum_{i}w_i\sum_{j\ne i} w_j\right)^2}\sigma^4_{\epsilon}G^2
        -\left<{\mathcal M}_{\rm ap}^2\right>^2 \ .  \label{eq:varest15b}
\end{align}
%


\section{Mask maps}\label{app:masks}

\begin{figure}[H]
  \centering {
    \includegraphics[width=7cm]{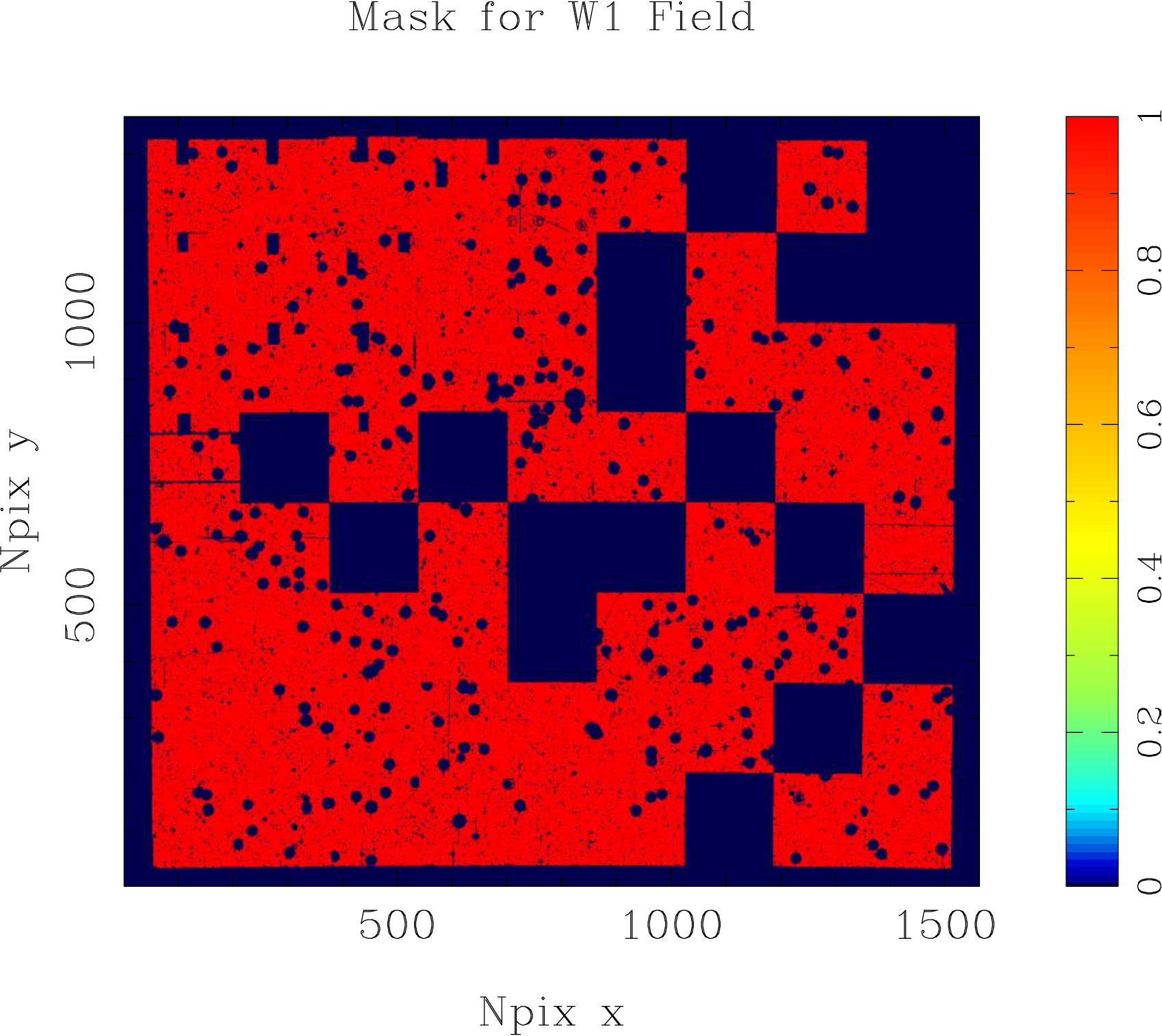}\hspace{0.2cm}
    \includegraphics[width=7cm]{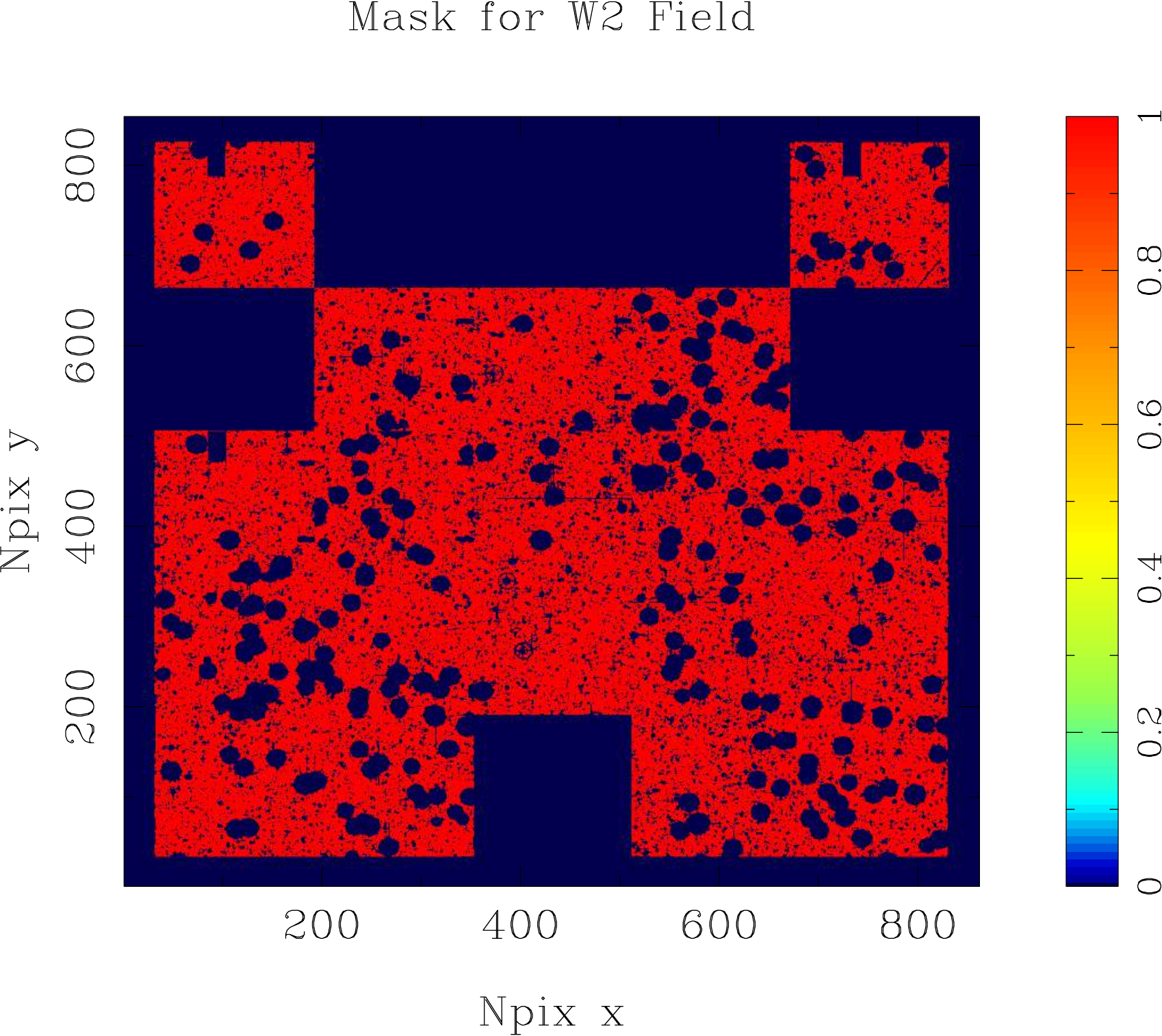}}
  \vspace{0.2cm}
  \centering {
    \includegraphics[width=7cm]{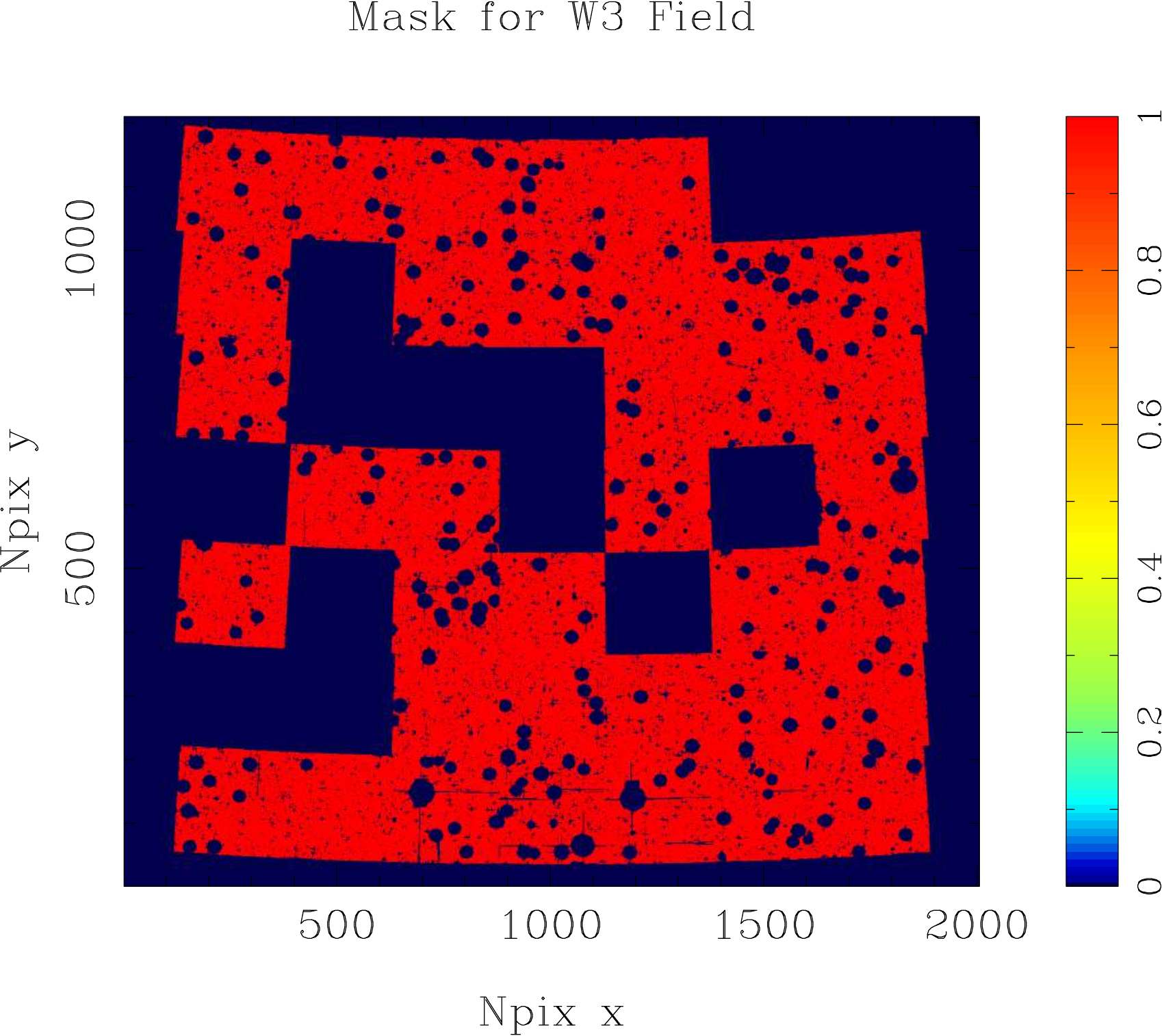}\hspace{0.2cm}
    \includegraphics[width=7cm]{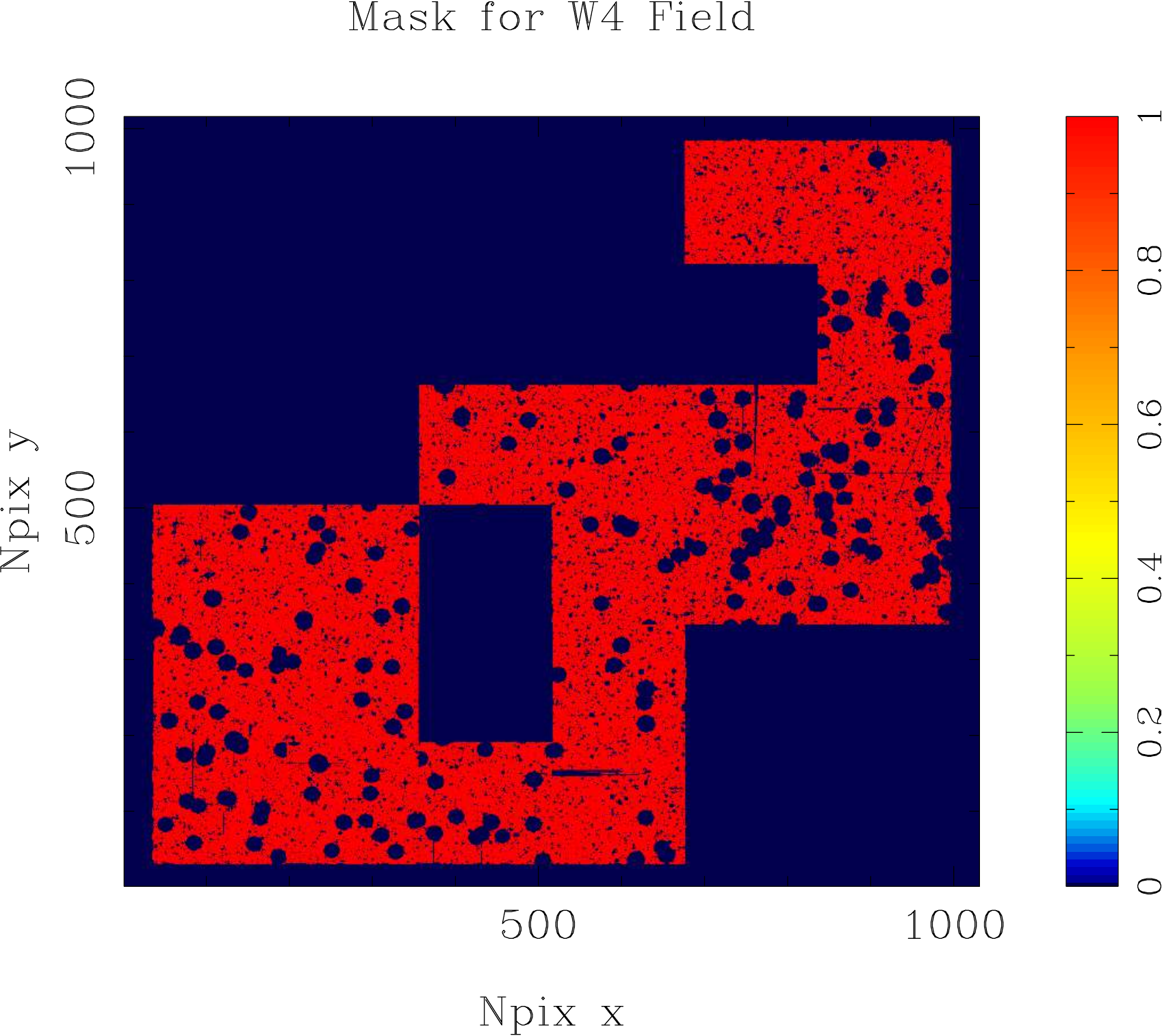}}
  \caption{\small{The pixel masks of the \cfhtl. The top
      left, top right, bottom left and bottom right show the masks for
      the W1, W2, W3 and W4 fields, respectively. The red area shows
      the observed footprint.  }\label{fig:masks}}
\end{figure}

In Figure~\ref{fig:masks} we show the survey masks for the W1, W2, W3
and W4 fields of the \cfhtl. This figure clearly illustrates
the problem with incomplete sky coverage due to the survey
boundaries, the holes drilled due to the diffraction effects of bright
stars and the gaps between chips. One can also notice that, while the
W1, W2 and W4 fields are fairly flat projections, the W3 field clearly
suffers more from the effects of the curved geometry of the sky.

\newpage

\section{Maps for the W2-W4 \cfhtls fields}\label{app:CFHTLOtherMaps}

In the following we show aperture maps for the remaining \cfhtls fields. The row are identically structured as in \Figs{fig:massmap1}{fig:massmap2}, but we only show the signal-to-noise map in the left column and the inverse variance weight map in the right column.

\newpage

\begin{figure*}
  \centering {
    \includegraphics[width=8.0cm]{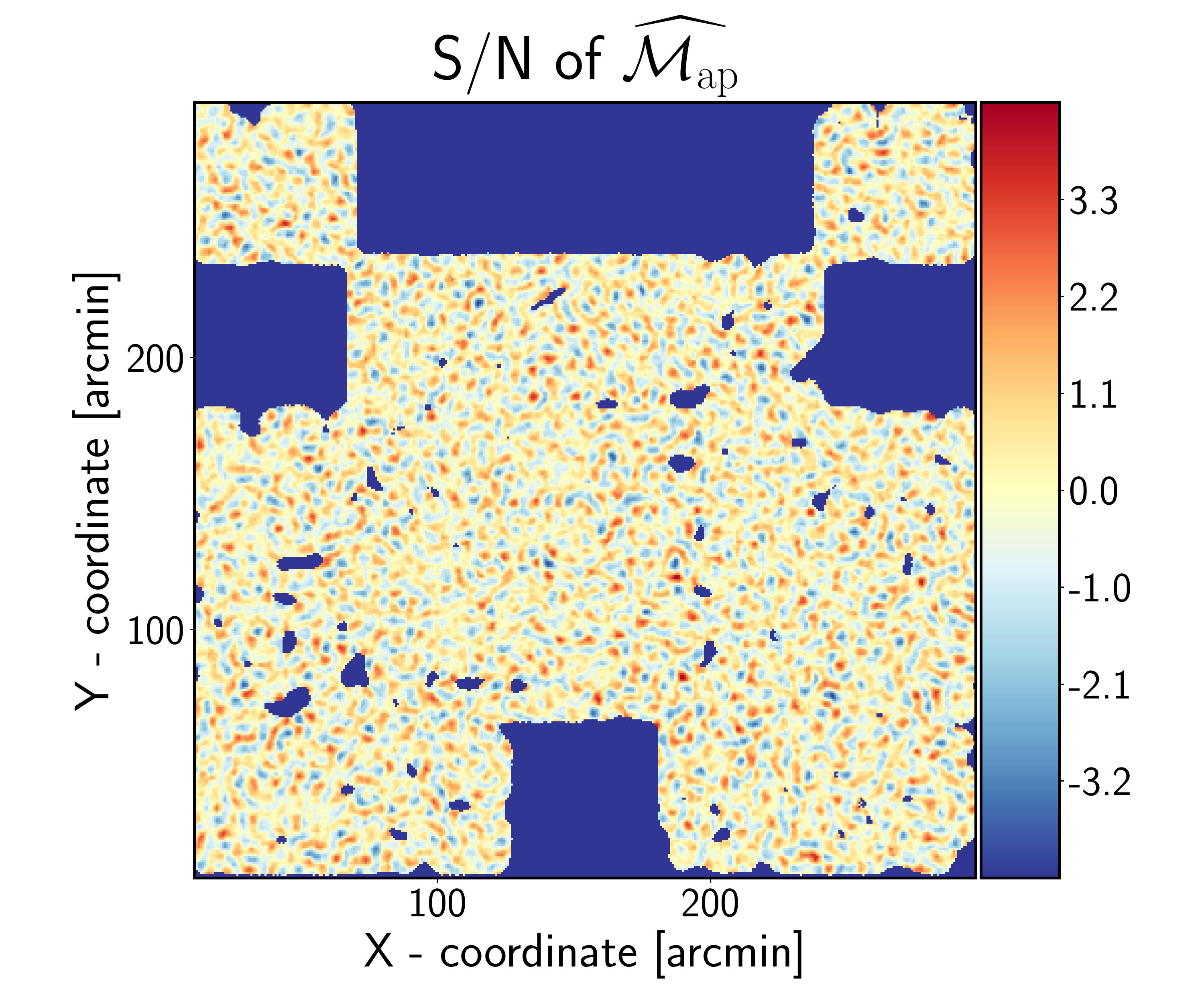}\hspace{0.2cm}
    \includegraphics[width=8.cm]{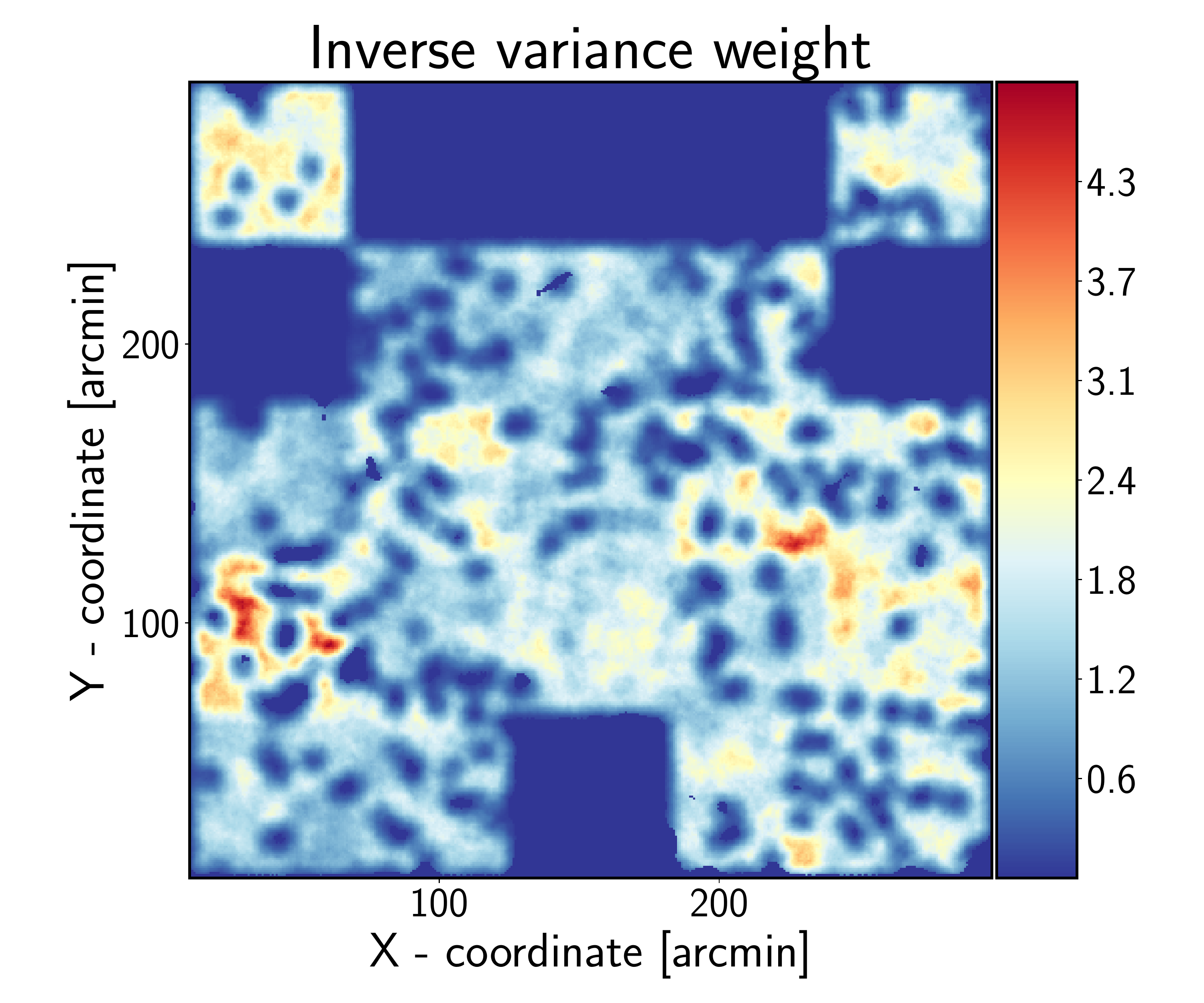}}
  \centering {
    \includegraphics[width=8.0cm]{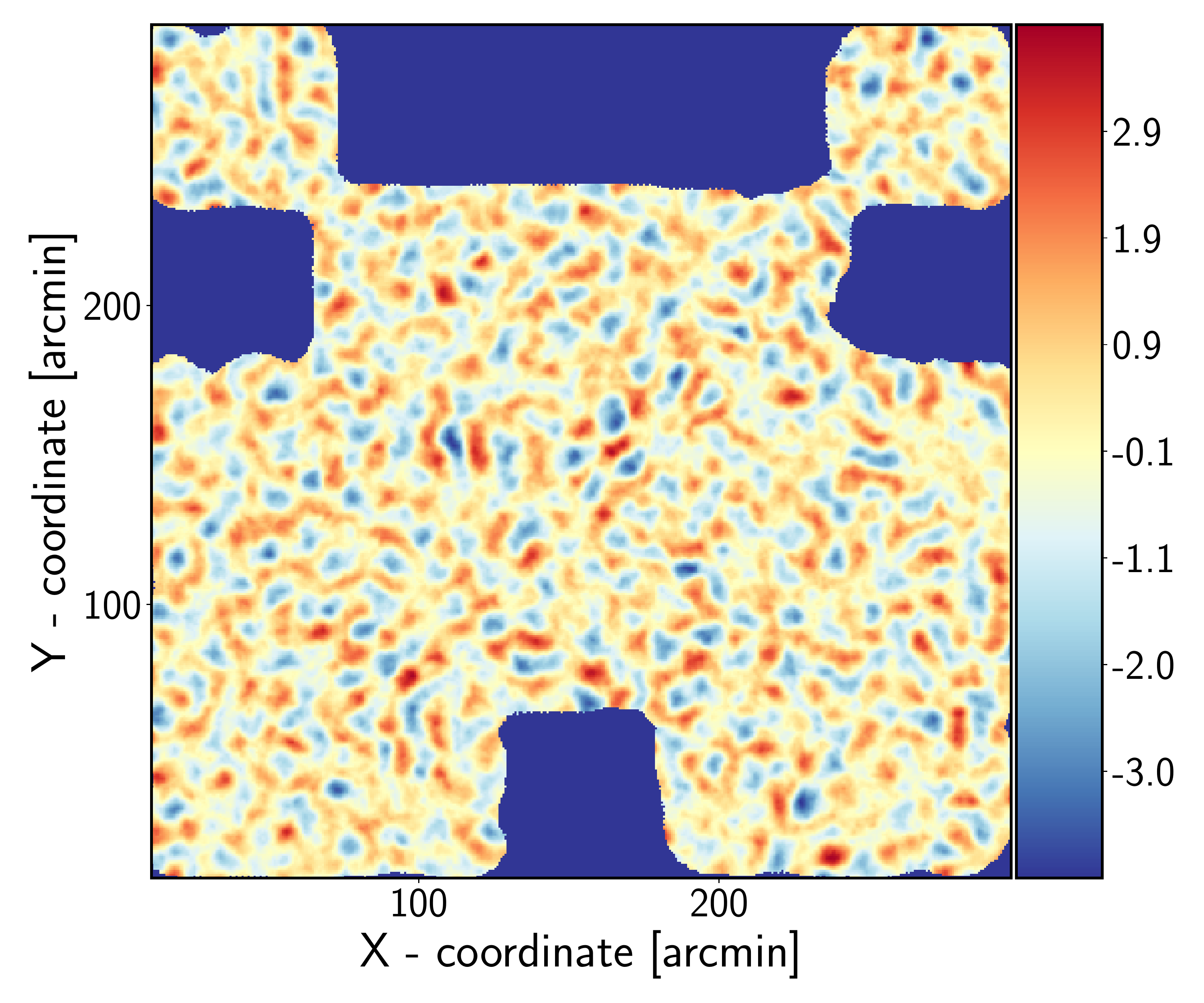}\hspace{0.2cm}
    \includegraphics[width=8.0cm]{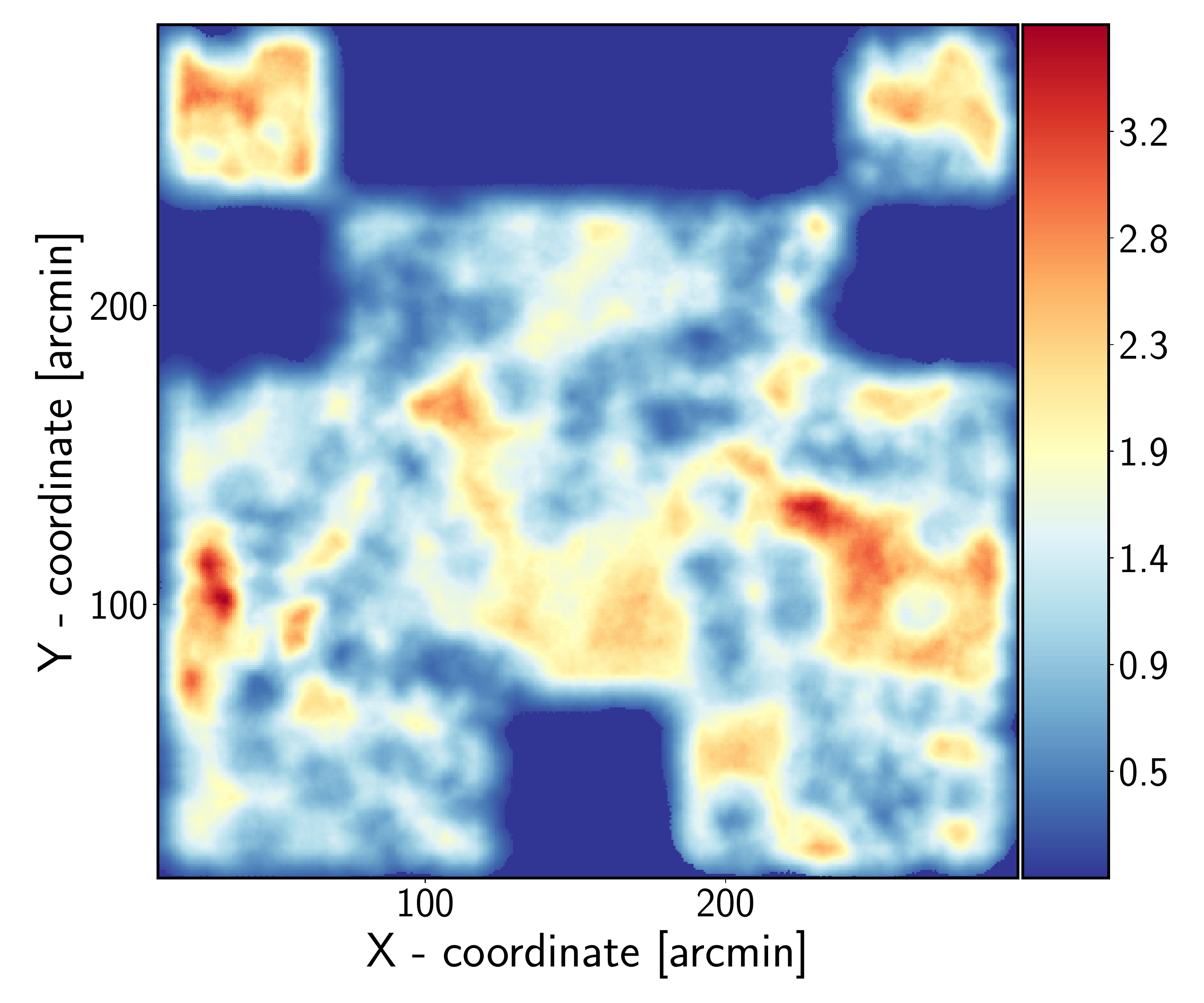}}
  \centering {
    \includegraphics[width=8.0cm]{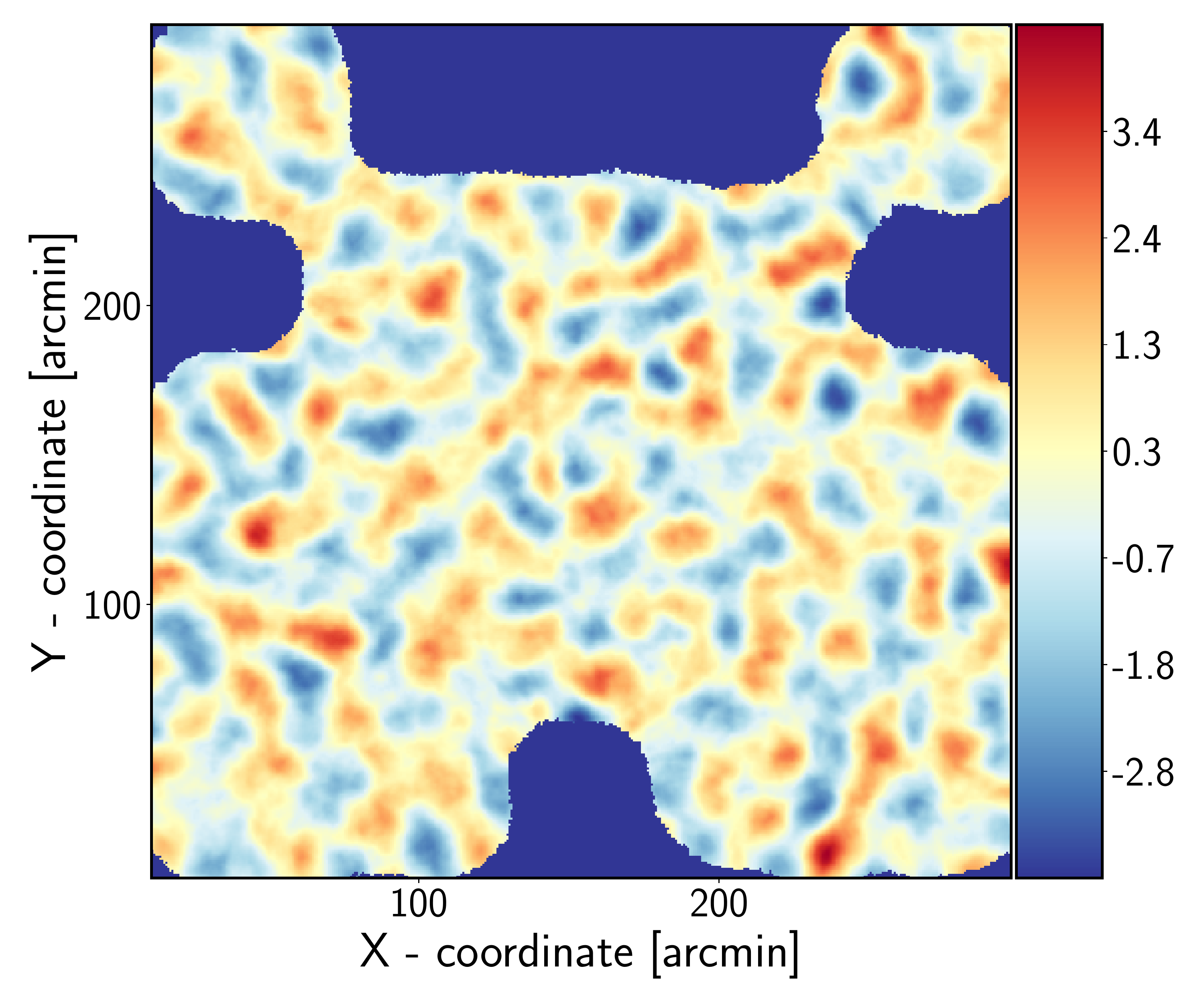}\hspace{0.2cm}
    \includegraphics[width=8.0cm]{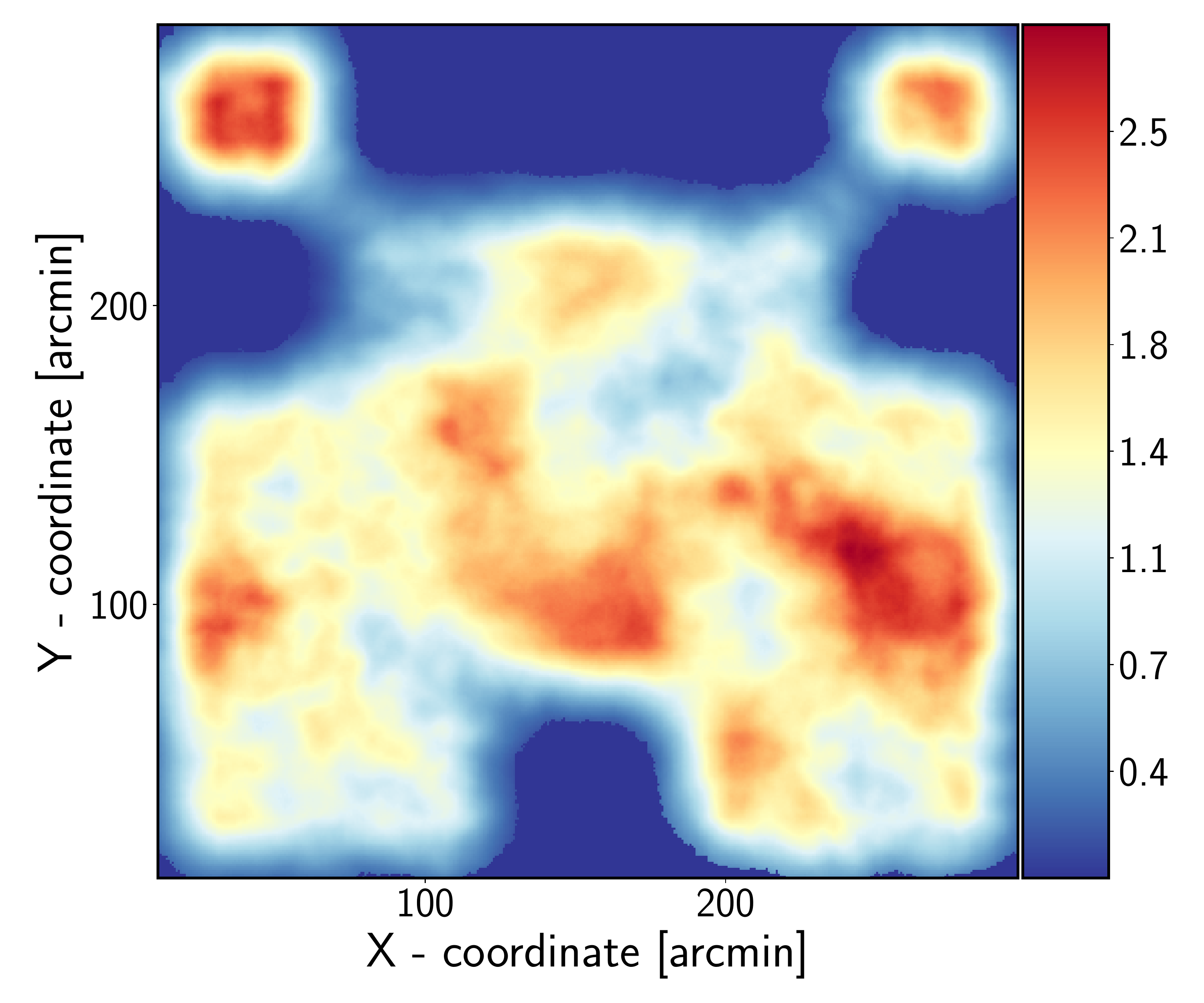}}
      \caption{\small{Signal-to-noise and aperture weight map of the W2 field of the \cfhtls
          data.}\label{fig:MapMapsW2}}
\end{figure*}

\begin{figure*}
  \centering {
    \includegraphics[width=8.5cm]{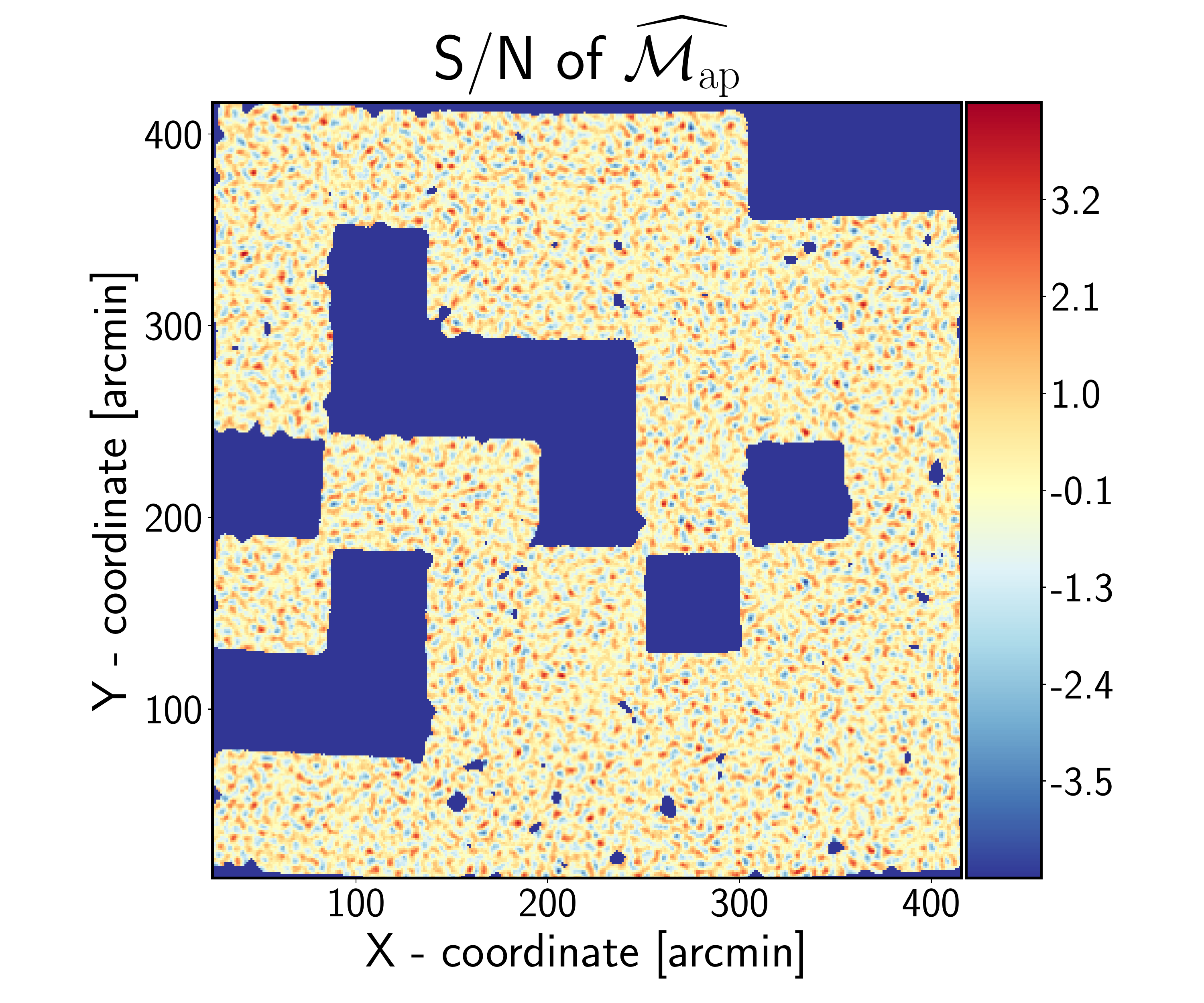}\hspace{0.2cm}
    \includegraphics[width=8.5cm]{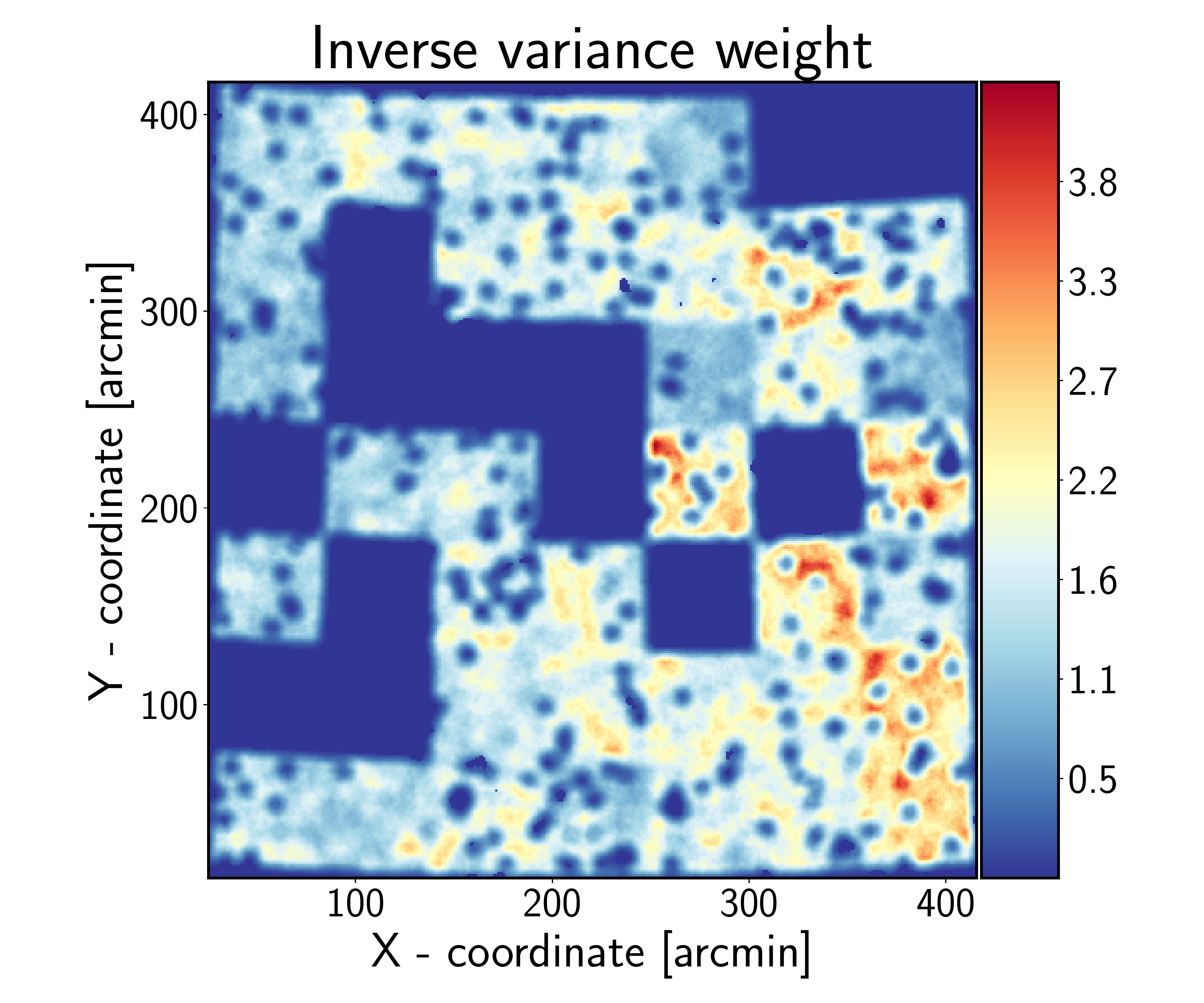}}
  \centering {
    \includegraphics[width=8.5cm]{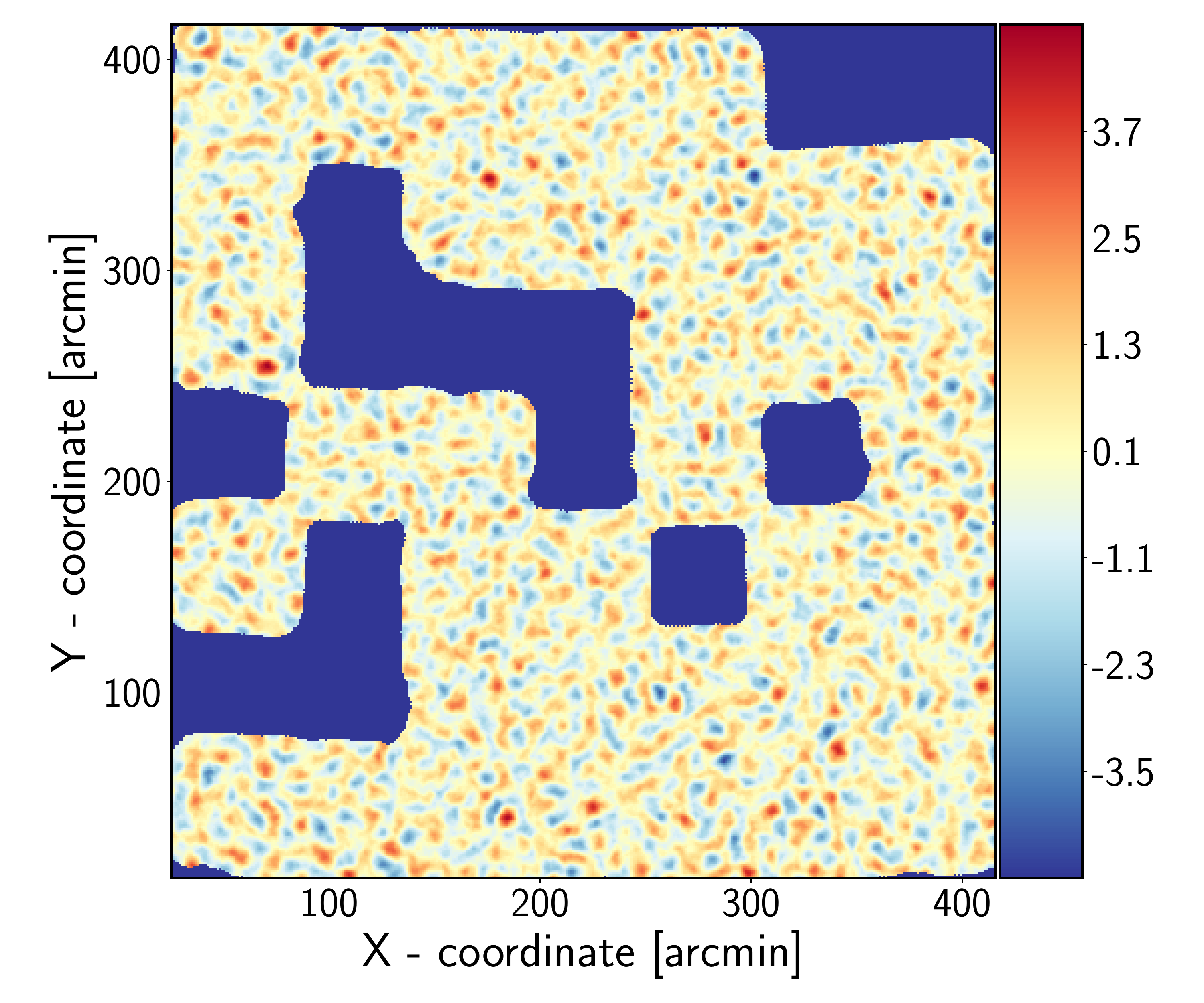}\hspace{0.2cm}
    \includegraphics[width=8.5cm]{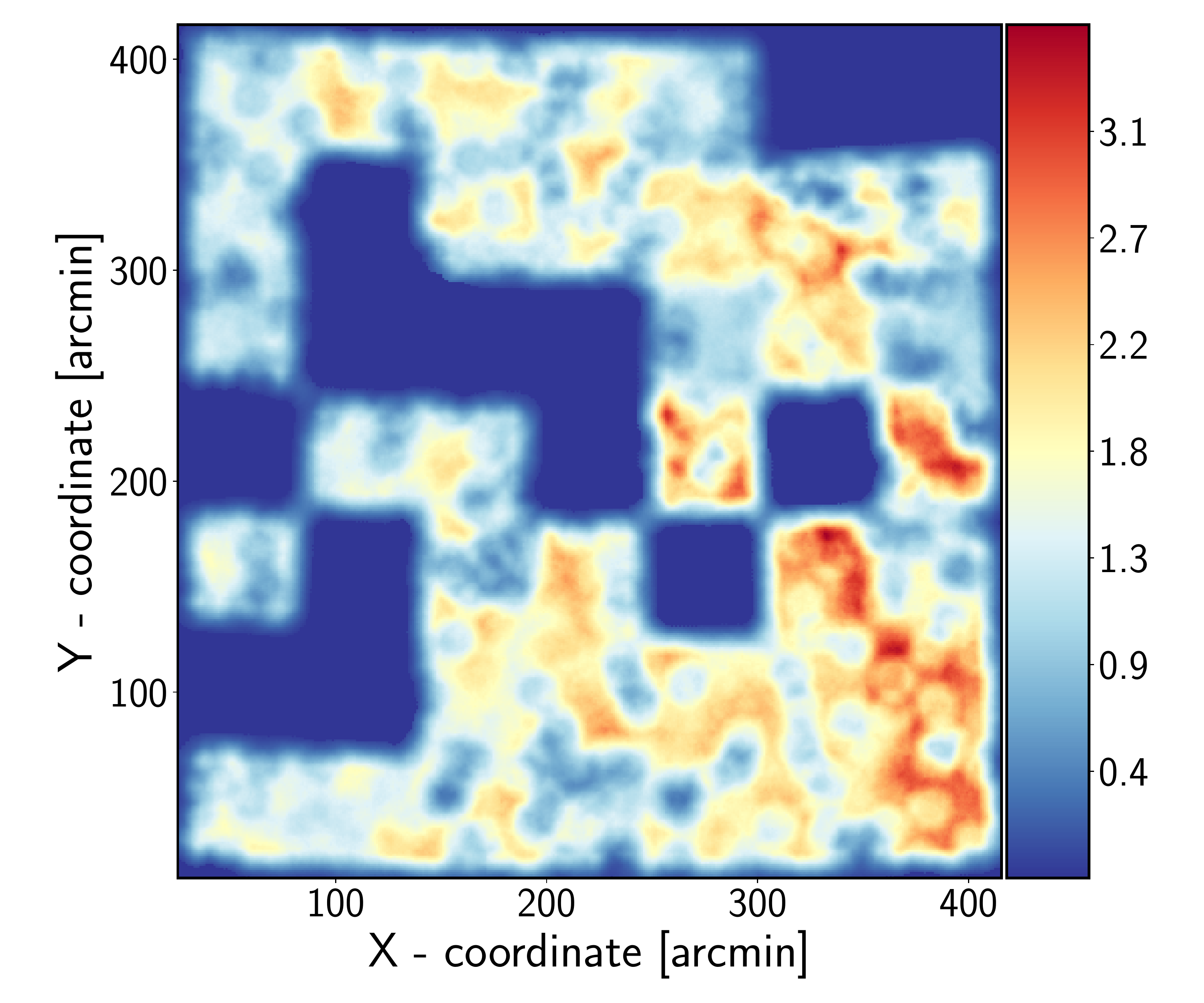}}
  \centering {
    \includegraphics[width=8.5cm]{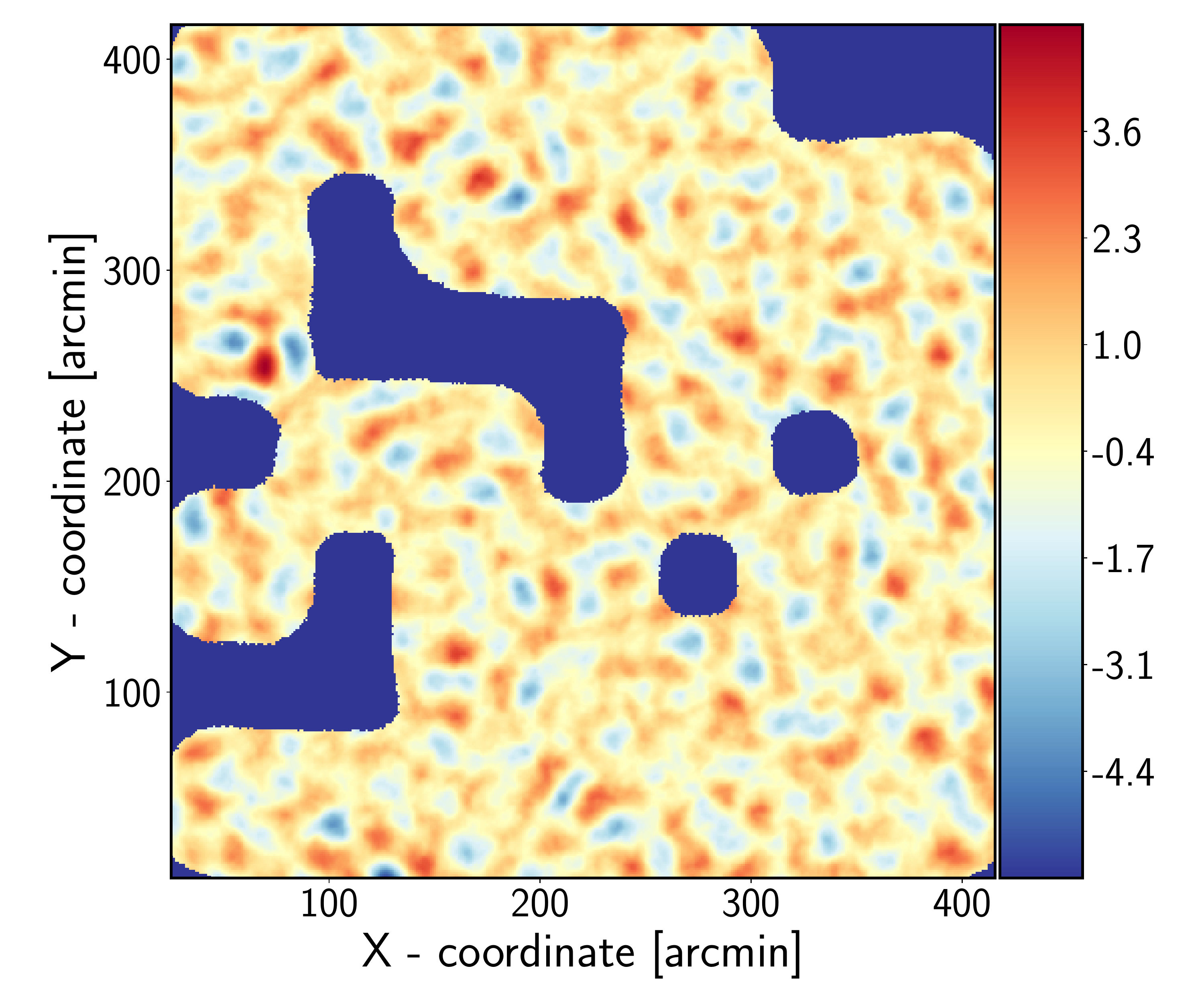}\hspace{0.2cm}
    \includegraphics[width=8.5cm]{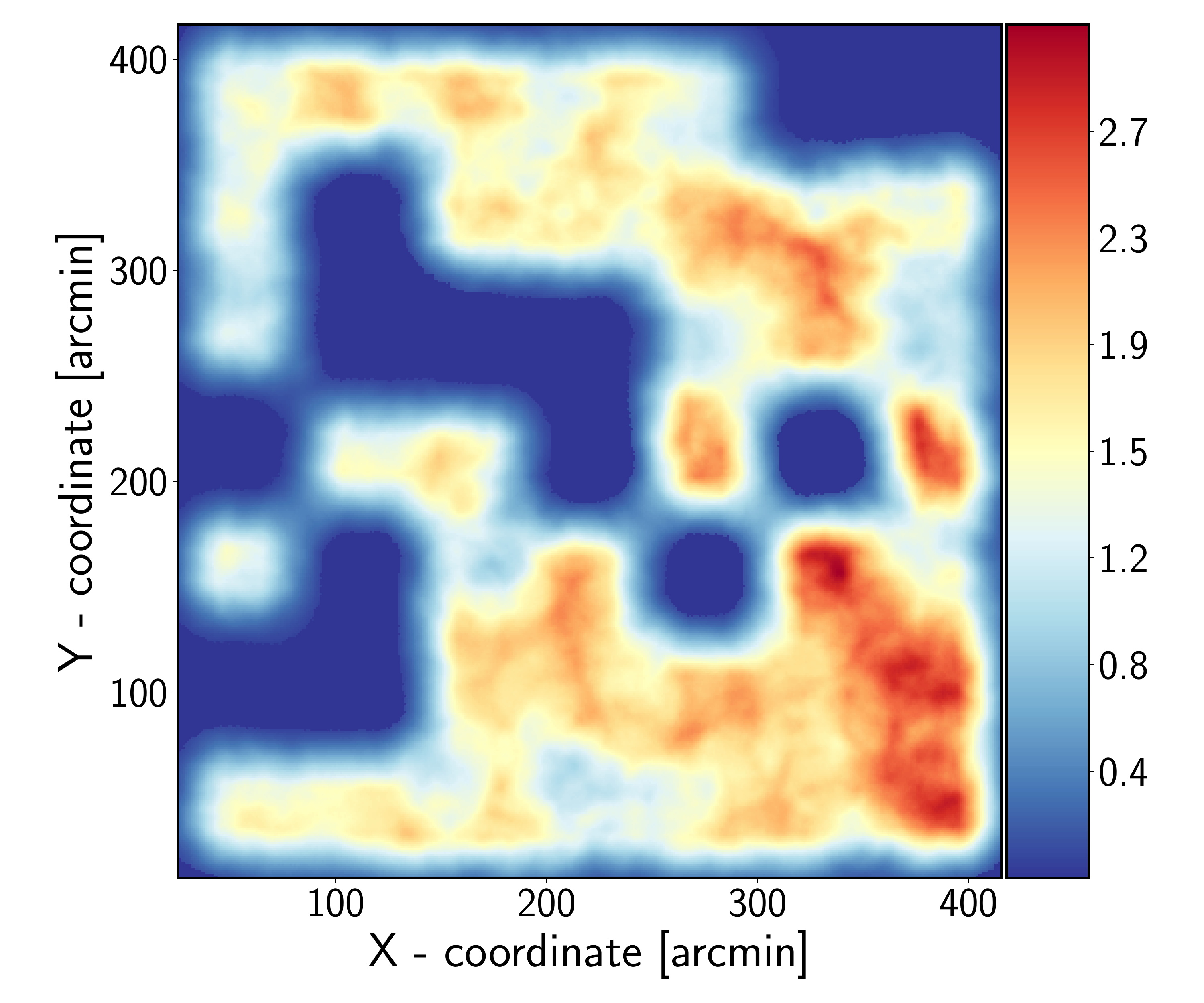}}
      \caption{\small{Signal-to-noise and aperture weight map of the W3 field of the \cfhtls
          data.}\label{fig:MapMapsW3}}
\end{figure*}

\newpage

\begin{figure*}
  \centering {
    \includegraphics[width=8.5cm]{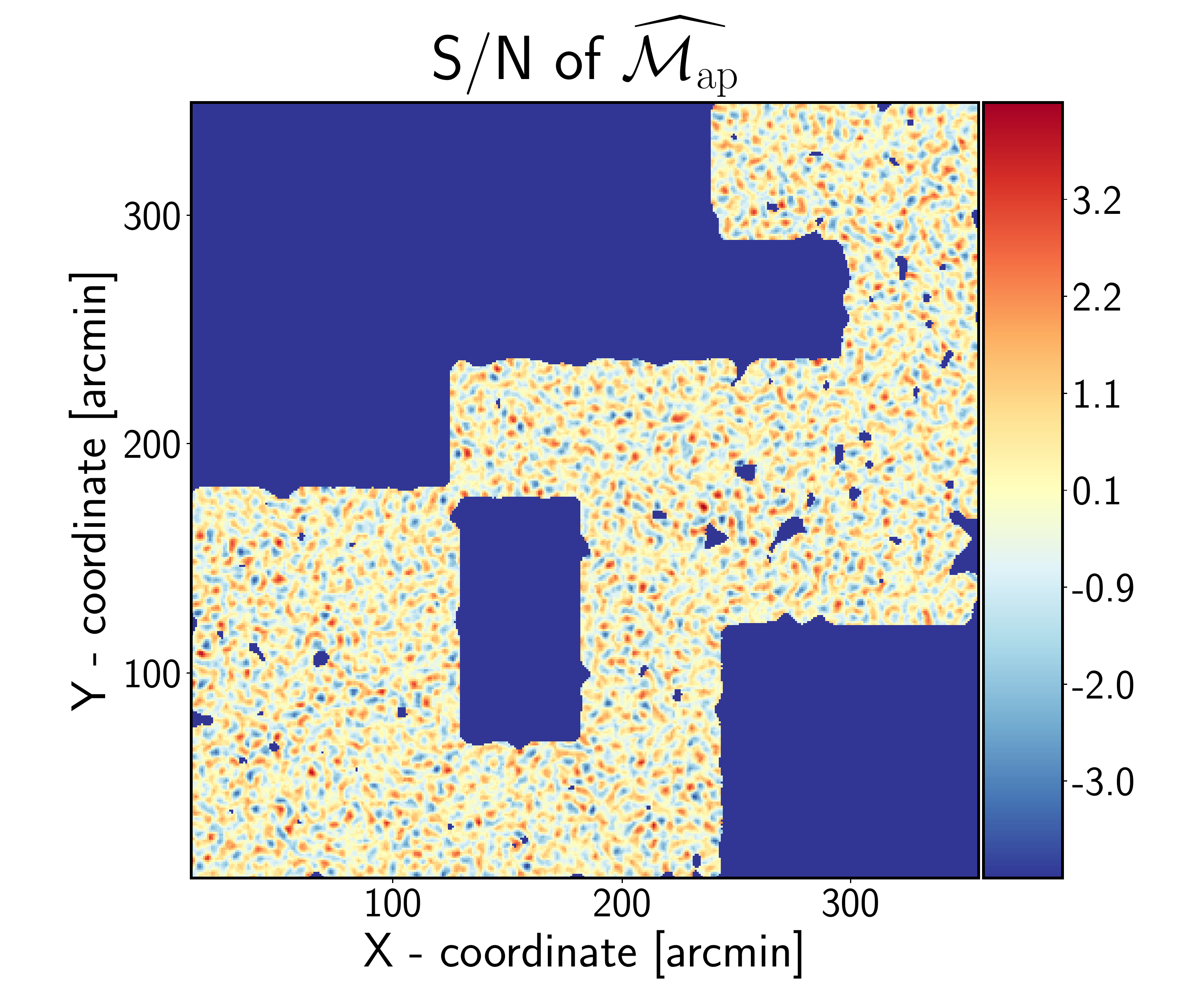}\hspace{0.2cm}
    \includegraphics[width=8.5cm]{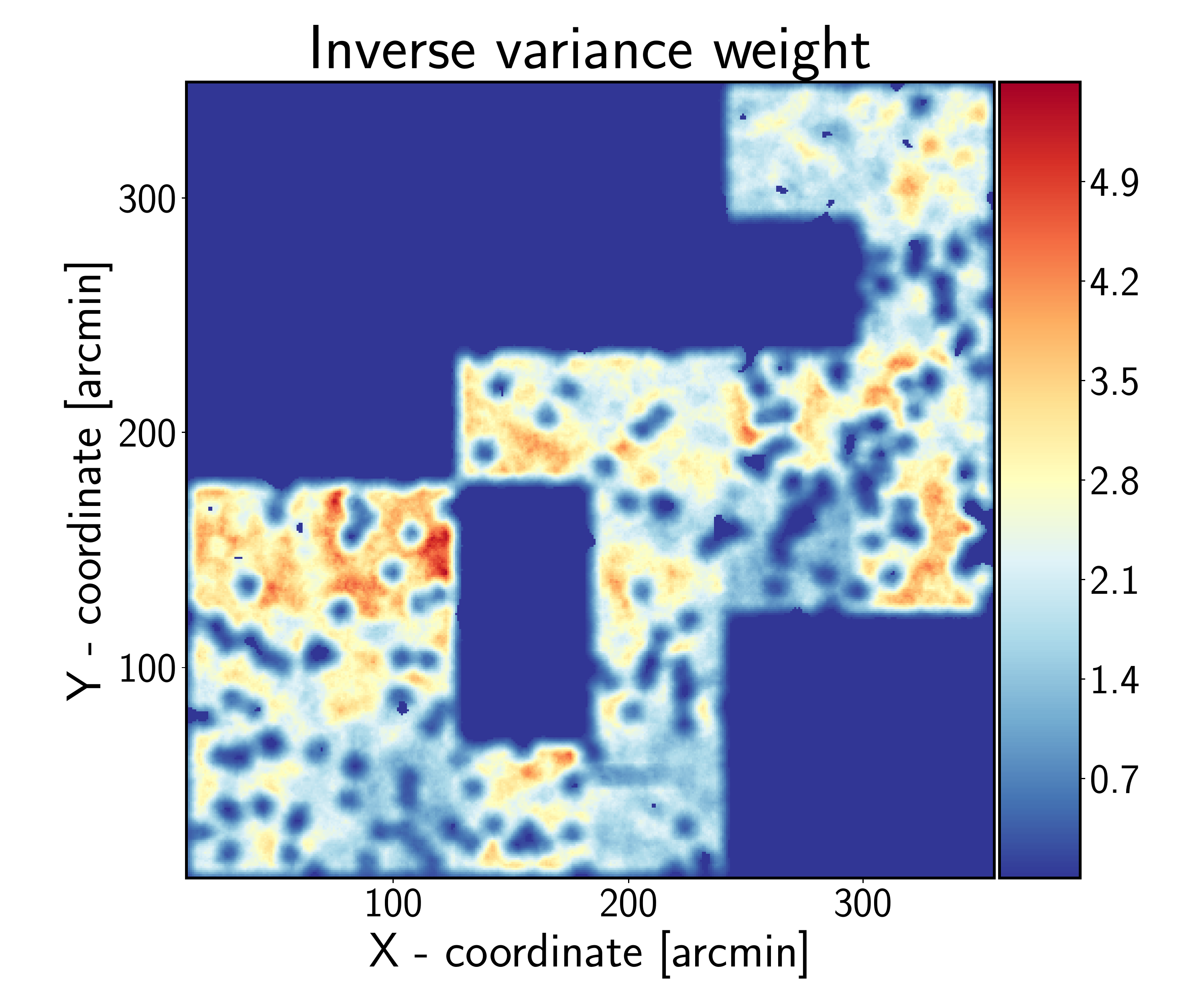}}
  \centering {
    \includegraphics[width=8.5cm]{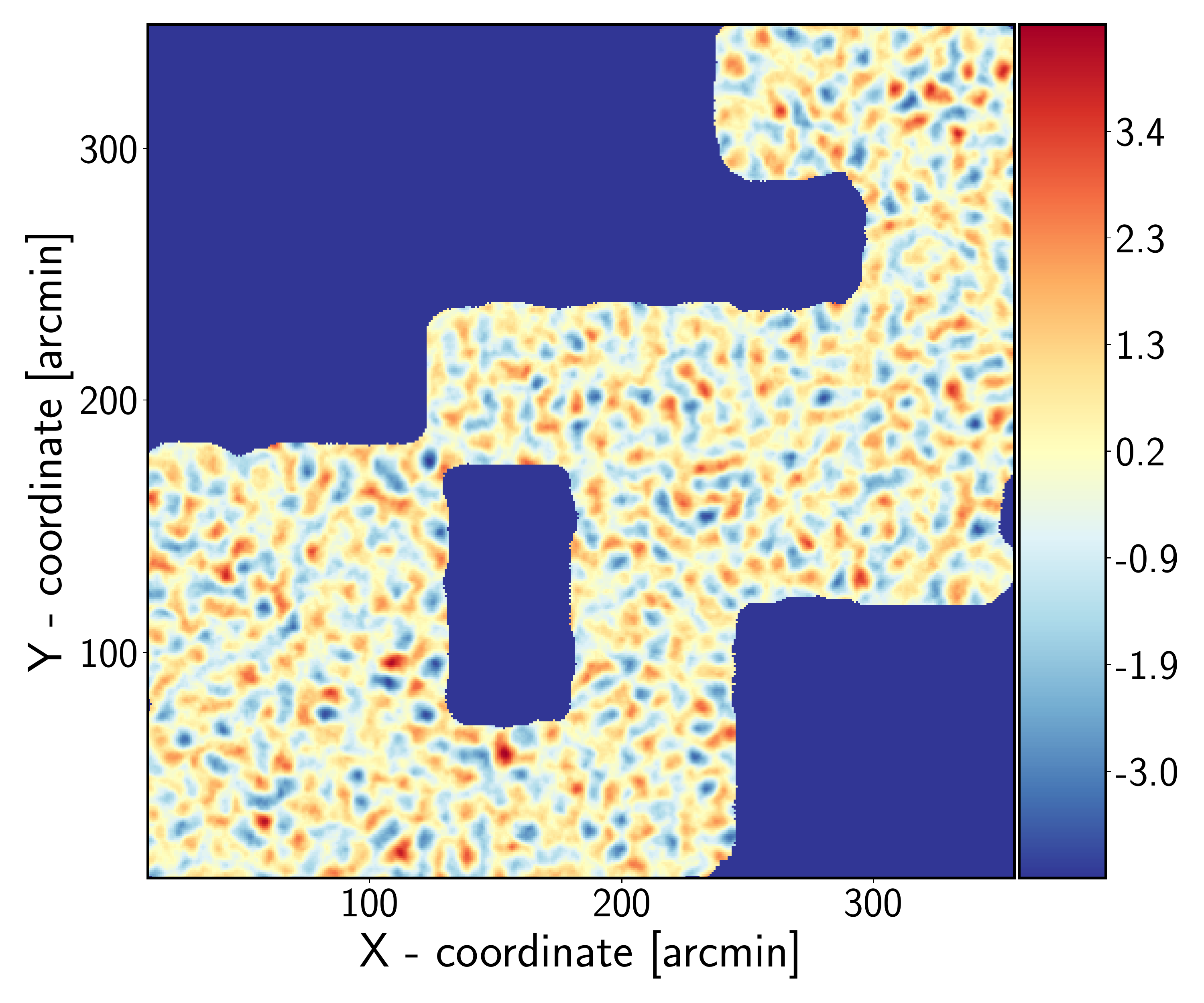}\hspace{0.2cm}
    \includegraphics[width=8.5cm]{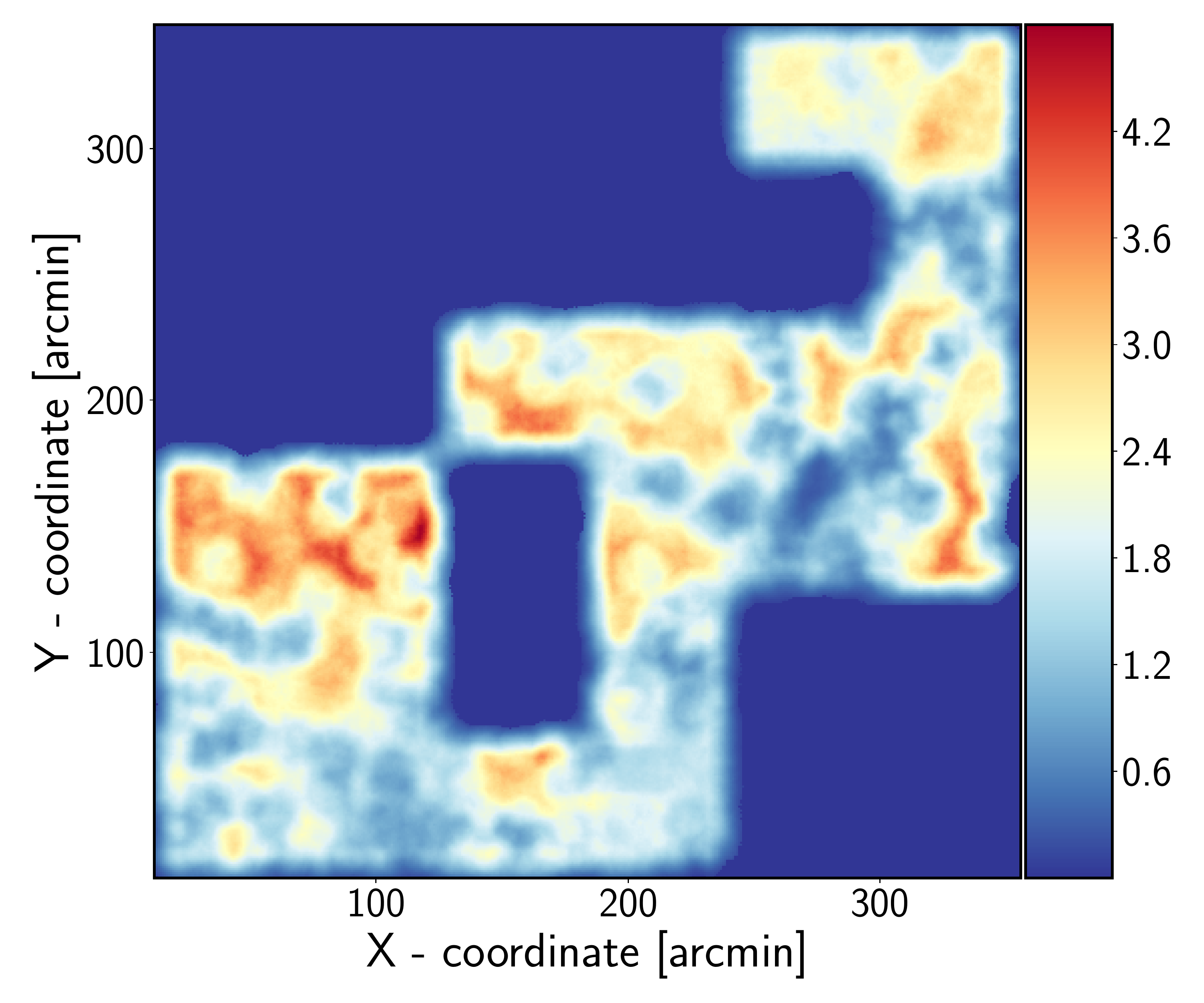}}
  \centering {
    \includegraphics[width=8.5cm]{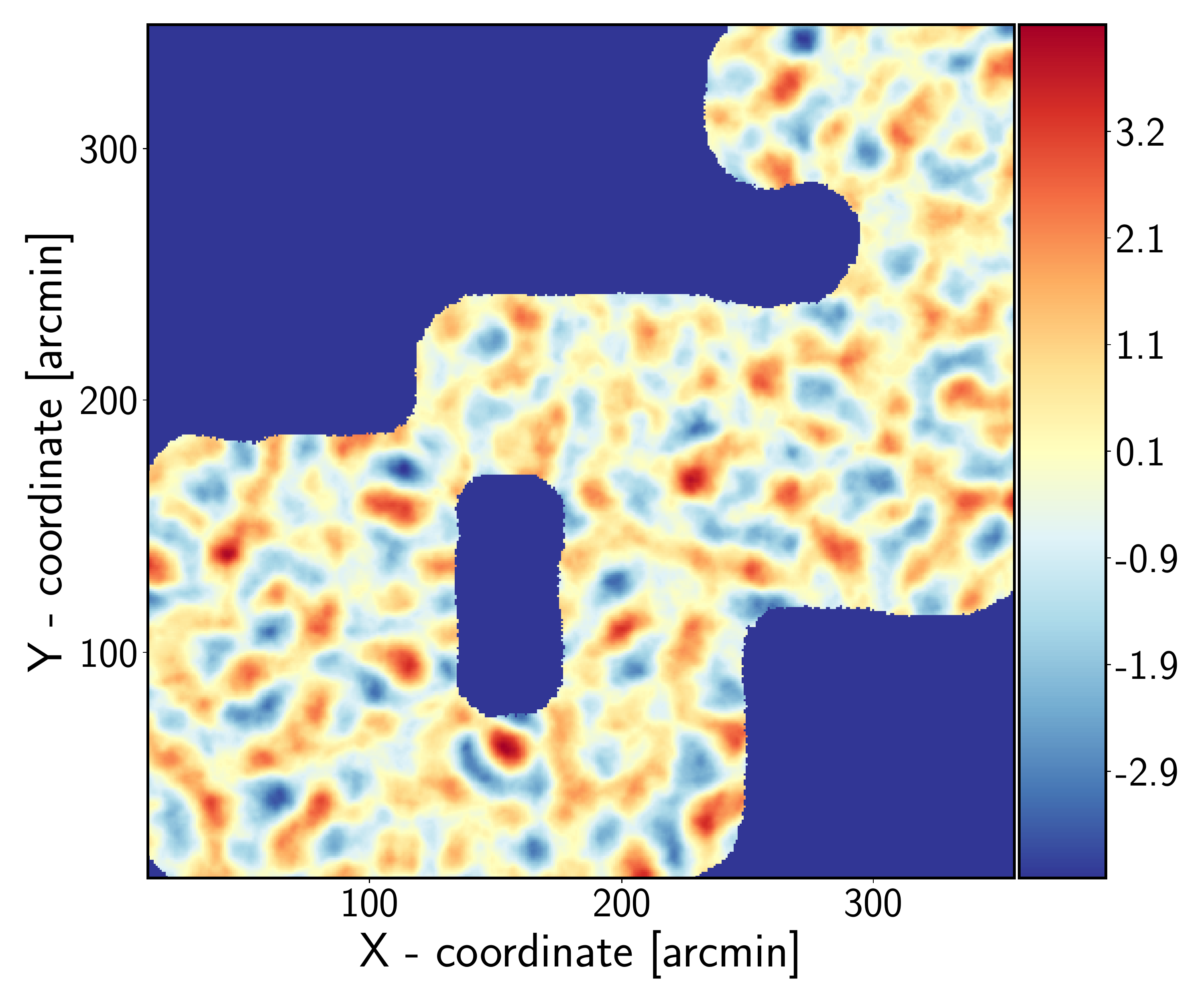}\hspace{0.2cm}
    \includegraphics[width=8.5cm]{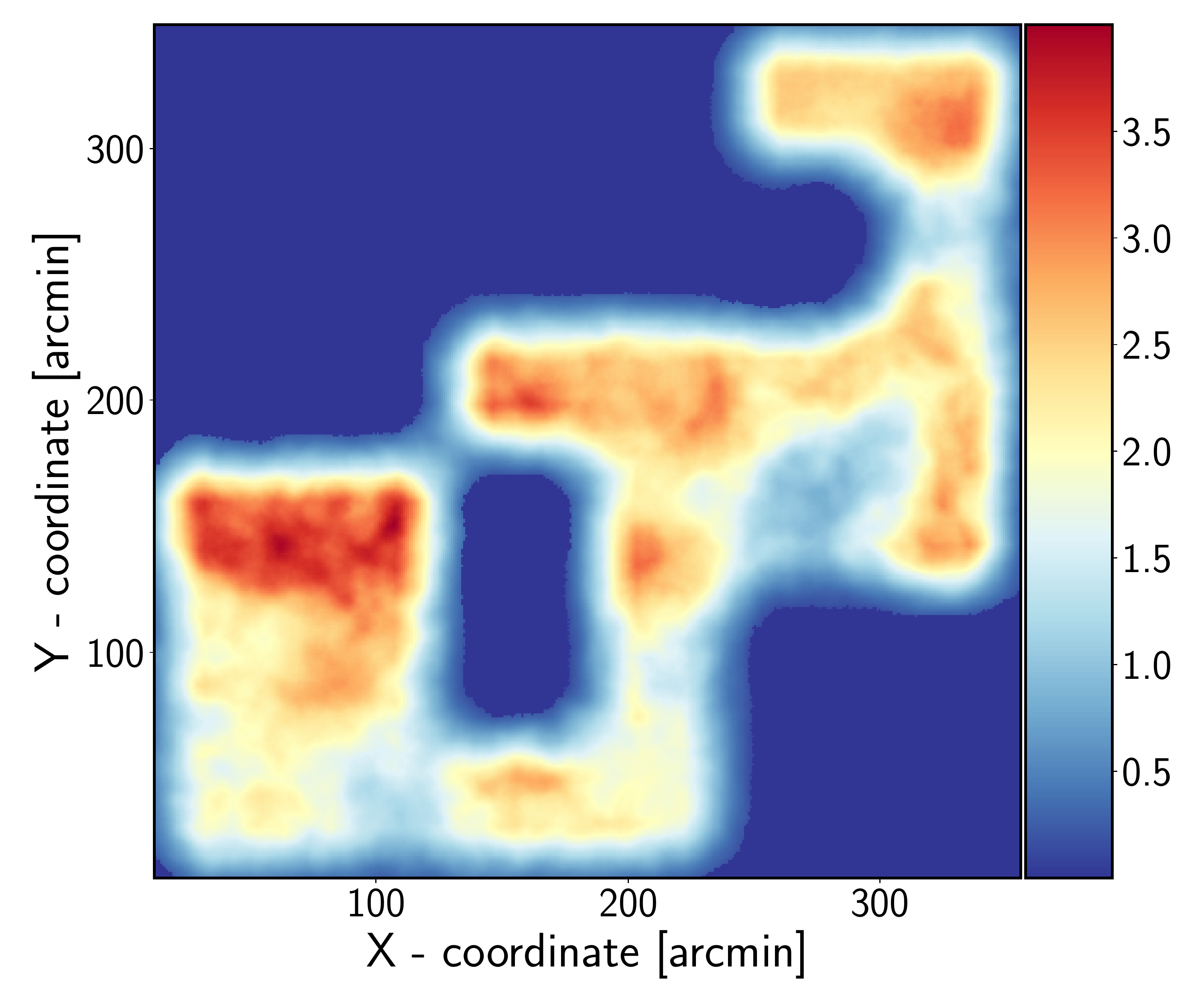}}
      \caption{\small{Signal-to-noise and aperture weight map of the W4 field of the \cfhtls
          data.}\label{fig:MapMapsW4}}
\end{figure*}


\label{lastpage}

\end{document}

%% file: defs.tex
\def\cfhtl{\mbox{CFHTLenS}}
\def\cfhtls{\mbox{CFHTLenS\ }}
\def\euclid{\emph{Euclid}}
\def\kids{\mbox{KiDS}}
\def\zhorizon{\texttt{zHORIZON}\ }
\def\treecorr{\mbox{TreeCorr\ }}
\def\lensfit{\emph{lens}fit }

\def\diff{{\rm d}}
\newcommand{\dSq}[1]{\diff ^2 {\bm {#1}} }

\newcommand{\tang}[2]{{#1}_{{\rm t}, {#2}}} 

\def\zH{z_{\rm H}}

\newcommand{\degt}{\mathrm{deg}}

\newcommand*\circled[1]{\tikz[baseline=(char.base)]{
            \node[shape=circle,draw,inner sep=2pt] (char) {#1};}}

\renewcommand{\vec}[1]{{\bm{#1}}}

\def\Map{\mathcal{M}_{\rm ap}}
\def\Mx{\mathcal{M}_{\times}}
\def\MapSq{\left<\mathcal{M}_{\rm ap}^2\right>}

\def\chiH{\chi_{\rm H}}

\newcommand\Eqn[1]     {Eq.\,(\ref{#1})}
\newcommand\Eqns[2]    {Eqs.\,(\ref{#1}) and~(\ref{#2})}

\newcommand\eqn[1]     {Eq.\,(\ref{#1})}

\newcommand\Fig[1]     {Fig.\,{\ref{#1}}}
\newcommand\Figs[2]    {Figs\,{\ref{#1}} and~{\ref{#2}}}

\newcommand\Sect[1]    {\S\ref{#1}}

\newcommand\nn         {\nonumber}

\newcommand{\be}{\begin{equation}}
\newcommand{\ee}{\end{equation}}

\newcommand{\ba}{\begin{eqnarray}}
\newcommand{\ea}{\end{eqnarray}}

\newcommand{\bes}{\begin{subequations}}
\newcommand{\ees}{\end{subequations}}

\def\ex{{\rm e}}

\def\dirac{\delta^{\rm D}}

\def\RR{{\mathbb{R}^2}}

\def\omr{\Omega_{\rm r}}
\def\oml{\Omega_{\Lambda}}
\def\omk{\Omega_{k}}

\def\omm0{\Omega_{\rm m,0}}
\def\omr0{\Omega_{\rm r,0}}
\def\oml0{\Omega_{\Lambda,0}}
\def\omk0{\Omega_{k,0}}

\def\kpc{\, h^{-1}{\rm kpc}}
\def\Mpc{\, h^{-1}{\rm Mpc}}

\def\bthet{{\bm\theta}}
\def\dthet{{{\rm d}^2\bm\theta}}

\def\dell{{{\rm d}^2\bm\ell}}
\def\bell{{\bm\ell}}

\def\1D{{\rm 1D}}
\def\2D{{\rm 2D}}
\def\3D{{\rm 3D}}

\def\by{{\bf y}}

\def\1Loop{{\rm 1Loop}}

\def\nbar{\bar{n}}

\def\wk{U}
\def\wg{Q}
\def\Map{{\mathcal M}_{\rm ap}}
\def\thetc{{\vartheta}}